\documentclass[useAMS,onecolumn,usenatbib,fleqn]{mnras}
\newif\ifAMStwofonts
\AMStwofontstrue

\usepackage{amsmath}	
\usepackage{amssymb}	
\usepackage{nccmath}

\def\pmb#1{\mbox{\boldmath$#1$}}
\def\gtsim {>\kern-1.2em\lower1.1ex\hbox{$\sim$}}
\def\ltsim {<\kern-1.2em\lower1.1ex\hbox{$\sim$}}
\def\gtsim {>\kern-1.2em\lower1.1ex\hbox{$\sim$}}
\def\ltsim {<\kern-1.2em\lower1.1ex\hbox{$\sim$}}

\def\be{\begin{equation}}
\def\ee{\end{equation}}
\def\be{\begin{eqnarray}}
\def\ee{\end{eqnarray}}

\def\Re{{\rm Re}}
\def\Im{{\rm Im}}

\usepackage{color}

\def\pmbmt#1{\pmb{\sf #1}}
\def\rmi{{\rm i}}
\def\rme{{\rm e}}



\usepackage{graphicx}
\usepackage{epsfig}
\begin{document}

\title{Axisymmetric second order perturbations of rotating main sequence stars}
\author[U. Lee]{ Umin Lee$^1$\thanks{E-mail: lee@astr.tohoku.ac.jp}
\\$^1$Astronomical Institute, Tohoku University, Sendai, Miyagi 980-8578, Japan}

\date{Typeset \today ; Received / Accepted}
\maketitle


\begin{abstract}
We calculate the second order perturbations driven by oscillation modes of rotating stars.
Assuming that the typical amplitude $a$ of oscillation modes is small,
we expand the perturbed quantities as $Q=Q^{(0)}+Q^{(1)}+Q^{(2)}+\cdots$ where 
$Q^{(0)}$ represents the equilibrium state and $Q^{(1)}$ and $Q^{(2)}$ are 
the first order and second order perturbations in $a$, respectively.
We assume that the first order perturbations are given by non-axisymmetric modes and the second order perturbations are axisymmetric.
For the second order perturbations, we derive a set of linear partial differential equations,
which have inhomogeneous terms due to the first order perturbations.
For low frequency $g$- and $r$-modes and overstable convective (OsC) modes of main sequence stars, we calculate the second order velocity field $\pmb{v}^{(2)}$ 
and find that prograde $g$-modes and OsC modes tend to accelerate and retrograde $r$-modes to decelerate $v_\phi^{(2)}$ in the surface equatorial regions where $v_\phi^{(2)}$ is the $\phi$ component of $\pmb{v}^{(2)}$.
Using the angular momentum conservation equation derived for waves,
we discuss that low frequency $g$- and $r$-modes transport angular momentum between
the inner and outer parts of the envelope. 
For OsC modes in the core resonantly coupled with envelope prograde $g$-modes,
we find that they can transport angular momentum from the core to the outer envelope so that
they tend to brake the core rotation.
We also suggest that the OsC modes provide the outer envelope of rotating stars with the torque enough to support a decretion disc.

\end{abstract}



\section{Introduction}

Angular momentum transport in rotating stars has long been one of the important topics studied in stellar physics (e.g., \citealt{AertsMathisRogers19}).
Carrying out evolution calculation of rotating stars, \citet{Quazzanietal2019} have recently attempted the confrontation between theory and observations for $\gamma$ Doradus stars where they employed the formalism given by \citet{Zahn1992} and \citet{MaederZahn1998} to follow the evolution of angular rotation velocity in the stars.
For example, \citet{MaederZahn1998} employ an advection-diffusion equation,
which takes account of the effects of meridional circulation for advection and of turbulent viscosity for diffusion of angular momentum.
\citet{Quazzanietal2019} found that the theory does not necessarily well explain observations, and they
suggested the existence of missing mechanisms for the braking of the core rotation before and along the main sequence evolution of stars.

Low frequency waves are known to be an efficient carrier of angular momentum in rotating stars. 
For example, \citet{Zahn1975,Zahn1977} assumed angular momentum transport by internal gravity waves excited in the stars in a binary system and argued that gravity waves excited by the tidal force due to the companion star propagate and damp in the interior, leading to synchronization between the stellar rotation and the orbital motion and to circularization of the binary orbit, by exchanging angular momentum between the waves and flows.
\citet{Zahn1975} calculated the time derivative of rotation speed of the stars estimating the torque onto the stars in a binary system.
\citet{LeeSaio93} discussed angular momentum transport by overstable convective (OsC) modes of rotating massive main sequence stars.
The OsC modes in the core excite low frequency $g$-modes in the envelope and are expected to transport angular momentum from the core to the surface (see also \citealt{Lee2021}).
Taking account of viscous effects associated with differential rotation,
\cite{TalonKumarZahn02} calculated shear layer oscillations in the radiative core of the sun.
They argued that
$g$-modes excited at the bottom of the convective envelope by the turbulent convection propagate inwards
carrying angular momentum to deposit in the radiative interior as the wave energy damps
because of radiative dissipation and viscous processes.
Hydrodynamical numerical 2D simulations have also been carried out by \citet{RogersGlatzmaier06} for solar $g$-modes which are excited by the convective plumes at the bottom of
the envelope convection zone and propagate inward to deposit excessive angular momentum to the mean flows.
More recently, \citet{TownsendGoldsteinZweibel18} carried out stellar evolution calculation of slowly pulsating B type stars assuming that angular frequency of rotation in the interior changes as a result of angular momentum transport by a single unstable (self-excited) prograde $g$-mode from the surface region to the deep interior.
They also took account of diffusive angular momentum transport due to convection, overshooting, and rotational instabilities for the evolution calculations.

There are several expressions used to describe angular momentum transport by waves and oscillation modes in rotating stars (e.g., \citealt{Ando83,LeeSaio93,KumarTalonZahn1999,Mathis09,Lee13,TownsendGoldsteinZweibel18}).
For example, using the Euler-Lagrange formalism, \citet{KumarTalonZahn1999} derived the equation
relating the energy density and the angular momentum density of the waves.
Using the energy flux of waves, they derived the expression for angular momentum flux carried by the waves
and applied the expression to $g$-modes excited by turbulent convection in the envelope of the Sun.
\citet{Mathis09} has formulated, under the quasi-adiabatic approximation,
angular momentum transport by internal gravity waves using
the Reynolds stresses, to which he added a modification suggested by \citet{Pantillonetal07}.
\citet{Lee13} derived the angular momentum conservation equation for linear oscillation modes based on the Lagrangian mean theory (e.g., \citealt{AndrewsMcintyre1978a,Dunkerton1980,Grimshaw1984}) and emphasised that the processes of angular momentum deposition to and extraction from the fluids are closely related to the thermal processes associated with the mode excitation and damping. 
\citet{TownsendGoldsteinZweibel18} considered the contributions from the Reynolds stresses to describe angular momentum transport by non-adiabatic $g$-modes assuming that the effects of rotation on the oscillation modes are negligible for slowly rotating stars.

For small amplitude waves, angular momentum transport takes place
as a second order process in wave amplitudes.
Such second order effects in wave amplitudes on fluid flows have long been investigated
in the field of fluid mechanics and geophysics (e.g., \citealt{AndrewsMcintyre1976a,AndrewsMcintyre1976b,AndrewsMcintyre1978a,AndrewsMcintyre1978b,Craik1985,
AndrewsHoltonLeovy1987,Buhler2014}).
We think it useful to derive, based on the second order perturbation treatment, a formula
that can be used to describe angular momentum transport in pulsating and rotating stars.
In this paper, we employ the Transformed Eulerian-Mean (TEM)
formalism (e.g., \citealt{AndrewsMcintyre1976a,AndrewsMcintyre1976b,Pedlosky1987,AndrewsHoltonLeovy1987,Buhler2014}) to derive a set of equations governing the second order perturbations which are driven by linear oscillation modes.
To derive a set of TEM equations for rotating stars, we expand the physical quantities in terms of the small parameter $a$ that represents the typical amplitudes of linear modes.
The zeroth order quantities correspond to the quantities in equilibrium.
Assuming that the first order perturbations are non-axisymmetric and the second order perturbations  axisymmetric, we employ the azimuthal averaging to pick up the second order perturbations from the basic equations of the fluid dynamics.
Here, we are interested in the second order velocity fields $\pmb{v}^{(2)}$ driven by linear oscillation modes.
In the TEM formalism, we employ the variable transformation for the meridional velocity components to derive the Eliassen-Palm flux $\pmb{F}^{\rm EP}$, which is known to play an important role in the angular momentum conservation equation
for waves.
We discuss the relation between $\partial v_\phi^{(2)}/\partial t$ and $\nabla\cdot\pmb{F}^{\rm EP}$ where $v_\phi^{(2)}$ is the $\phi$-component of  $\pmb{v}^{(2)}$.
We also derive the Lagrangian mean angular momentum conservation equation for waves to see 
the relation to the Eulerian mean equation.

In \S 2 we derive the governing equations for the second order perturbations.
\S 3 gives numerical results for the second order perturbations driven by low frequency $g$-modes and $r$-modes of $4M_\odot$ main sequence models.
Discussions are given in \S 4, where we derive the equations of angular momentum conservation for non-axisymmetric waves and we also touch the problem of angular momentum transport by OsC modes in rapidly rotating main sequence stars. 
We conclude in \S 5.

\section{Method of calculation}

\subsection{Basic equations}

We use the basic equations for inviscid fluids written
in spherical polar coordinates $(r,\theta,\phi)$.
The three components of the equation of motion are in the inertial frame
\begin{align}
 \rho \left( {\frac{{dv _r }}{{dt}} - \frac{{v _\theta ^2  + v _\phi ^2 }}{r}} \right) = -\rho{\partial\Phi\over\partial r}   - \frac{{\partial p}}{{\partial r}} ,
\end{align}
\begin{align}
 \rho \left( {\frac{{dv _\theta  }}{{dt}} + \frac{{v _r v _\theta  }}{r} - \frac{{v _\phi ^2 \cot \theta }}{r}} \right) = -\rho {1\over r}{\partial\Phi\over\partial\theta}   - \frac{1}{r}\frac{{\partial p}}{{\partial \theta }} ,
 \end{align}
\begin{align}
 \rho \left( {\frac{{dv _\phi  }}{{dt}} + \frac{{v _r v _\phi  }}{r} + \frac{{v _\theta  v _\phi  \cot \theta }}{r}} \right) = -\rho {1\over r\sin\theta}{\partial\Phi\over\partial\phi}   - \frac{1}{{r\sin \theta }}\frac{{\partial p}}{{\partial \phi }} ,
\label{eq:eom_phi}
 \end{align}
where $v_r$, $v_\theta$, and $v_\phi$ are the three components of the velocity vector $\pmb{v}$, and
$\rho$, $p$, and $\Phi$ are respectively the mass density, pressure, and 
gravitational potential, and
\be
\frac{{dv _j }}{{dt}} = \frac{{\partial v _j }}{{\partial t}} + \pmb{v} \cdot \nabla v _j ,
\ee
for $j=r,\theta,\phi$.
The continuity equation is
\be
{\partial\rho\over\partial t}=-\nabla\cdot\left(\rho\pmb{v}\right).
\ee
For the time change rate of the specific entropy $s$, we use
\be
\rho T{ds\over dt}=\rho\epsilon_N -\nabla\cdot\pmb{F},
\ee
where $\epsilon_{N}$ is the nuclear energy generation rate, and 
$\pmb{F}$ is the energy flux given by the sum of the radiative energy flux $\pmb{F}_{\rm rad}$ and the
convective energy flux $\pmb{F}_{\rm conv}$, which is determined by the mixing length theory.
The radiative energy flux is given by
\be
\pmb{F}_{\rm rad}=-\lambda\nabla T,
\ee
where 
\be
\lambda={16\sigma_{\rm SB}T^3\over 3\kappa\rho},
\ee
and $T$ is the temperature, $\kappa$ is the opacity, and $\sigma_{\rm SB}$ is the Stefan-Boltzmann constant.

\subsection{Linearized equations}

By substituting
$\pmb{v}=\pmb{v}^{(0)}+\pmb{v}'$, $p=p^{(0)}+p'$, $\rho=\rho^{(0)}+\rho^\prime$ and so on into the basic equations
where $\pmb{v}^{(0)}$, $p^{(0)}$ and $\rho^{(0)}$ represent the equilibrium quantities and
$\pmb{v}'$, $p'$ and $\rho^\prime$ are the perturbations,
we obtain a set of linearized equations for the perturbations.
If we assume $v_r^{(0)}=v_\theta^{(0)}=0$ and $v_\phi^{(0)}=r\sin\theta\Omega(r,\theta)$ and
the Cowling approximation in which $\Phi'=0$, 
the equations for the perturbations are given by
\be
{\cal D}v_r^\prime-2{v_\phi^{(0)}v_\phi^\prime\over r}=-{1\over\rho^{(0)}}{\partial p'\over\partial r}+{\rho'\over\rho^2}{\partial p^{(0)}\over\partial r},
\label{eq:dvr}
\ee
\begin{align}
{\cal D}v_\theta^\prime
-2{v_\phi^{(0)}v_\phi^\prime\over r}\cot\theta=
-{1\over \rho^{(0)}}{1\over r}{\partial p^\prime\over\partial\theta},
\label{eq:dvtheta}
\end{align}
\begin{align}
{\cal D}v_\phi^\prime
+v_r^\prime{\partial v_\phi^{(0)}\over\partial r}
+v_\theta^\prime{1\over r}{\partial v_\phi^{(0)}\over\partial \theta}
+{v'_rv_\phi^{(0)}\over r}
+{v'_\theta v_\phi^{(0)}\over r}\cot\theta
=-{1\over\rho^{(0)}}{1\over r\sin\theta}{\partial p'\over\partial\phi},
\label{eq:dvphi}
\end{align}
\be
{\cal D}\rho'=-\nabla\cdot\left(\rho^{(0)}\pmb{v}'\right),
\label{eq:drho_cont}
\ee
\be
{\cal D}s'+v_r^\prime{\partial s^{(0)}\over\partial r}
=\left({J\over\rho T}\right)^\prime,
\label{eq:vrs}
\ee
\be
\pmb{F}_{\rm rad}'=-\lambda'\nabla T^{(0)}-\lambda^{(0)}\nabla T',
\ee
where
\be
J=\rho\epsilon_N-\nabla\cdot\pmb{F},
\ee
and
\be
{\cal D}={\partial\over\partial t}+\Omega{\partial\over\partial\phi}.
\ee
For the perturbation of the convective energy flux, we assume $\nabla\cdot\pmb{F}_{\rm conv}^\prime=0$ in this paper.
Since
\be\pmb{v}^\prime=\delta\pmb{v}-\pmb{\xi}\cdot\nabla\pmb{v}^{(0)},
\ee
where $\pmb{\xi}$ is the displacement vector,
we obtain
\be
v_r^\prime={\cal D}\xi_r, \quad
v_\theta^\prime={\cal D}\xi_\theta, \quad 
v_\phi^\prime={\cal D}\xi_\phi-r\sin\theta\pmb{\xi}\cdot\nabla\Omega.
\label{eq:vdxi}
\ee
Note that to derive the perturbed equations we have used the solutions to the equilibrium structure given by
\be
\nabla p^{(0)}=-\rho^{(0)}\nabla\Phi^{(0)}, \quad \nabla^2\Phi^{(0)}=4\pi G\rho^{(0)}, \quad ds^{(0)}/dt=0,
\ee
where we have neglected rotational deformation.

Assuming that the equilibrium state is time-independent and axisymmetric, we give
the time dependence of the perturbations by 
${\rm e}^{\rmi\sigma t}$ with $\sigma$ being the oscillation frequency in the inertial frame
and the angular dependence by 
series expansion using spherical harmonic functions $Y_l^m(\theta,\phi)$ for a given $m$.
For example, the three components of the displacement vector $\pmb{\xi}=\xi_r\pmb{e}_r+\xi_\theta\pmb{e}_\theta+\xi_\phi\pmb{e}_\phi$ are given by 
\be
{\xi_r\over r}=\sum_j^{j_{\rm max}}S_{l_j}(r)Y_{l_j}^m{\rm e}^{\rmi\sigma t},
\ee
\be
{\xi_\theta\over r}=\sum_j^{j_{\rm max}}\left[H_{l_j}(r){\partial\over\partial\theta}Y_{l_j}^m+T_{l'_j}(r){1\over\sin\theta}{\partial\over\partial\phi}Y_{l'_j}^m\right]{\rm e}^{\rmi\sigma t}, 
\ee
\be
{\xi_\phi\over r}=\sum_j^{j_{\rm max}}\left[H_{l_j}(r){1\over\sin\theta}{\partial\over\partial\phi}Y_{l_j}^m-T_{l'_j}(r){\partial\over\partial\theta}Y_{l'_j}^m\right]{\rm e}^{\rmi\sigma t},
\ee
and the pressure perturbation by
\be
p'=\sum_{j}^{j_{\rm max}}p^\prime_{l_j}(r)Y_{l_j}^m{\rm e}^{\rmi\sigma t},
\ee
where $\pmb{e}_r$, $\pmb{e}_\theta$, and $\pmb{e}_\phi$ are the unit vectors along the $r$, $\theta$ and $\phi$ coordinates, and $l_j=|m|+2(j-1)$ and $l'_j=l_j+1$ for even modes and 
$l_j=|m|+2j-1$ and $l'_j=l_j-1$ for odd modes, and $j_{\rm max}$ is the expansion length for the linear modes.
The relation between $\pmb{v}'$ and $\pmb{\xi}$ is given by
\be
v'_r=\rmi\omega\xi_r, 
\quad
v'_\theta=\rmi\omega\xi_\theta, 
\quad
v'_\phi=\rmi\omega\xi_\phi-r\sin\theta\pmb{\xi}\cdot\nabla\Omega,
\label{eq:vxi}
\ee
where
$
\omega=\sigma+m\Omega
$
is the frequency in the local co-rotating frame.
Using equations (\ref{eq:vxi}), we find
that equation (\ref{eq:drho_cont}) reduces to
\be
\rho'=-\nabla\cdot(\rho^{(0)}\pmb{\xi}).
\ee
For uniformly rotating stars, non-axisymmetric modes are separated into prograde and retrograde modes
when observed in the co-rotating frame 
and in the convention employed in this paper prograde modes have $m\omega_{\rm R}<0$ and retrograde modes
$m\omega_{\rm R}>0$ where $\omega_{\rm R}={\rm Re}(\omega)$.

Imposing boundary conditions at the centre and the surface of the stars, we solve the set of linearized equations as eigenvalue problem for $\sigma$
(e.g., \citealt{LeeSaio86,LeeSaio87}).
Non-adiabatic modes have complex eigenfrequencies $\sigma=\sigma_{\rm R}+\rmi\sigma_{\rm I}$.
Since the time dependence is given by the factor $\rme^{\rmi\sigma t}$,
unstable modes have $\sigma_{\rm I}<0$.
Non-adiabatic calculations of linear modes can give the luminosity variations $\delta L_{\rm rad}/L_{\rm rad}$ at the surface of the star and hence the magnitude variation $\delta m\approx -{2.5\over \ln 10}\delta L_{\rm rad}/L_{\rm rad}\approx-1.1\delta L_{\rm rad}/L_{\rm rad}$ where $\delta L_{\rm rad}$ is the Lagrangian variation of the surface luminosity $L_{\rm rad}$ and is given by $\delta L_{\rm rad}=\sum_l\delta L_{{\rm rad},l}Y_l^m{\rm e}^{\rmi\sigma t}$
for rotating stars.

For the amplitude normalization of modes, we introduce the parameter $\epsilon_E$ defined by
\be
\epsilon_E={{1\over 2}\int_0^R\rho |\omega_{\rm R}|^2\pmb{\xi}^*\cdot\pmb{\xi}dV\over GM^2/R},
\ee
where $G$ is the gravitational constant, and $M$ and $R$ are the mass and radius of the star.
If we assume $\epsilon_E=10^{-10}$, for example, we have $|\delta m|$ of order of $10^{-3}$
for unstable $g$-modes and $r$-modes of the main sequence model considered in this paper.

\subsection{Equations for the second order perturbations}

Let us introduce the parameter $a\sim O(|\pmb{\xi}|)$ to represent the typical amplitude of 
oscillation modes of a star.
Assuming the amplitude $a$ is small,
we calculate axisymmetric perturbations of order of $a^2$ which are
driven by non-axisymmetric oscillation modes of order of $a$.
To derive the set of equations governing the axisymmetric second order perturbations, we expand the physical 
quantities in terms of the small parameter $a$.
For example, we expand the velocity field as
\be
v_r=v_r^{(1)}+v_r^{(2)}, \quad v_\theta=v_\theta^{(1)}+v_\theta^{(2)}, \quad
v_\phi=v_\phi^{(0)}+v_\phi^{(1)}+v_\phi^{(2)}, 
\ee
and the pressure as
\be
p=p^{(0)}+p^{(1)}+p^{(2)},
\ee
where $v_\phi^{(0)}$, $v_\phi^{(1)}$, and $v_\phi^{(2)}$ are respectively the zeroth-order, first-order, and second-order velocity perturbations in $a$, and we have assumed that
$v_r^{(0)}=v_\theta^{(0)}=0$, and $v_\phi^{(0)}=r\sin\theta\Omega(r,\theta)$.
The zeroth-order quantities are those in the equilibrium state and the first-order
perturbations satisfy the linearized equations discussed in the previous subsection.

To derive the governing equations for the axisymmetric second order perturbations, we use
the zonal averaging of a quantity $Q(\pmb{x},t)$ defined as
\be
\overline{Q}={1\over2\pi}\int_0^{2\pi}Q(\pmb{x},t)d\phi.
\ee
If $Q^{(1)}$ is non-axisymmetric, proportional to $\rme^{\rmi m\phi}$ with $m$ being a non-zero integer,
and $Q^{(0)}$ and $Q^{(2)}$ are axisymmetric in the expansion $Q=Q^{(0)}+Q^{(1)}+Q^{(2)}$,
we obtain
\be
\overline{Q}=Q^{(0)}+Q^{(2)}, \quad \overline{Q^{(1)}}=0.
\ee
If we apply the zonal averaging to $\pmb{v}$ and $p$, for example, we have
\be
\overline{v_r}=\overline{v_r^{(2)}}=v_r^{(2)}, \quad 
\overline{v_\theta}=\overline{v_\theta^{(2)}}=v_\theta^{(2)}, \quad 
\overline{v_\phi}
=v_\phi^{(0)}+v_\phi^{(2)},
\ee
and 
\be
\overline{p}=p^{(0)}+p^{(2)}.
\ee

Substituting these expansions into the equation of motion, we obtain for the second-order
quantities
\begin{align}
{\partial v_r^{(2)}\over \partial t}&+g{d\ln\rho^{(0)}gr\over d\ln r}{p^{(2)}\over \rho^{(0)}gr}+g{\rho^{(2)}\over\rho^{(0)}}+gr{\partial\over\partial r}{p^{(2)}\over\rho^{(0)}gr}
-{2 v_\phi^{(0)}\over r}v_\phi^{(2)}
={\rho^{(1)}\over(\rho^{(0)})^2}{\partial p^{(1)}\over\partial r}-\pmb{v}^{(1)}\cdot\nabla v_r^{(1)}+{(v_\theta^{(1)})^2+(v_\phi^{(1)})^2\over r},
\label{eq:vr_second}
\end{align}
\begin{align}
{\partial v_\theta^{(2)}\over\partial t}&-{2v_\phi^{(0)}\over r}\cot\theta v_\phi^{(2)}+g{\partial\over\partial\theta}{p^{(2)}\over\rho^{(0)}gr}
=g{\rho^{(1)}\over\rho^{(0)}}{\partial \over\partial\theta}{p^{(1)}\over\rho^{(0)}gr}
-\pmb{v}^{(1)}\cdot\nabla v_\theta^{(1)}-{v_r^{(1)}v_\theta^{(1)}\over r}+{(v_\phi^{(1)})^2\over r}\cot\theta,
\label{eq:vtheta_second}
\end{align}
\begin{align}
{\partial v_\phi^{(2)}\over\partial t} 
+{\cal P}v_r^{(2)}+{\cal Q}v_\theta^{(2)}
+v_r^{(1)}{1\over r}{\partial\over\partial r}rv_\phi^{(1)}
+v_\theta^{(1)}{1\over r\sin\theta}{\partial\over\partial \theta}\sin\theta v_\phi^{(1)}
  +v_\phi^{(1)}{1\over r\sin\theta}{\partial\over\partial \phi}v_\phi^{(1)}
=g{\rho^{(1)}\over\rho^{(0)}}{1\over \sin\theta}{\partial \over\partial\phi}{p^{(1)}\over\rho^{(0)}gr},
\label{eq:vphi_second}
\end{align}
where $g=GM_r/r^2$ with $M_r=\int_0^r4\pi r^2\rho^{(0)}dr$, and 
\be
{\cal P}={1\over r}{\partial rv_\phi^{(0)}\over\partial r},
\quad
{\cal Q}={1\over r\sin\theta}{\partial\over\partial\theta}\sin\theta v_\phi^{(0)},
\ee
which become for $v_\phi^{(0)}=r\sin\theta\Omega(r,\theta)$
\be
{\cal P}=2\sin\theta\Omega f_r, \quad
{\cal Q}=2\cos\theta\Omega f_\theta,
\ee
where
\be f_r=1+{1\over 2}{\partial\ln\Omega\over\partial\ln r}, \quad f_\theta=1+{1\over 2}{\sin\theta\over\cos\theta}{\partial\ln \Omega\over\partial\theta}.
\ee
Using $l^{(1)}=r\sin\theta v_\phi^{(1)}$, we rewrite equation (\ref{eq:vphi_second}) as
\begin{align}
{\partial v_\phi^{(2)}\over\partial t} 
+{\cal P}v_r^{(2)}+{\cal Q}v_\theta^{(2)}
+{1\over r\sin\theta \rho^{(0)}}\left(\nabla\cdot\pmb{F}^l+l^{(1)}{\cal D}\rho^{(1)}\right)
-{1\over 2}{1\over r\sin\theta}{\partial\over\partial\phi}\left[(v_r^{(1)})^2+(v_\theta^{(1)})^2\right]
=g{\rho^{(1)}\over\rho^{(0)}}{1\over \sin\theta}{\partial \over\partial\phi}{p^{(1)}\over\rho^{(0)}gr},
\label{eq:vphi_second_B}
\end{align}
where $\pmb{F}^l\equiv\rho^{(0)}l^{(1)}\pmb{v}^{(1)}$.
The continuity equation of second order is given by
\be
\nabla\cdot\left(\rho^{(0)}\pmb{v}^{(2)}+\rho^{(2)}\pmb{v}^{(0)}\right)=-{\partial\over\partial t}\rho^{(2)}
-\nabla\cdot\left(\rho^{(1)}\pmb{v}^{(1)}\right).
\label{eq:cont_second}
\ee
The thermal equations of second-order are
\be
{\partial s^{(2)}\over\partial t}+v_r^{(2)}{\partial s^{(0)}\over\partial r}
+{1\over\rho^{(0)}}\nabla\cdot\left(\rho^{(0)}s^{(1)}\pmb{v}^{(1)}\right)-{s^{(1)}\over\rho^{(0)}}\nabla\cdot\left(\rho^{(0)}\pmb{v}^{(1)}\right)=\left({J\over\rho T}\right)^{(2)},
\label{eq:ds_second}
\ee
and
\be
\pmb{F}_{\rm rad}^{(2)}=-\left(\lambda\nabla T\right)^{(2)},
\label{eq:Frad_second}
\ee
and we do not consider in this paper second order interactions between convective motions and pulsations.

For the meridional velocity field defined as
$\widetilde{\pmb{v}}
=(\rho^{(0)})^{-1}\nabla\times(A_\phi\pmb{e}_\phi)$, the two components $\widetilde v_r$ and $\widetilde v_\theta$ are given by
\be
\widetilde v_r={1\over \rho^{(0)}}{1\over r\sin\theta}{\partial\over\partial\theta}\sin\theta A_\phi, \quad
\widetilde v_\theta=-{1\over\rho^{(0)}}{1\over r}{\partial\over\partial r}rA_\phi,
\label{eq:vtilde}
\ee
for which 
\be
{\cal P}\widetilde v_r+{\cal Q}\widetilde v_\theta={1\over r\sin\theta\rho^{(0)}}
\left[{1\over r^2}{\partial\over\partial r}r^2r\sin\theta(-{\cal Q})A_\phi
+{1\over r\sin\theta}{\partial\over\partial\theta}\sin\theta r\sin\theta{\cal P}A_\phi\right]\equiv
-{1\over r\sin\theta\rho^{(0)}}\nabla\cdot\widetilde{\pmb{F}},
\ee
where
\be
\widetilde{\pmb{F}}={\cal Q}r\sin\theta A_\phi\pmb{e}_r-{\cal P}r\sin\theta A_\phi\pmb{e}_\theta.
\ee
Since the velocity field $\widetilde{\pmb{v}}$ satisfies $\nabla\cdot(\rho^{(0)}\widetilde{\pmb{v}})=0$,
$A_\phi$ may be regarded as a stream function. For the function \footnote{In geophysics it may be more common to use $A_\phi=\rho^{(0)}(ds^{(0)}/dr)^{-1}\overline{s^{(1)}v_\theta^{(1)}}$, which reduces to equation (\ref{eq:aphi_svtheta})
for adiabatic modes satisfying $\delta s^{(1)}=s^{(1)}+\xi_rds^{(0)}/dr=0$.}
\be
A_\phi=-\rho^{(0)}\overline{\xi_rv^{(1)}_\theta},
\label{eq:aphi_svtheta}
\ee
we introduce the meridional velocity field $\hat{\pmb{v}}=\hat v_r\pmb{e}_r+\hat v_\theta\pmb{e}_\theta$ defined by
\be
\hat v_r=v_r^{(2)}+\widetilde v_r, \quad
\hat v_\theta=v_\theta^{(2)}+\widetilde v_\theta.
\label{eq:vhat}
\ee

Taking the zonal average of equations (\ref{eq:vr_second}), (\ref{eq:vtheta_second}), (\ref{eq:vphi_second}),
(\ref{eq:cont_second}), (\ref{eq:ds_second}), and (\ref{eq:Frad_second}), and replacing $v_r^{(2)}$ and $v_\theta^{(2)}$
by $\hat v_r-\widetilde v_r$ and $\hat v_\theta-\widetilde v_\theta$, we obtain
\be
{1\over r^2}{\partial\over\partial r}r^2\rho^{(0)}\hat v_r+{1\over r\sin\theta}{\partial\over\partial\theta}\sin\theta \rho^{(0)}\hat v_\theta=-{\partial\over\partial t}\rho^{(2)}
-\overline{\nabla\cdot\left(\rho^{(1)}\pmb{v}^{(1)}\right)},
\label{eq:cont_star}
\ee
\begin{align}
{\partial \hat v_r\over \partial t}+g{d\ln\rho^{(0)}gr\over d\ln r}{p^{(2)}\over \rho^{(0)}gr}+g{\rho^{(2)}\over\rho^{(0)}}+gr{\partial\over\partial r}{p^{(2)}\over\rho^{(0)}gr}
-{2v_\phi^{(0)}\over r} v_\phi^{(2)}=\Xi ,
\label{eq:vr_star}
\end{align}
\begin{align}
{\partial \hat v_\theta\over\partial t}&-{2v_\phi^{(0)}\cot\theta\over r} v_\phi^{(2)}+g{\partial\over\partial\theta}{p^{(2)}\over\rho^{(0)}gr}=\Theta,
\label{eq:vtheta_star}
\end{align}
\begin{align}
{\partial v_\phi^{(2)}\over\partial t}+{\cal P}\hat v_r+{\cal Q}\hat v_\theta
= \Psi,
\label{eq:vphi_epf}
\end{align}
\be
{\partial s^{(2)}\over\partial t}+\hat v_r{\partial s^{(0)}\over \partial r}=\left({J\over\rho T}\right)^{(2)}
-{1\over\rho^{(0)}}\nabla\cdot\left(\rho^{(0)}\overline{\delta s^{(1)}\pmb{v}^{(1)}}\right)
+{1\over\rho^{(0)}}{1\over r^2}{\partial\over\partial r}r^2\rho^{(0)}{ds^{(0)}\over dr}\overline{\xi_rv_r^{(1)}}
+{1\over\rho^{(0)}}\overline{s^{(1)}\nabla\cdot\left(\rho^{(0)}\pmb{v}^{(1)}\right)},
\label{eq:ent_star_b}
\ee
\be
F_{{\rm rad},r}^{(2)}=-{\lambda^{(2)}\over\lambda^{(0)}}\lambda^{(0)}{\partial\over\partial r} T^{(0)}-\lambda^{(0)}{\partial\over\partial r} T^{(2)}
-\overline{\lambda^{(1)}{\partial\over\partial r} T^{(1)}},
\ee
\be
F_{{\rm rad},\theta}^{(2)}=-\lambda^{(0)}{1\over r}{\partial\over\partial\theta}T^{(2)}-\overline{\lambda^{(1)}{1\over r}{\partial\over\partial\theta} T^{(1)}},
\label{eq:fh}
\ee 
where $\Xi$, $\Theta$, and $\Psi$ are inhomogeneous terms given by
\begin{align}
\Xi=\overline{{\rho^{(1)}\over(\rho^{(0)})^2}{\partial p^{(1)}\over\partial r}}-\overline{\pmb{v}^{(1)}\cdot\nabla v_r^{(1)}}+\overline{{(v_\theta^{(1)})^2+
(v_\phi^{(1)})^2\over r}}
+{\partial\widetilde v_r\over\partial t},
\end{align}
\begin{align}
\Theta=\overline{{\rho^{(1)}\over(\rho^{(0)})^2}{1\over r}{\partial p^{(1)}\over\partial\theta}}
-\overline{\pmb{v}^{(1)}\cdot\nabla v_\theta^{(1)}}-\overline{{v_r^{(1)}v_\theta^{(1)}\over r}}
+\overline{{(v_\phi^{(1)})^2\over r}}\cot\theta
+{\partial\widetilde v_\theta\over\partial t},
\end{align}
\begin{align}
\Psi={1\over r\sin\theta \rho^{(0)}}\left(-\nabla\cdot\pmb{ F}^{\rm EP}
+\overline{{\rho^{(1)}\over\rho^{(0)}}{\partial p^{(1)}\over\partial\phi}}-\overline{l^{(1)}{\cal D}\rho^{(1)}}\right),
\label{eq:right_psi}
\end{align}
and 
\begin{align}
\pmb{F}^{\rm EP}=r\sin\theta \rho^{(0)}\left(\overline{v_\phi^{(1)}v_r^{(1)}}-{\cal Q}\overline{\xi_rv^{(1)}_\theta} \right)\pmb{e}_r
+r\sin\theta \rho^{(0)}\left(\overline{v_\phi^{(1)}v_\theta^{(1)}}+{\cal P}\overline{\xi_rv^{(1)}_\theta}\right)\pmb{e}_\theta\equiv F^{\rm EP}_r\pmb{e}_r+F^{\rm EP}_\theta\pmb{e}_\theta,
\label{eq:EPflux}
\end{align}
and $\pmb{F}^{\rm EP}(=\overline{\pmb{F}^l}+\widetilde{\pmb{F}})$ is called the Eliassen-Palm flux
(\citealt{EliassenPalm1961,Pedlosky1987,Buhler2014}).
The components $F_r^{\rm EP}$ and $F_\theta^{\rm EP}$ depend on $r$ and $z=R\cos\theta$. 
We find that
$F_r^{\rm EP}(r,-z)=F_r^{\rm EP}(r,z)$ and $\sin\theta F_\theta^{\rm EP}(r,-z)=-\sin\theta F_\theta^{\rm EP}(r,z)$ so that $\nabla\cdot\pmb{F}^{\rm EP}$ is symmetric about the equatorial plane.
Note that we have made the approximation given by
\be
\left({J\over\rho T}\right)^{(2)}={J^{(2)}\over(\rho T)^{(0)}}-{J^{(1)}\over(\rho T)^{(0)}}{(\rho T)^{(1)}\over(\rho T)^{(0)}}
\approx {-\nabla\cdot\pmb{F}^{(2)}\over(\rho T)^{(0)}}-{\rmi\omega \delta s^{(1)}}{(\rho T)^{(1)}\over(\rho T)^{(0)}}.
\ee
The set of equations given above are linear equations for the second order perturbations and contain
inhomogeneous terms, which are also second order in the amplitude parameter $a$ and are given by the sum of products of the first order perturbations.

The first order perturbations such as $\pmb{v}^{(1)}$ and $p^{(1)}$ are in this paper
given by the eigenfunctions of linear oscillation modes such that $\pmb{v}^{(1)}=\pmb{v}'$, $p^{(1)}=p'$, and so on.
For a product of the eigenfunctions $a'$ and $b'$ of a mode where $a'$ and $b'$ are assumed
proportional to ${\rm e}^{\rmi m\phi+\rmi\sigma t}$,
we may define the zonal averaging of the product as
\be
\overline{a'b'}={1\over 2\pi}\int_0^{2\pi}\Re(a')\Re(b')d\phi={a'^*b'+a'b'^*\over 4}={1\over 2}\Re(a'^*b')={1\over 2}\Re(a'b'^*),
\ee
where the asterisk ${}^*$ implies complex conjugation.
Note that $\overline{a'a'}=|a'|^2/2$.
For complex $\omega=\omega_{\rm R}+\rmi\omega_{\rm I}$ where $\omega=\sigma+m\Omega$, we have
\be
\overline{a'{\cal D}b'}=-{1\over 2}\omega_{\rm R}\Im(a'^*b')-{1\over 2}\omega_{\rm I}\Re(a'^*b'),
\ee
and hence
\be
\overline{a'{\cal D}a'}=-{1\over 2}\omega_{\rm I}|a'|^2={1\over 4}{\partial\over\partial t}|a'|^2={1\over 2}{\cal D}\overline{a'a'},
\ee
which vanishes for $\omega_{\rm I}=0$, that is, for steady perturbations with constant amplitudes.
Since $\Im(a'^*b')=-\Im(a'b'^*)$, we have
\be
\overline{a'{\cal D}b'}+\overline{({\cal D}a')b'}=-\omega_{\rm I}\Re(a'^*b')=-2\omega_{\rm I}\overline{a^\prime b^\prime}={\partial \over\partial t}\overline{a^\prime b^\prime}={\cal D}\overline{a^\prime b^\prime},
\ee
and 
$
\overline{b'{\cal D}a'}=-\overline{a'{\cal D}b'}
$
when $\omega_{\rm I}=0$.

\subsection{Equations for the second order perturbations in vector form}

To represent the second order axisymmetric perturbations, we use series expansions in terms of $Y_l^0$.
For example, the second order velocity field is given by
\be
{\hat v_r\over\sigma_0 r}=\sum_{j=1}^{k_{\rm max}} S_{l_j}^{(2)}(r,t)Y_{l_j}^0, \quad 
{\hat v_\theta\over\sigma_0r}=\sum_{j=1}^{k_{\rm max}} H_{l_j}^{(2)}(r,t){\partial Y_{l_j}^0\over\partial\theta}, \quad
{v_\phi^{(2)}\over\sigma_0 r}=-\sum_{j=1}^{k_{\rm max}} T_{l'_j}^{(2)}(r,t){\partial Y_{l'_j}^0\over\partial\theta}, 
\label{eq:vrthetaphi}
\ee
and the pressure perturbation by
\be
p^{(2)}=\sum_{j=1}^{k_{\rm max}}p_{l_j}^{(2)}(r,t)Y_{l_j}^0,
\ee
where $l_j=2(j-1)$ and $l'_j=l_j+1$, and $\sigma_0=\sqrt{GM/R^3}$.
Note that the expansion length ${k_{\rm max}}$ for the second order perturbations is not necessarily 
the same as $j_{\rm max}$ for the first order perturbations.
We usually use $k_{\rm max}=2j_{\rm max}$ in this paper.

Substituting these expansions into equations (\ref{eq:cont_star}) to (\ref{eq:fh}), and
using the dependent variables $\pmb{y}_1^{(2)}$, $\pmb{y}_2^{(2)}$, $\pmb{h}^{(2)}$, $\pmb{t}^{(2)}$, $\pmb{l}^{(2)}$, and
$\pmb{s}^{(2)}$ in vector form whose $j$-th components are defined by
\begin{align}
&\left(\pmb{y}_1^{(2)}\right)_j=S_{l_j}^{(2)}(r,t), \quad \left(\pmb{y}_2^{(2)}\right)_j={p_{l_j}^{(2)}(r,t)\over \rho g r},
\quad \left(\pmb{h}^{(2)}\right)_j=H_{l_j}^{(2)}(r,t), \quad \left(\pmb{t}^{(2)}\right)_j=T_{l'_j}^{(2)}(r,t),\nonumber\\
&\left(\pmb{l}^{(2)}\right)_j={L_{{\rm rad},r,l_j}^{(2)}(r,t)\over L_{\rm rad}^{(0)}}, \quad \left(\pmb{s}^{(2)}\right)_j={s^{(2)}_{l_j}(r,t)\over c_p},
\end{align}
we obtain
\be
r{\partial\over\partial r}\pmb{y}_1^{(2)}=-\left(3+rA-{V\over\Gamma_1}\right)\pmb{y}_1^{(2)}
+\widetilde{\pmbmt{\Lambda}}\pmb{h}^{(2)}-{V\over\Gamma_1}{\partial \pmb{y}_2^{(2)}\over \partial \tau}+\alpha_T{\partial \pmb{s}^{(2)}\over \partial \tau}+\pmb{Y}_c,
\label{eq:dy1_vec}
\ee
\begin{align}
r{\partial\over\partial r}\pmb{y}_2^{(2)}
=-c_1{\partial\over\partial\tau}\pmb{y}_1^{(2)}+\left(1-rA-U\right)\pmb{y}_2^{(2)}
+\alpha_T\pmb{s}^{(2)}
- 2c_1\bar\Omega\pmbmt{C}_0\pmb{t}^{(2)}+\pmb{Y}_r,
\label{eq:dy2_vec}
\end{align}
\be
\widetilde{\pmbmt{C}}_1{\partial\over\partial\tau}\pmb{h}^{(2)}+2\bar\Omega\pmbmt{B}_1\pmb{t}^{(2)}+\pmbmt{C}_1{\pmb{y}_2^{(2)}\over c_1}
={\pmb{Y}_\theta\over c_1},
\label{eq:eom_theta_vec}
\ee
\be
-\pmbmt{C}_0{\partial\over\partial\tau}\pmb{t}^{(2)}+2\bar\Omega\widetilde{\pmbmt{B}}_0\pmb{h}^{(2)}
+2\bar\Omega f_r\pmbmt{A}_0\pmb{y}_1^{(2)}
={\pmb{Y}_\phi\over c_1},
\label{eq:eom_phi_vec}
\ee
\begin{align}
r{\partial\over \partial r}\pmb{l}^{(2)}
=c_2{rA\over\alpha_T}\pmb{y}_1^{(2)}-{\nabla_{\rm ad}\over\nabla}{\pmbmt{\Lambda}}{\pmb{y}_2^{(2)}}-{d\ln L_{\rm rad}^{(0)}\over d\ln r}\pmb{l}^{(2)}-{\pmbmt{\Lambda}\over V\nabla}\pmb{s}^{(2)}-c_2{\partial\over\partial\tau}\pmb{s}^{(2)}
+c_2{\pmb{Y}}_s
+{\pmb{f}\over V\nabla},
\label{eq:dy3_vec}
\end{align}
\begin{align}
{1\over V\nabla_{}}r{\partial\over \partial r}\pmb{s}^{(2)}
={\nabla_{\rm ad}\over\nabla}c_1{\partial\over\partial\tau}\pmb{y}_1^{(2)}
&+C_{42}{\pmb{y}_2^{(2)}}-\pmb{l}^{(2)}+\left[4-\kappa_s-\left({\nabla_{\rm ad}\over\nabla}-1\right)\alpha_T\right]\pmb{s}^{(2)}
+{\nabla_{\rm ad}\over\nabla}2c_1\bar\Omega\pmbmt{C}_0\pmb{t}^{(2)}
-{\nabla_{\rm ad}\over\nabla}\pmb{Y}_r-{\pmb{X}\over V\nabla},
\label{eq:ds_vec}
\end{align}
where $\bar\Omega=\Omega/\sigma_0$, $\tau=\sigma_0 t$, 
\be
C_{42}=4V\nabla_{\rm ad}-V\kappa_{\rm ad}-{\nabla_{\rm ad}\over\nabla}V
+\left({\nabla_{\rm ad}\over\nabla}
-1\right){V\over\Gamma_1}-{\nabla_{\rm ad}\over\nabla}{d\ln\nabla_{\rm ad}\over d\ln r},
\ee
and the $j$-th component of the inhomogeneous terms $\pmb{Y}_c$, $\pmb{Y}_r$, $\pmb{Y}_\theta$, $\pmb{Y}_\phi$, $\pmb{Y}_s$, and $\pmb{X}$ are
\be
\left(\pmb{Y}_c\right)_j=-{1\over \rho^{(0)}\sigma_0}\int  Y_{l_j}^0\overline{\nabla\cdot(
\rho^{(1)}\pmb{v}^{(1)})}\;do,
\ee
\begin{align}
\left(\pmb{Y}_r\right)_j&=\int Y_{l_j}^0{\Xi\over g } do,
\end{align}
\begin{align}
\left(\pmb{Y}_\theta\right)_{j}&=\int Y_{l_j+1}^0\sin\theta{\Theta\over g}do,
\end{align}
\be
\left(\pmb{Y}_\phi\right)_j=\int Y_{l_j}^0\sin\theta{\Psi\over g}do={1\over\rho^{(0)}gr}\int Y_{l_j}^0\left[-\nabla\cdot\pmb{F}^{\rm EP}+\overline{{\rho^{(1)}\over\rho^{(0)}}{\partial p^{(1)}\over\partial\phi}}+r\sin\theta\overline{v_\phi^{(1)}\nabla\cdot\left(\rho^{(0)}\pmb{v}^{(1)}\right)}\right]do,
\ee
\begin{align}
\left(\pmb{Y}_s\right)_j
=\int Y_{l_j}^0\Biggl[\overline{-\rmi\bar\omega{\delta s^{(1)}\over c_p}{(\rho T)^{(1)}\over(\rho T)^{(0)}}}
-{1\over\rho^{(0)}c_p}{1\over r^2}{\partial\over\partial r}r^3\rho^{(0)}c_p\overline{{v_r^{(1)}\over r\sigma_0}{s^{(1)}\over c_p}}
-{1\over \sin\theta}{\partial\over\partial\theta}\sin\theta\overline{{\delta s^{(1)}\over c_p}{v_\theta^{(1)}\over r\sigma_0}}
-{1\over\sigma_0}\overline{{s^{(1)}\over c_p}{\cal D}{\rho^{(1)}\over\rho^{(0)}}}\Biggr]do, 
\end{align}
\be
(\pmb{X})_j=\int  Y_{l_j}^0\overline{{\lambda^{(1)}\over\lambda^{(0)}}\left(r{d\over dr}{T^{(1)}\over T^{(0)}}-V\nabla{T^{(1)}\over T^{(0)}}\right)}do,
\ee
\be
\left(\pmb{f}\right)_j=\int Y_{l_j}^0\left(\overline{{\lambda^{(1)}\over\lambda^{(0)}}{1\over\sin\theta}{\partial\over\partial\theta}\sin\theta
{\partial\over\partial\theta}{T^{(1)}\over T^{(0)}}}\right)do,
\ee
where $l_j=2(j-1)$ for $j=1,\>2,\>\cdots,k_{\rm max}$ and $do=\sin\theta d\theta d\phi$, and we have assumed
\be
{s^{(2)}\over c_p}={T^{(2)}\over T^{(0)}}-\nabla_{\rm ad}{p^{(2)}\over p^{(0)}}={1\over\alpha_T}\left({1\over\Gamma_1}{p^{(2)}\over p^{(0)}}-{\rho^{(2)}\over\rho^{(0)}}\right),
\ee
\begin{align}
{\lambda^{(2)}\over\lambda^{(0)}}&=3{T^{(2)}\over T^{(0)}}-{\rho^{(2)}\over\rho^{(0)}}
-\kappa_{\rm ad}{p^{(2)}\over p^{(0)}}-\kappa_T{s^{(2)}\over c_p},
\end{align}
and
\be
\Gamma_1=\left({\partial\ln p\over\partial\ln\rho}\right)_{\rm ad}, \quad
\alpha_T=-\left({\partial\ln\rho\over\partial\ln T}\right)_p, \quad
\nabla_{\rm ad}=\left({\partial\ln T\over\partial\ln T}\right)_{\rm ad},
 \quad \kappa_{\rm ad}=\left({\partial\ln\kappa\over\partial\ln p}\right)_s, \quad
\kappa_T=c_p\left({\partial\ln\kappa\over\partial s}\right)_p.
\ee
Note that we have assumed $f_\theta=1$ for simplicity.
Other symbols are defined as
\be
rA={d\ln\rho\over d\ln r}-{1\over\Gamma_1}{d\ln p\over d\ln r}, \quad V=-{d\ln p\over d\ln r}, \quad U={d\ln M_r\over d\ln r}, \quad c_1={(r/R)^3\over M_r/M}, \quad c_2={4\pi r^3\rho T c_p\over L_{{\rm rad},r}}\sigma_0,\quad
\nabla={d\ln T\over d\ln p},
\ee
where $\rho=\rho^{(0)}$, $p=p^{(0)}$, and $T=T^{(0)}$ and so on.
The non-zero elements of the matrices $\pmbmt{\Lambda}$, $\pmbmt{A}_0$, $\pmbmt{B}_0$, $\pmbmt{B}_1$,
$\pmbmt{C}_0$, and $\pmbmt{C}_1$ are
\be
(\pmbmt{\Lambda})_{jj}=l_j(l_j+1),
\ee
\be
(\pmbmt{A}_0)_{jj}=1-(J_{l_j}^m)^2-(J_{l_j+1}^m)^2, \quad 
(\pmbmt{A}_0)_{j,j+1}=-J_{l_j+1}^mJ_{l_j+2}^m,\quad
(\pmbmt{A}_0)_{j+1,j}=-J_{l_j+1}^mJ_{l_j+2}^m,
\ee
\be
(\pmbmt{B}_0)_{jj}=l_j(J_{l_j+1}^m)^2-(l_j+1)(J_{l_j}^m)^2,\quad
(\pmbmt{B}_0)_{j,j+1}=-(l_j+3)J_{l_j+1}^mJ_{l_j+2}^m,\quad
(\pmbmt{B}_0)_{j+1,j}=l_jJ_{l_j+1}^mJ_{l_j+2}^m,
\ee
\be
(\pmbmt{B}_1)_{jj}=(l_j+1)(J_{l_j+2}^m)^2-(l_j+2)(J_{l_j+1}^m)^2,\quad
(\pmbmt{B}_1)_{j,j+1}=-(l_j+4)J_{l_j+2}^mJ_{l_j+3}^m,\quad
(\pmbmt{B}_1)_{j+1,j}=(l_j+1)J_{l_j+2}^mJ_{l_j+3}^m,
\ee
\be
(\pmbmt{C}_0)_{jj}=-(l_j+2)J_{l_j+1}^m,\quad (\pmbmt{C}_0)_{j+1,j}=(l_j+1)J_{l_j+2}^m,\quad
(\pmbmt{C}_1)_{jj}=l_jJ_{l_j+1}^m, \quad (\pmbmt{C}_1)_{j,j+1}=-(l_j+3)J_{l_j+2}^m,
\ee
where
$
J_l^m=\sqrt{(l^2-m^2)/ (4l^2-1)}
$
for $l\ge |m|$ and $J_l^m=0$ otherwise.
The $ij$-element of the matrix $\widetilde{\pmbmt{\Lambda}}$, for example, is given by
\be
(\widetilde{\pmbmt{\Lambda}})_{ij}=(\pmbmt{\Lambda})_{i,j+1}.
\ee

The outer boundary conditions we use are $\overline{\delta p^{(2)}}=0$ and 
$4\delta T^{(2)}/T^{(0)}=\delta L^{(2)}_{\rm rad}/L_{\rm rad}^{(0)}$ at the surface,
where $\overline{\delta p^{(2)}}$ is the mean Lagrangian pressure perturbation of second order  
given by
\be
\overline{\delta p^{(2)}}=\overline{p^{(2)}}+\overline{\pmb{\xi}\cdot\nabla p^{(1)}}
+{1\over 2}\left[\overline{\xi_r\xi_r}{\partial^2 p^{(0)}\over\partial r^2}
+\left(\overline{\xi_\theta\xi_\theta}+\overline{\xi_\phi\xi_\phi}\right){1\over r}{\partial p^{(0)}\over\partial r}\right].
\ee
For the expansion $p^{(2)}=\sum_l p^{(2)}_lY_l^0$, 
the surface boundary condition $\overline{\delta p^{(2)}}=0$ leads to
\be
\overline{p^{(2)}_l}=-\int Y_l^0\left\{\overline{\pmb{\xi}\cdot\nabla p^{(1)}}
+{1\over 2}\left[\overline{\xi_r\xi_r}{\partial^2 p^{(0)}\over\partial r^2}
+\left(\overline{\xi_\theta\xi_\theta}+\overline{\xi_\phi\xi_\phi}\right){1\over r}{\partial p^{(0)}\over\partial r}\right]\right\}do ,
\ee
which gives inhomogeneous boundary conditions.
The inner boundary conditions we use at the centre are that the variables $x\pmb{y}^{(2)}_1$ and $x^2\pmb{y}^{(2)}_2$ are regular and that $\partial\pmb{s}^{(2)}/\partial\ln r=0$ where $x=r/R$.

Since the inhomogeneous terms $\pmb{Y}_i$ are given by the sum of products of the eigenfunctions of oscillation modes,
it may be convenient to write the term $\pmb{Y}_i$ more explicitly as $\pmb{Y}_i^{jk}=\pmb{Y}_i(\pmb{\xi}_j,\pmb{\xi}_k)$ where $\pmb{\xi}_j$ and $\pmb{\xi}_k$
represent the eigenfunctions of the oscillation modes with the eigenfrequencies $\sigma_j$ and $\sigma_k$, respectively. 
If the first order perturbation that excites second order perturbations is given by a linear combination of oscillation modes such that 
\be
\pmb{\xi}=\sum_ic_i\pmb{\xi}_i,
\ee
where $c_i$ denotes the expansion coefficient, the terms $\pmb{Y}_i$ for the perturbation are
given as
\be
\pmb{Y}_i=\sum_{j,k}c_j^*c_k\pmb{Y}_i^{jk}.
\ee
If the first order perturbation is given by a single mode, the terms $\pmb{Y}_i$ have time dependency given by
${\rm e}^{-2\sigma_{\rm I}t}$, which is usually slowly growing with time for an unstable mode with $\sigma_{\rm I}<0$.
But when the first order perturbation is given by a linear combination of several unstable oscillation modes, the time dependency of the inhomogeneous terms can be much more complex.
For example, if the perturbation contains
two different modes $\pmb{\xi}_j$ and $\pmb{\xi}_k$, the time dependency of the terms 
$\pmb{Y}_i$ is given by ${\rm e}^{-\rmi(\sigma_{j}^*-\sigma_{k})t}
={\rm e}^{-\rmi(\sigma_{j,{\rm R}}-\sigma_{k,{\rm R}})t}{\rm e}^{-(\sigma_{j,{\rm I}}+\sigma_{k,{\rm I}})t}$.
Since $|\sigma_{\rm R}|\gg|\sigma_{\rm I}|$ in general, unless $\sigma_{j,{\rm R}}\approx \sigma_{k,{\rm R}}$ the factor ${\rm e}^{-\rmi(\sigma_{j,{\rm R}}-\sigma_{k,{\rm R}})t}$ rapidly oscillates with time compared with the factor ${\rm e}^{-(\sigma_{j,{\rm I}}+\sigma_{k,{\rm I}})t}$. 
Thus, the terms proportional to this oscillating factor could be ignored when we discuss a long time-scale, of order of ${\rm min}(1/|\sigma_{j,{\rm I}}|,1/|\sigma_{k,{\rm I}}|)$, development of the second order perturbatipons.
If this is the case, the inhomogeneous terms may be simply given by
\be
\pmb{Y}_i=\sum_j|c_j|^2\pmb{Y}_i^{jj}.
\ee

\subsection{Formal discussion of the solutions of the second order perturbations}

The set of partial differential equations for the second order perturbations 
derived above is formally written as
\be
\left(\pmbmt{K}(r){\partial\over\partial \tau}+\pmbmt{L}(r)\right)\pmb{z}^{(2)}(r,\tau)=\pmb{Y}^{(2)}(r,\tau),
\label{eq:formaleq}
\ee
where $\pmbmt{K}(r)$ is a coefficient matrix, and $\pmbmt{L}(r)$ is the time-independent linear differential operator containing $r \partial/\partial r$, and $\pmb{z}^{(2)}$
is the dependent variable defined as
\be
\pmb{z}^{(2)}=\left(\begin{array}{c}\pmb{y}_1^{(2)}\\ \pmb{h}^{(2)}\\ \pmb{t}^{(2)} \\ \pmb{y}_2^{(2)} \\ \pmb{l}^{(2)} \\ \pmb{s}^{(2)}\end{array}\right),
\ee
and $\pmb{Y}^{(2)}$ represents the inhomogeneous terms such as $\pmb{Y}_i$, $\pmb{X}$, and $\pmb{f}$.
Equation (\ref{eq:formaleq}) is solved for an initial condition
\be
\pmb{z}^{(2)}(r,0)=\pmb{z}_0^{(2)}(r),
\ee
and for outer and inner boundary conditions given by
\be
\pmbmt{G}_k\pmb{z}^{(2)}=\pmb{g}_k,
\ee
where $\pmbmt{G}_k$ represent
coefficient matrices and the boundary conditions are homogeneous if
$\pmb{g}_k=0$ or inhomogeneous if $\pmb{g}_k\not=0$.
General solutions to the inhomogeneous linear differential equations may be given by the sum of
complementary solutions and particular solutions.

We may look for complementary solutions of the form $\pmb{z}^{(2)}=\hat{\pmb{z}}^{(2)}(r){\rm e}^{\alpha \tau}$,
assuming $\pmb{Y}^{(2)}=0$ and $\pmbmt{G}_k{\pmb{z}}^{(2)}=0$ so that we obtain a set of homogeneous linear differential equations.
Substituting the expression $\pmb{z}^{(2)}=\hat{\pmb{z}}^{(2)}(r){\rm e}^{\alpha \tau}$ into (\ref{eq:formaleq}), we obtain $(\alpha\pmbmt{K}+\pmbmt{L})\hat{\pmb{z}}^{(2)}=0$, and $\det(\alpha\pmbmt{K}+\pmbmt{L})=0$ determines the eigenvalues $\alpha_i$ and eigenfunctions $\hat{\pmb{z}}_i^{(2)}$ that satisfy $\pmbmt{L}\hat{\pmb{z}}_i^{(2)}=-\alpha_i\pmbmt{K}\hat{\pmb{z}}_i^{(2)}$.
A complementary solution is given by $\pmb{z}_c^{(2)}(r,\tau)=\sum_ic_i\hat{\pmb{z}}_i^{(2)}(r){\rm e}^{\alpha_i t}$ where $c_i$ are coefficients determined so that the initial condition $\pmb{z}_c^{(2)}(r,0)=\pmb{z}_0^{(2)}(r)$ is satisfied.

If the time-dependence of the forcing terms $\pmb{Y}^{(2)}$ is given as $\pmb{Y}^{(2)}(r,\tau)=\sum_i\pmb{Y}^{(2)}_i(r){\rm e}^{\gamma_i\tau}$, we may look for
a particular solution of the form $\pmb{z}_p^{(2)}(r,\tau)=\sum_i\widetilde{\pmb{z}}_i^{(2)}(r){\rm e}^{\gamma_i\tau}$.
For each $\gamma_i$, we calculate the function $\widetilde{\pmb{z}}_i^{(2)}(r)$ that satisfies
\be
\left(\gamma_i\pmbmt{K}(r)+\pmbmt{L}(r)\right)\widetilde{\pmb{z}}_i^{(2)}(r)=\pmb{Y}_i^{(2)}(r),
\label{eq:formaleq_inhomo}
\ee
and the boundary conditions $\pmbmt{G}_k\widetilde{\pmb{z}}_i^{(2)}=\pmb{g}_k$.
The general solution is now given by $\pmb{z}_g^{(2)}=\pmb{z}_c^{(2)}+\pmb{z}_p^{(2)}$.

\subsection{Differential equations for particular solutions of the form $\pmb{z}^{(2)}(r,\tau)=\pmb{z}^{(2)}(r){\rm e}^{\gamma \tau}$}

If the time dependence of the forcing terms and the second order perturbations is given by
the factor ${\rm e}^{\gamma \tau}$, $\partial/\partial\tau$ in the differential equations
can be replaced by $\gamma$.
To obtain the set of differential equations we numerically solve,
we start by writing equations (\ref{eq:eom_theta_vec}) and (\ref{eq:eom_phi_vec}) as
\begin{align}
\gamma \pmbmt{C}\left(\begin{array}{c}\pmb{h}^{(2)}\\\pmb{t}^{(2)}\end{array}\right)+2\bar\Omega\pmbmt{B}\left(\begin{array}{c}\pmb{h}^{(2)}\\\pmb{t}^{(2)}\end{array}\right)=\pmbmt{A}\left(\begin{array}{c}\pmb{y}_1^{(2)}\\\pmb{y}_2^{(2)}/c_1\end{array}\right)
+\left(\begin{array}{c}\pmb{Y}_\theta /c_1\\ \pmb{Y}_\phi/c_1\end{array}\right),
\end{align}
where
\be
\pmbmt{C}=\left(\begin{array}{cc} \widetilde{\pmbmt{C}}_1 & 0 \\0 &-\pmbmt{C}_0\end{array}\right), \quad
\pmbmt{B}=\left(\begin{array}{cc}0 & \pmbmt{B}_1\\ {}\widetilde{\pmbmt{B}}_0 & 0\end{array}\right), \quad
\pmbmt{A}=\left(\begin{array}{cc}0 & -\pmbmt{C}_1\\-2\bar\Omega f_r\pmbmt{A}_0 & 0\end{array}\right).
\ee
We then obtain
\begin{align}
\left(\begin{array}{c}\pmb{h}^{(2)}\\\pmb{t}^{(2)}\end{array}\right)=\pmbmt{F}\left(\begin{array}{c}\pmb{y}_1^{(2)}\\\pmb{y}_2^{(2)}/c_1\end{array}\right)+\left(\begin{array}{c}\pmb{Z}_\theta/c_1\\ \pmb{Z}_\phi/c_1\end{array}\right),
\label{eq:h2t2}
\end{align}
where
\begin{align}
\pmbmt{F}=\left(\gamma\pmbmt{C}+2\bar\Omega\pmbmt{B}\right)^{-1}\pmbmt{A}, 
\end{align}
\begin{align}
\left(\begin{array}{c}\pmb{Z}_\theta\\ \pmb{Z}_\phi\end{array}\right)=\left(\gamma\pmbmt{C}+2\bar\Omega\pmbmt{B}\right)^{-1}\left(\begin{array}{c}\pmb{Y}_\theta \\ \pmb{Y}_\phi\end{array}\right).
\end{align}
The second order perturbations $\pmb{h}^{(2)}$ and $\pmb{t}^{(2)}$ are represented by using $\pmb{y}_1^{(2)}$ and $\pmb{y}_2^{(2)}$ and the inhomogeneous terms.
Substituting $\pmb{h}^{(2)}$ and $\pmb{t}^{(2)}$ as given by (\ref{eq:h2t2}) into equations (\ref{eq:dy1_vec}), (\ref{eq:dy2_vec}),
(\ref{eq:dy3_vec}), and (\ref{eq:ds_vec}), we obtain
\be
r{\partial\over\partial r}\pmb{y}_1^{(2)}=\left[-\left(3+rA-{V\over\Gamma_1}\right)\pmbmt{1}+\widetilde{\pmbmt{\Lambda}}\pmbmt{F}_{11}\right]\pmb{y}_1^{(2)}
+\left(-\gamma{V\over\Gamma_1}\pmbmt{1}+{\widetilde{\pmbmt{\Lambda}}\pmbmt{F}_{12}\over c_1}\right)
{\pmb{y}_2^{(2)}}+\gamma\alpha_T\pmb{s}^{(2)}+\pmb{J}_1,
\label{eq:dy1_vec_gamma}
\ee
\begin{align}
r{\partial\over\partial r}\pmb{y}_2^{(2)}=-c_1\left(\gamma\pmbmt{1}+2\bar\Omega\pmbmt{C}_0\pmbmt{F}_{21}\right)\pmb{y}_1^{(2)}+\left[(1-rA-U)\pmbmt{1}-2\bar\Omega\pmbmt{C}_0\pmbmt{F}_{22}\right]\pmb{y}_2^{(2)}+\alpha_T\pmb{s}^{(2)}
+\pmb{J}_2,
\label{eq:dy2_vec_gamma}
\end{align}
\begin{align}
r{\partial\over \partial r}\pmb{l}^{(2)}
=c_2{rA\over\alpha_T}\pmb{y}_1^{(2)}-{\nabla_{\rm ad}\over\nabla}{\pmbmt{\Lambda}}{\pmb{y}_2^{(2)}}-{d\ln L_r\over d\ln r}\pmb{l}^{(2)}-\left({\pmbmt{\Lambda}\over V\nabla}+\gamma c_2\pmbmt{1}\right)\pmb{s}^{(2)}
+\pmb{J}_3,
\label{eq:dy3_vec_gamma}
\end{align}
\begin{align}
{1\over V\nabla_{}}r{\partial \over \partial r}\pmb{s}^{(2)}
&={\nabla_{\rm ad}\over\nabla}c_1\left(\gamma\pmbmt{1}+2\bar\Omega\pmbmt{C}_0\pmbmt{F}_{21}\right)\pmb{y}_1^{(2)}+\left(C_{42}\pmbmt{1}+{\nabla_{\rm ad}\over\nabla}2\bar\Omega\pmbmt{C}_0\pmbmt{F}_{22}\right){\pmb{y}_2^{(2)}}-\pmb{l}^{(2)}+\left[4-\kappa_s-\left({\nabla_{\rm ad}\over\nabla}-1\right)\alpha_T\right]\pmb{s}^{(2)}
+\pmb{J}_4,
\label{eq:ds_vec_gamma}
\end{align}
where 
\be
\pmb{J}_1=\pmb{Y}_c+{\widetilde{\pmbmt{\Lambda}}\pmb{Z}_\theta\over c_1}, \quad
\pmb{J}_2=\pmb{Y}_r-2\bar\Omega\pmbmt{C}_0\pmbmt{Z}_\phi, \quad
\pmb{J}_3=c_2{\pmb{Y}}_s+{\pmb{f}\over V\nabla}, \quad
\pmb{J}_4=-{\nabla_{\rm ad}\over\nabla}\pmb{J}_2-{\pmb{X}\over V\nabla},
\ee
and we have rewritten
\begin{align}
\pmbmt{F}=\left(\begin{array}{cc}\pmbmt{F}_{11} & \pmbmt{F}_{12}\\\pmbmt{F}_{21} &\pmbmt{F}_{22}\end{array}\right). 
\end{align}
The set of partial differential equations now reduces to the set of ordinary differential equations (\ref{eq:dy1_vec_gamma}) to (\ref{eq:ds_vec_gamma}), which can be formally written as
\be
r{\partial\pmb{y}^{(2)}\over\partial r}=\pmbmt{\cal L}(r,\gamma)\pmb{y}^{(2)}+\pmb{J}(r,\gamma),
\ee
where $\pmbmt{\cal L}$ is the coefficient matrix, and
\be
\pmb{y}^{(2)}=\left(\begin{array}{c}\pmb{y}_1^{(2)}\\ \pmb{y}_2^{(2)}\\ \pmb{l}^{(2)} \\ \pmb{s}^{(2)} \end{array}\right),
\quad 
\pmb{J}=\left(\begin{array}{c}\pmb{J}_1\\ \pmb{J}_2\\ \pmb{J}_3 \\ V\nabla \pmb{J}_4 \end{array}\right).
\ee
Regarding $\gamma$ as a parameter, we solve the set of differential equations (\ref{eq:dy1_vec_gamma}) to (\ref{eq:ds_vec_gamma}) imposing boundary conditions at the
center and surface of the star.

Steady states of the second order perturbations are obtained by setting $\gamma=0$, for which
\be
\pmbmt{F}={1\over 2\bar\Omega}\left(\begin{array}{cc} 0 & \widetilde{\pmbmt{B}}_0^{-1} \\
\pmbmt{B}_1^{-1} & 0\end{array}\right)\pmbmt{A}={1\over 2\bar\Omega}\left(\begin{array}{cc}-2\bar\Omega f_r\widetilde{\pmbmt{B}}_0^{-1}\pmbmt{A}_0 & 0\\ 0 & -\pmbmt{B}_1^{-1}\pmbmt{C}_1\end{array}\right),
\ee
that is,
$
\pmbmt{F}_{11}=-f\widetilde{\pmbmt{B}}_0^{-1}\pmbmt{A}_0$, $\pmbmt{F}_{12}=0$, $\pmbmt{F}_{21}=0$, 
$\pmbmt{F}_{22}=-\pmbmt{B}_1^{-1}\pmbmt{C}_1/2\bar\Omega$,
and
\be
\left(\begin{array}{c} \pmb{Z}_\theta \\ \pmb{Z}_\phi\end{array}\right)={1\over 2\bar\Omega}\left(\begin{array}{c}\widetilde{\pmbmt{B}}_0^{-1}\pmb{Y}_\phi \\ \pmbmt{B}_1^{-1}\pmb{Y}_\theta\end{array}\right).
\ee
We thus obtain
\be
r{\partial\over\partial r}\pmb{y}_1^{(2)}=\left[-\left(3+rA-{V\over\Gamma_1}\right)\pmbmt{1}+\widetilde{\pmbmt{\Lambda}}\pmbmt{F}_{11}\right]\pmb{y}_1^{(2)}
+\pmb{J}_1,
\label{eq:dy1_vec_gamma0}
\ee
\begin{align}
r{\partial\over\partial r}\pmb{y}_2^{(2)}=\left[(1-rA-U)\pmbmt{1}-2\bar\Omega\pmbmt{C}_0\pmbmt{F}_{22}\right]\pmb{y}_2^{(2)}+\alpha_T\pmb{s}^{(2)}
+\pmb{J}_2,
\label{eq:dy2_vec_gamma0}
\end{align}
\begin{align}
r{\partial\over \partial r}\pmb{l}^{(2)}
=c_2{rA\over\alpha_T}\pmb{y}_1^{(2)}-{\nabla_{\rm ad}\over\nabla}{\pmbmt{\Lambda}}{\pmb{y}_2^{(2)}}-{d\ln L_r\over d\ln r}\pmb{l}^{(2)}-{\pmbmt{\Lambda}\over V\nabla}\pmb{s}^{(2)}
+\pmb{J}_3,
\label{eq:dy3_vec_gamma0}
\end{align}
\begin{align}
{1\over V\nabla_{}}r{\partial \over \partial r}\pmb{s}^{(2)}
&=\left(C_{42}\pmbmt{1}+{\nabla_{\rm ad}\over\nabla}2\bar\Omega\pmbmt{C}_0\pmbmt{F}_{22}\right){\pmb{y}_2^{(2)}}-\pmb{l}^{(2)}+\left[4-\kappa_s-\left({\nabla_{\rm ad}\over\nabla}-1\right)\alpha_T\right]\pmb{s}^{(2)}
+\pmb{J}_4,
\label{eq:ds_vec_gamma0}
\end{align}
and the horizontal components $\pmb{h}^{(2)}$ and $\pmb{t}^{(2)}$ are given by
\be
2\bar\Omega\pmbmt{B}_1\pmb{t}^{(2)}=-\pmbmt{C}_1\pmb{y}_2^{(2)}/c_1+\pmb{Y}_\theta/c_1, 
\label{eq:b1t2nad}
\ee
\be
2\bar\Omega\widetilde{\pmbmt{B}}_0\pmb{h}^{(2)}=-2\bar\Omega f_r\pmbmt{A}_0\pmb{y}_1^{(2)}+\pmb{Y}_\phi/c_1.
\label{eq:b0h2nad}
\ee

\section{Numerical Results}

We calculate axisymmetric second order perturbations driven by non-axisymmetric unstable low frequency modes of rotating stars.
We first compute linear oscillation modes and
calculate the inhomogeneous terms using their eigenfrequencies and eigenfunctions.
We then solve the inhomogeneous set of differential equations for the second order perturbations.
As the background models we use $4M_\odot$ main sequence models with $X=0.7$ and $Y=0.28$.
The models are produced by Paczynski code (\citealt{Paczynski1970}) and
no effects of rotational deformation on the models are considered.

\subsection{Linear mode calculations}

\begin{figure*}
\resizebox{0.45\columnwidth}{!}{
\includegraphics{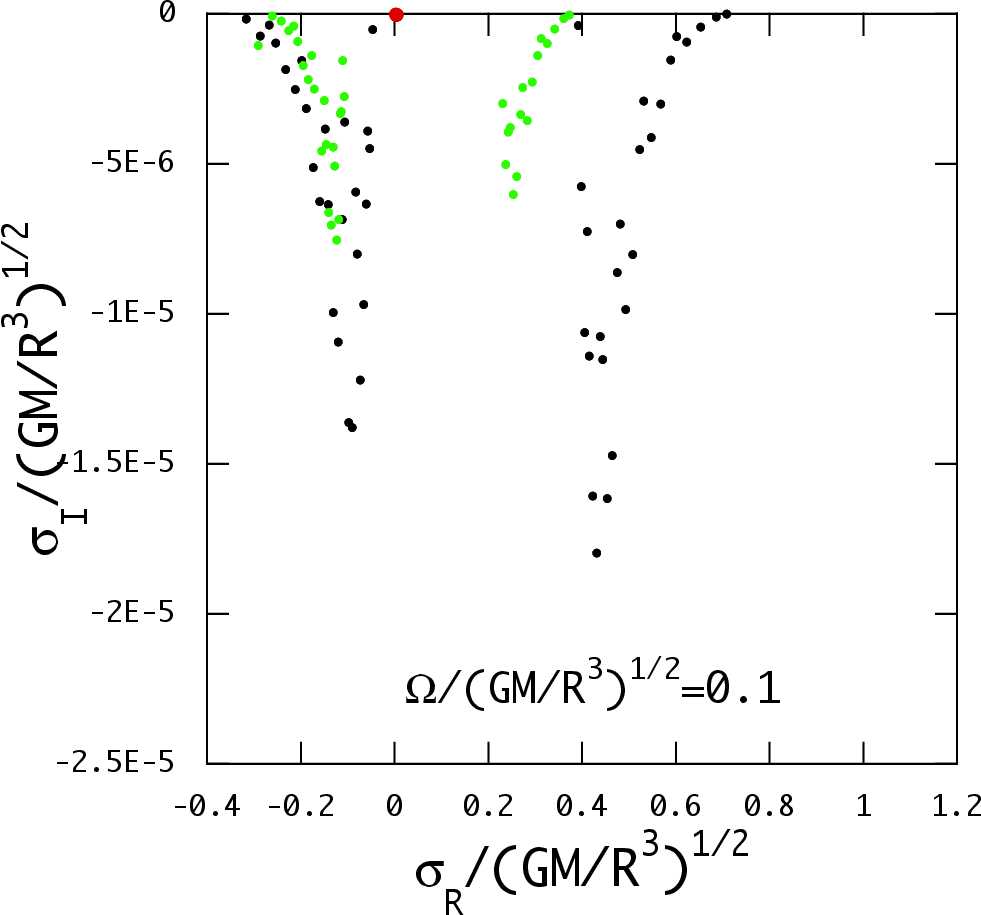}}
\hspace{0.3cm}
\resizebox{0.45\columnwidth}{!}{
\includegraphics{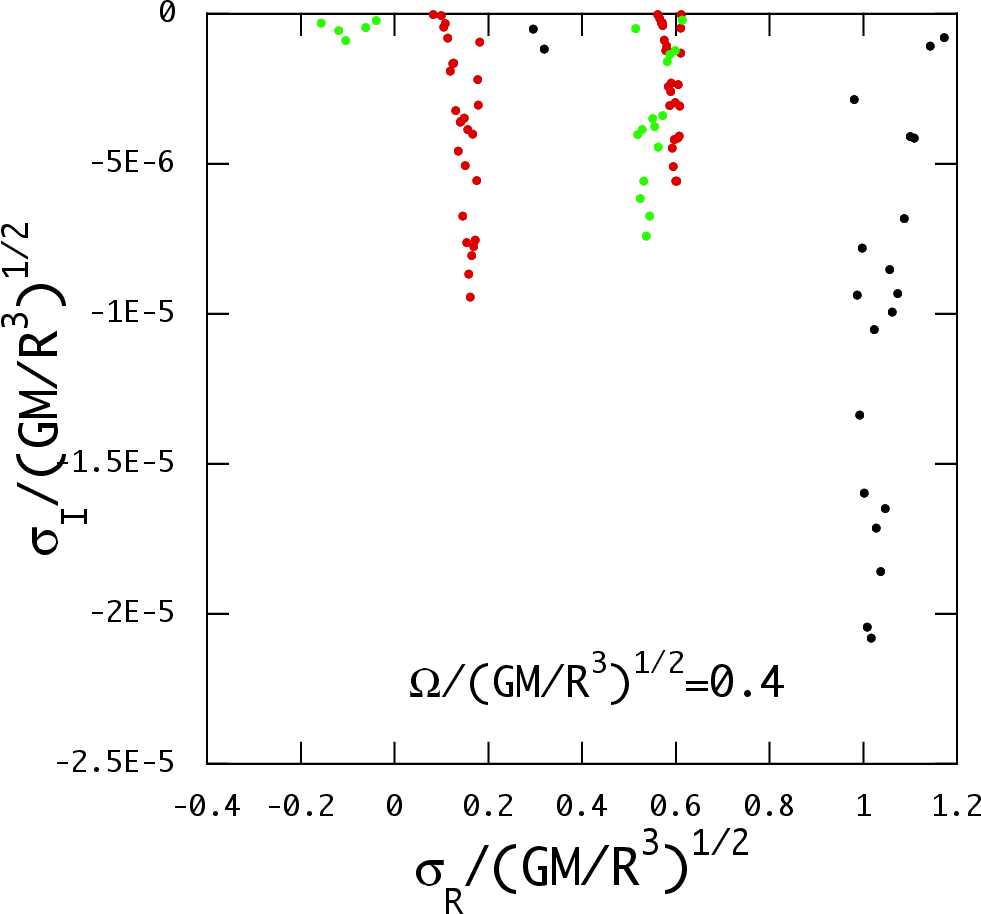}}
\caption{$\bar\sigma_{\rm I}$ versus $\bar\sigma_{\rm R}$ for even parity $g$-modes and odd parity $r$-modes of the $4M_\odot$ main sequence model at $\bar\Omega=\Omega/\sqrt{GM/R^3}=0.1$ (left panel) and $\bar\Omega=0.4$ (right panel), where the black dots and green dots respectively stand for $g$-modes of $m=-2$ and $-1$ and the red dots for
$r$-modes of $m=-2$ and $-1$. Note that low radial order $r$-modes of odd parity have inertial frame frequency $\sigma_{\rm R}\approx -{(|m|+2)(|m|-1)\over|m|(|m|+1)}m\Omega$ where $\sigma=\sigma_{\rm R}+\rmi\sigma_{\rm I}$ signifies the complex eigenfrequency in the inertial frame.
The modes of $\sigma_{\rm I}<0$ are unstable.
For $m<0$, $\sigma_{\rm R}$ of retrograde $g$-modes are less than $\sigma_{\rm R}$ of prograde $g$-modes.}
\label{fig:sisr_nonad}
\end{figure*}

Assuming uniform rotation, we carried out non-adiabatic calculations of $g$-modes and $r$-modes for rotation rates $\bar\Omega=\Omega/\sigma_0=0.1$ and $\bar\Omega=0.4$ and obtained the eigenfrequencies $\sigma$ and eigenfunctions $\pmb{\xi}$, where the luminosity $L$ and effective temperature $T_{\rm eff}$ of the model are given by $\log(L/L_\odot)=2.496$ and $\log T_{\rm eff}=4.129$, and the hydrogen content at the center is $X_c=0.356$.
For the mode calculations we used the method of calculation by \citet{LeeBaraffe95}.
The results of non-adiabatic analyses are summarized in Fig. \ref{fig:sisr_nonad}, in which we plot $\bar\sigma_{\rm I}$ of even parity $g$-modes and odd parity $r$-modes versus $\bar\sigma_{\rm R}$ for $m=-1$ and $-2$.
Note that modes whose pressure perturbation $p'$ is symmetric (anti-symmetric) about the equator are called even (odd) parity modes in this paper.
In the co-rotating frame of the star, $g$-modes are separated into prograde modes and retrograde modes and the modal properties can be significantly different between
prograde and retrograde modes when $|2\Omega/\omega_{\rm R}|\gtrsim 1$.
For $m<0$, in our convention, prograde (retrograde) $g$-modes have positive (negative) $\omega_{\rm R}$, and hence $\sigma_{\rm R}({\rm retrograde})<\sigma_{\rm R}({\rm prograde})$. 
On the other hand, $r$-modes are retrograde modes and have $\omega_{\rm R}<0$ in the co-rotating frame.
The frequency of low radial order $r$-modes whose angular dependence of the eigenfunction $\xi_\phi$ is dominated by $\partial Y_{l'}^m(\theta,\phi)/\partial\theta$ is approximately given by
$\omega_{\rm R}\approx 2m\Omega/l'(l'+1)<0$ and hence $\sigma_{\rm R}=\omega_{\rm R}-m\Omega\approx -{(l'+2)(l'-1)\over l'(l'+1)}m\Omega>0$ for $m<0$ where $l'=|m|+1$ for even modes and $l'=|m|$ for odd modes.
Note that $\sigma_{\rm R}\approx 0$ for odd parity $r$-modes of $l'=|m|=1$.
Fig. \ref{fig:sisr_nonad} shows that for $\bar\Omega=0.1$ many prograde and retrograde $g$-modes of $m=-1$ and $-2$ are pulsationally unstable.
For $\bar\Omega=0.4$, however, although both prograde and retrograde $g$-modes of $m=-1$ are unstable,
only prograde $g$-modes of $m=-2$ are found unstable .
This is an example of differences found between prograde and retrograde $g$-modes, that is,
retrograde $g$-modes are more likely to be stabilized than prograde $g$-modes.
Fig. \ref{fig:sisr_nonad} also shows that for $\bar\Omega=0.1$ only $m=-1$ $r$-modes are unstable, 
but both $m=-1$ and $m=-2$ $r$-modes become unstable for $\bar\Omega=0.4$.
Note that we find no unstable $r$-modes of even parity whose angular dependence of $\xi_\phi$ is dominated by $\partial Y_{|m|+1}^m(\theta,\phi)/\partial\theta$.

\subsection{Second order perturbations}

As discussed in \S 2.3, to obtain axisymmetric second order perturbations of rotating stars, we solve the
inhomogeneous set of linear differential equations, solutions to which are given by the sum of complementary solutions and particular solutions.
Complementary solutions are given by a linear combination of the eigen-solutions $\alpha_i$ and $\hat{\pmb{z}}_i^{(2)}$ to equation (\ref{eq:formaleq}) with $\pmb{Y}^{(2)}=0$ and $\pmb{g}_k=0$.
Under the assumptions $v_r^{(0)}=v_\theta^{(0)}=0$ and $v_\phi^{(0)}=r\sin\theta\Omega$ with constant $\Omega$, no eigen-solutions with ${\rm Re}(\alpha_i)>0$ are numerically found, suggesting that complementary solutions all decay with time.
We therefore assume that only particular solutions which are computed for the forcing terms of the form $\pmb{Y}^{(2)}(r,\tau)=\pmb{Y}^{(2)}(r) {\rm e}^{\gamma\tau}$ are relevant.
The time dependence of the oscillation mode is given by ${\rm e}^{\rmi\sigma t}$ with $\sigma=\sigma_{\rm R}+\rmi \sigma_{\rm I}$. 
For an unstable mode with $\sigma_{\rm I}<0$, the forcing term is proportional to 
${\rm e}^{-2\sigma_{\rm I}t}$ and hence we have $\gamma=-2\sigma_{\rm I}>0$.
For adiabatic modes, we usually have $\sigma_{\rm I}=0$ and hence $\gamma=0$, that is, both the first order perturbations and the second order perturbations are stationary.

\subsubsection{Second order perturbations for adiabatic modes}

We assume $\delta s^{(2)}=0$ for adiabatic second order perturbations, where $\delta s^{(2)}$ is
the Lagrangian entropy perturbation of second order given by
\be
\delta s^{(2)}=s^{(2)}+\pmb{\xi}\cdot\nabla s^{(1)}
+{1\over 2}\left[\xi_r\xi_r{\partial^2 s^{(0)}\over\partial r^2}
+\left(\xi_\theta\xi_\theta+\xi_\phi\xi_\phi\right){1\over r}{\partial s^{(0)}\over\partial r}\right].
\ee
From $\overline{\delta s^{(2)}}=0$, we have
\be
s_l^{(2)}=-\int do Y_l^0\left\{\overline{\pmb{\xi}\cdot\nabla s^{(1)}}
+{1\over 2}\left[\overline{\xi_r\xi_r}{\partial^2 s^{(0)}\over\partial r^2}
+\left(\overline{\xi_\theta\xi_\theta}+\overline{\xi_\phi\xi_\phi}\right){1\over r}{\partial s^{(0)}\over\partial r}\right]\right\}\equiv-c_pJ_l^s,
\ee
where $s_l^{(2)}=\int do Y_l^0s^{(2)}$.
Substituting $s_l^{(2)}/c_p=-J_l^s$ into equations (\ref{eq:dy1_vec_gamma}) and
(\ref{eq:dy2_vec_gamma}), we obtain for
adiabatic second order perturbations driven by adiabatic modes 
\be
r{\partial\over\partial r}\pmb{y}_1^{(2)}=\left[-\left(3+rA-{V\over\Gamma_1}\right)\pmbmt{1}+\widetilde{\pmbmt{\Lambda}}\pmbmt{F}_{11}\right]\pmb{y}_1^{(2)}
+\left(-\gamma{V\over\Gamma_1}\pmbmt{1}+{\widetilde{\pmbmt{\Lambda}}\pmbmt{F}_{12}\over c_1}\right)
{\pmb{y}_2^{(2)}}+\pmb{J}_1-\gamma\alpha_T\pmb{J}^s,
\label{eq:dy1_vec_gamma_ad}
\ee
\begin{align}
r{\partial\over\partial r}\pmb{y}_2^{(2)}=-c_1\left(\gamma\pmbmt{1}+2\bar\Omega\pmbmt{C}_0\pmbmt{F}_{21}\right)\pmb{y}_1^{(2)}+\left[(1-rA-U)\pmbmt{1}-2\bar\Omega\pmbmt{C}_0\pmbmt{F}_{22}\right]\pmb{y}_2^{(2)}
+\pmb{J}_2-\alpha_T\pmb{J}^s,
\label{eq:dy2_vec_gamma_ad}
\end{align}
where $(\pmb{J}^s)_j=J_{l_j}^s$.

It is interesting to note that $\pmb{Y}_c=\pmb{Y}_\phi=0$
for adiabatic modes with real eigenfrequencies $\sigma$ (i.e., the amplitudes stay constant), and hence
$\pmb{J}_1={\widetilde{\pmbmt{\Lambda}}\pmb{Z}_\theta / c_1}$.
However, it is also important to note that even for adiabatic modes if their eigenfrequencies are complex
such that $\sigma=\sigma_{\rm R}+\rmi\sigma_{\rm I}$ and the amplitudes grow or decay with time,
the inhomogeneous terms $\pmb{Y}_c$ and $\pmb{Y}_\phi$ do not
necessarily vanish identically.
Such an example may be given by overstable convective (OsC) modes 
in rotating main sequence stars which can be unstable and have complex eigenfrequencies even under the adiabatic approximation (e.g., \citealt{LeeSaio20,Lee2021}).

For steady states ($\gamma=0$) of adiabatic perturbations of second order, 
$\pmb{J}_1-\gamma\alpha_T\pmb{J}^s=\widetilde{\pmbmt{\Lambda}}\widetilde{\pmbmt{B}}_0^{-1}\pmb{Y}_\phi/(2c_1\bar\Omega)=0$, and 
we can show that
$\hat v_r=v_r^{(2)}+\widetilde v_r=0$ and $ \hat v_\theta=v_\theta^{(2)}+\widetilde v_\theta=0$,
and we obtain meridional flows of second order given by 
\be
v_r^{(2)}=-\widetilde v_r, \quad v_\theta^{(2)}=-\widetilde v_\theta.
\ee
But, since $A_\phi=-\rho^{(0)}\overline{\xi_rv_\theta}=-\rho^{(0)}{\rm Re}\left(\xi_r^*\rmi\omega\xi_\theta\right)/2=0$ and hence $\widetilde v_r=\widetilde v_\theta=0$, we simply have $v_r^{(2)}=v_\theta^{(2)}=0$
for adiabatic modes having real $\omega$.
On the other hand, $v_\phi^{(2)}$ is determined by the geostrophic balance
with the pressure perturbation $\pmb{y}_2^{(2)}$ as represented by equation (\ref{eq:h2t2}), and
$\pmb{y}_2^{(2)}$ itself is given by the particular solution to (\ref{eq:dy2_vec_gamma_ad}) with the non-zero inhomogeneous term $\pmb{J}_2-\alpha_T\pmb{J}^s$.
Note that $\pmbmt{F}_{21}=\pmbmt{F}_{12}=0$ for $\gamma=0$.

\begin{figure*}
\resizebox{0.4\columnwidth}{!}{
\includegraphics{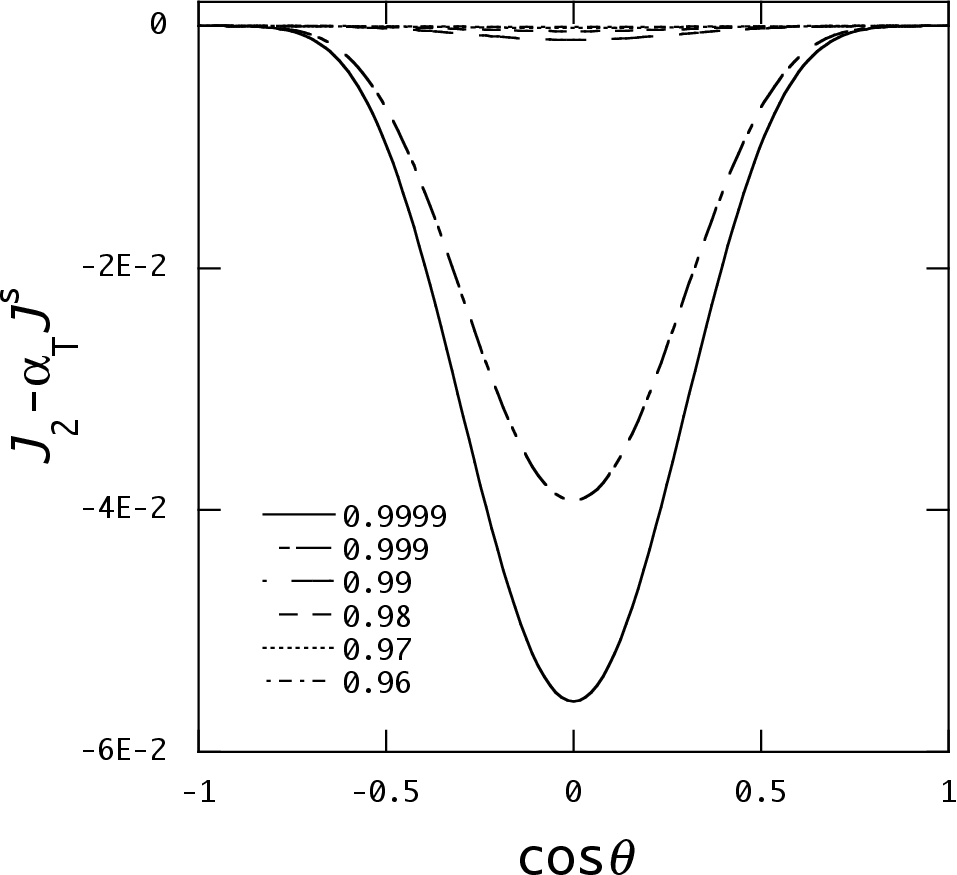}}
\hspace{0.3cm}
\resizebox{0.4\columnwidth}{!}{
\includegraphics{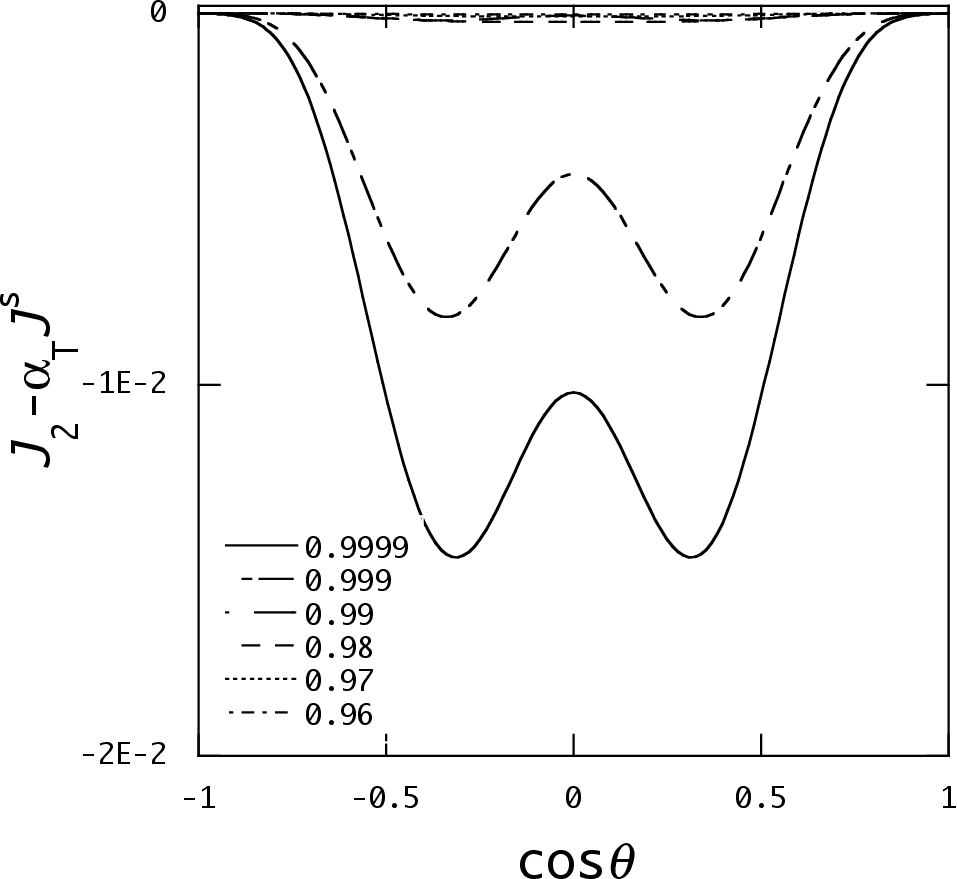}}
\caption{Inhomogeneous term $J_2-\alpha_TJ^s$ at different fractional radii is plotted versus $\cos\theta$ for the even prograde $g_{30}$-mode
(left panel) and for the odd $r_{30}$-mode (right panel) of $m=-2$ at $\bar\Omega=0.4$ where the linear modes are adiabatic and their amplitudes are determined by $\epsilon_E=10^{-10}$.
The numbers in the legend indicate the fractional radii.
}
\label{fig:j2mjs_ad}
\end{figure*}

\begin{figure*}
\resizebox{0.4\columnwidth}{!}{
\includegraphics{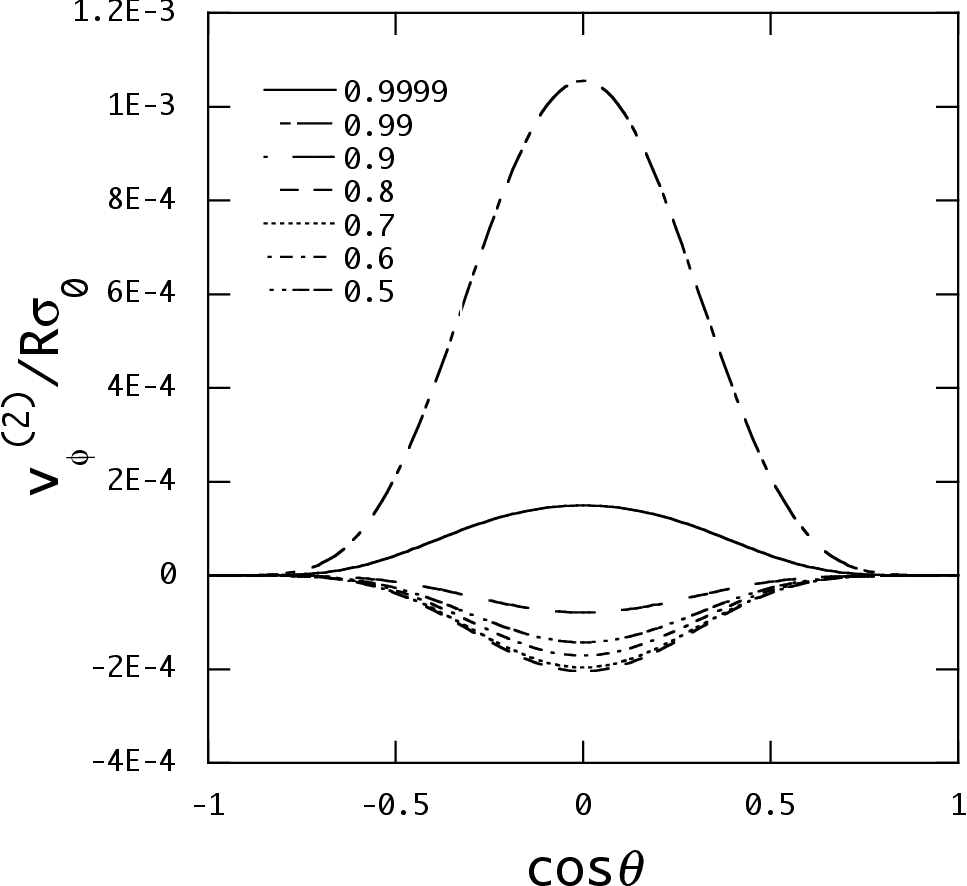}}
\hspace{0.3cm}
\resizebox{0.4\columnwidth}{!}{
\includegraphics{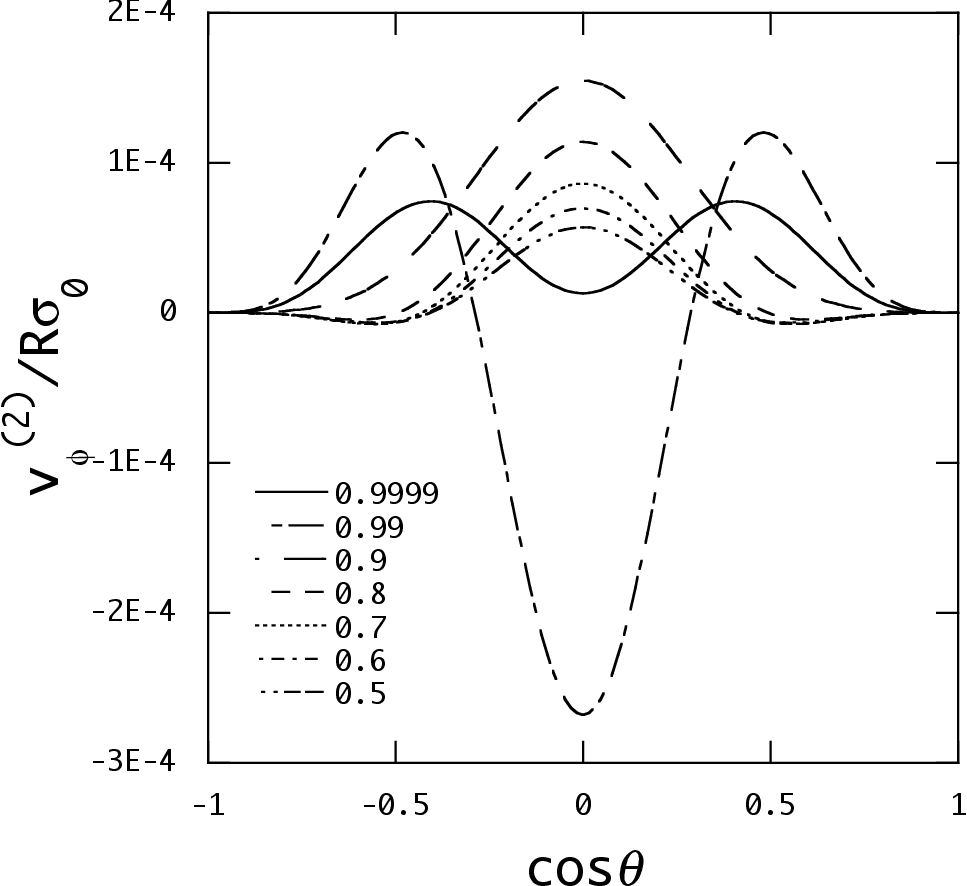}}
\caption{Same as Fig. \ref{fig:j2mjs_ad} but for $v_\phi^{(2)}/R\sigma_0$.}
\label{fig:vphi_ad}
\end{figure*}

We use the even prograde $g_{30}$-mode and the odd $r_{30}$-mode of $m=-2$ at $\bar\Omega=0.4$ for the oscillation modes that drive axisymmetric flows of second order where
the two modes are found unstable in non-adiabatic analyses.
Since $\pmb{Y}_c=\pmb{Y}_\phi=0$ for adiabatic modes with real eigenfrequencies, only the function $J_2-\alpha_TJ^s$ is non-zero inhomogeneous term and is plotted against $\cos\theta$
for several different fractional radii $r/R$
for the $g$-mode (left panel) and $r$-mode (right panel) in Fig. \ref{fig:j2mjs_ad}, where
\be
J_2-\alpha_TJ^s=\sum_{j=1}^{k_{\rm max}}\left(\pmb{J}_2-\alpha_T\pmb{J}^s\right)_jY_{l_j}^0,
\ee
and $l_j=2(j-1)$.
To compute the inhomogeneous term, we use the eigenfunctions of the modes whose amplitudes are determined by $\epsilon_E=10^{-10}$.
Fig. \ref{fig:j2mjs_ad} shows that the inhomogeneous term has large amplitudes only in a very thin surface layer and that the $\cos\theta$ dependence of the term is symmetric about the equator $\cos\theta=0$ 
for both of the modes but the dependence of the $g$-mode on $\cos\theta$ is different from that of the $r$-mode, that is,
the former has a single peak at the equator but the latter has two peaks at mid latitudes.
With the non-zero inhomogeneous term $J_2-\alpha_TJ^s$, we solve the adiabatic second order perturbation equations to obtain
$v_\phi^{(2)}(r,\theta)$ and $p^{(2)}(r,\theta)$ as given by equation (\ref{eq:vrthetaphi}).
In Fig. \ref{fig:vphi_ad}, we plot $v_\phi^{(2)}(r,\theta)$ at different fractional radii against $\cos\theta$ for the $g_{30}$-mode (left panel) and $r_{30}$-mode (right panel).
We find that the confinement of the amplitudes of $v_\phi^{(2)}(r,\theta)$ into the surface layers is not as strong as that of $J_2-\alpha_TJ^s$.
We also find that the $\theta$ dependence of $v_\phi^{(2)}(r,\theta)$ reflects
that of $J_2-\alpha_TJ^s$ although the maximum of $v_\phi^{(2)}(r,\theta)$ does not occur at $r/R=1$.

\begin{figure*}
\resizebox{0.45\columnwidth}{!}{
\includegraphics{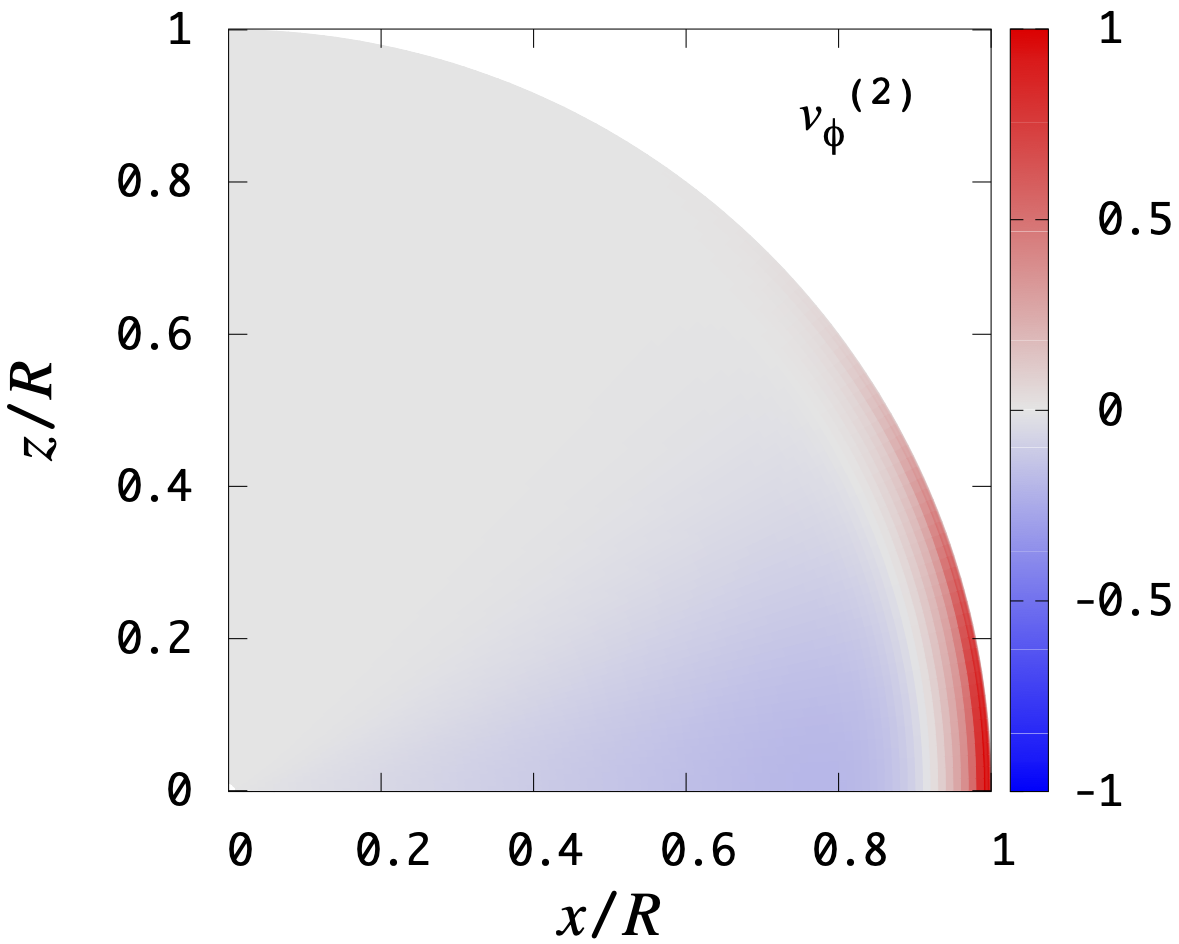}}
\hspace{0.3cm}
\resizebox{0.45\columnwidth}{!}{
\includegraphics{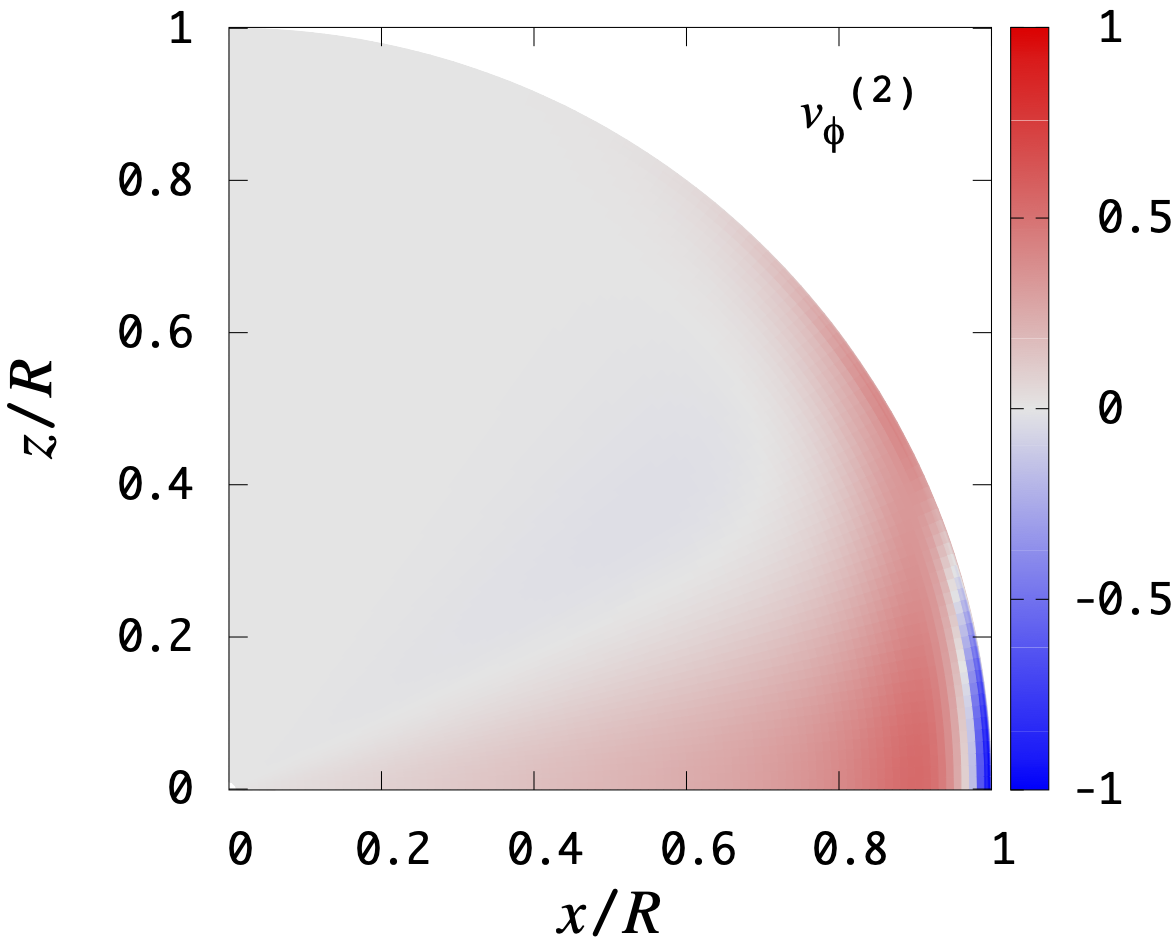}}
\caption{Color maps of $v_\phi^{(2)}$ for the adiabatic even prograde $g_{30}$-mode (left panel) and odd $r_{30}$-mode (right panel) of $m=-2$ at $\bar\Omega=0.4$, where 
$v_\phi^{(2)}$ is normalized by its maximum value.
Here, $\gamma=0$ is assumed.
}
\label{fig:colormap_vphi_ad}
\end{figure*}

Fig. \ref{fig:colormap_vphi_ad} shows the color maps of the function $v_\phi^{(2)}(r,\theta)$,
which tends to have large amplitudes in the surface equatorial regions.
We note that $v_\phi^{(2)}$ of the prograde $g_{30}$-mode is positive in the surface region and negative in the deep interior.
On the other hand, $v_\phi^{(2)}$ of the retrograde $r_{30}$-mode is negative in the surface equatorial region but positive both in the deep interior and in the surface layers at mid latitudes.

\begin{figure*}
\resizebox{0.4\columnwidth}{!}{
\includegraphics{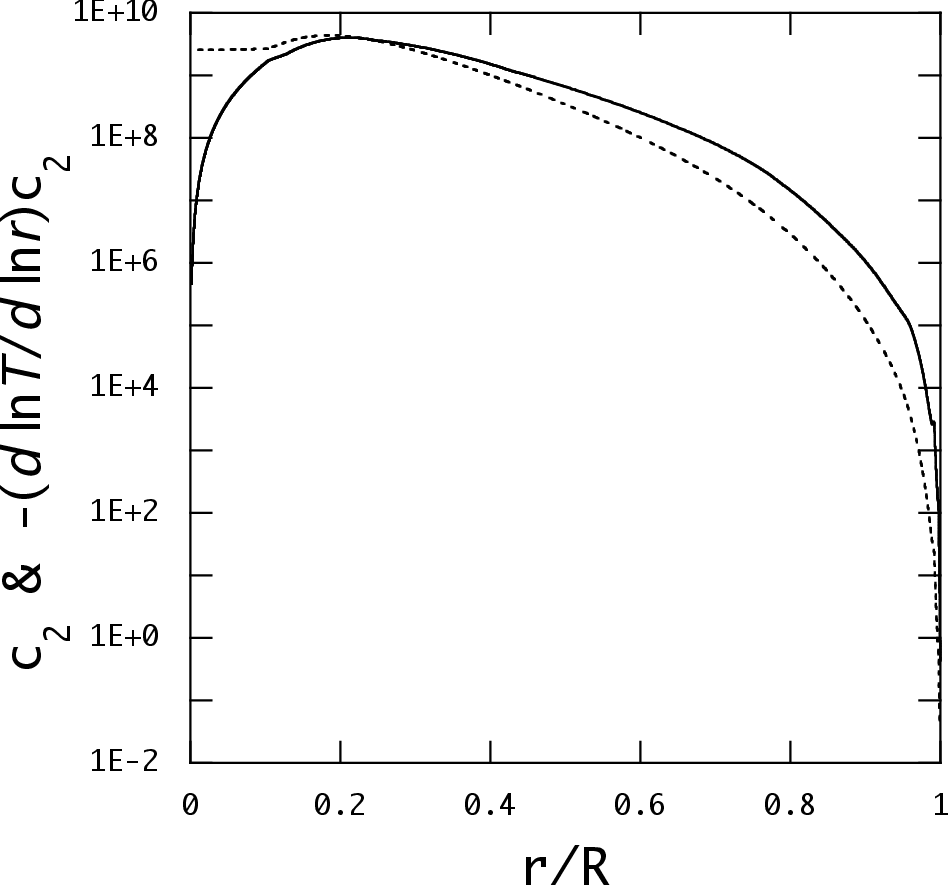}}
\caption{$-(d\ln T/d\ln r)c_2=V\nabla c_2$ (solid curve) and $c_2=4\pi r^3\rho Tc_p\sigma_0/L_{{\rm rad},r}$ (dotted curve) versus $r/R$ for the $4M_\odot$ model.}
\label{fig:c2}
\end{figure*}

\subsubsection {Second order perturbations for non-adiabatic modes}

Non-adiabatic second order perturbations driven by non-adiabatic modes are governed by equations
(\ref{eq:dy1_vec_gamma}) to (\ref{eq:ds_vec_gamma}).
It is important to point out that $v_\phi^{(2)}(\gamma)$ for $\gamma<0$ can be significantly
different from that for $\gamma>0$.
To understand this,
we rewrite equation (\ref{eq:ds_vec_gamma}) as $\pmb{l}^{(2)}=-(V\nabla)^{-1}r{\partial\pmb{s}^{(2)}/\partial r}+\cdots$, which we substitute into equation (\ref{eq:dy3_vec_gamma}) to obtain
\be
{\partial^2\pmb{s}^{(2)}\over\partial(\ln r)^2}=\left({\pmbmt{\Lambda} }+\gamma c_2V\nabla\pmbmt{1}\right)\pmb{s}^{(2)}+\cdots.
\label{eq:d2sdrr}
\ee
Fig. \ref{fig:c2} shows that the dimensionless quantities $c_2$ and $c_2V\nabla $ can be very large in the deep interior of the star.
The largeness of the quantity $c_2V\nabla$ in the interior suggests that
unless $|\gamma|$ is much smaller than $10^{-9}$ for this model, we can safely ignore the term $\pmbmt{\Lambda}$ for low $l$ components of $\pmb{s}^{(2)}$ in most regions of the radiative envelope and we can approximate equation (\ref{eq:d2sdrr}) to $\partial^2\pmb{s}^{(2)}/\partial(\ln r)^2\approx\gamma c_2 V\nabla\pmb{s}^{(2)}+\cdots$.
This equation suggests that, since $c_2V\nabla$ is large and positive, $\pmb{s}^{(2)}$ 
shows spatial oscillations with very short wavelengths $1/k_r=1/\sqrt{-\gamma c_2V\nabla}$ for negative $\gamma$.
On the other hand, the amplitudes of $\pmb{s}^{(2)}$ rapidly decay or grow monotonically over the length scale $1/\kappa =1/\sqrt{\gamma c_2V\nabla}$ for positive $\gamma$ and the amplitudes tend to be confined to a thin layer in the envelope.
That is, non-adiabatic second order perturbations for $\gamma<0$ can be very different from
those for $\gamma>0$.
In this paper, we consider the cases of positive $\gamma$ since only unstable oscillation modes can be
responsible for axisymmetric flows of second order when no external forces exist.

\begin{figure*}
\resizebox{0.4\columnwidth}{!}{
\includegraphics{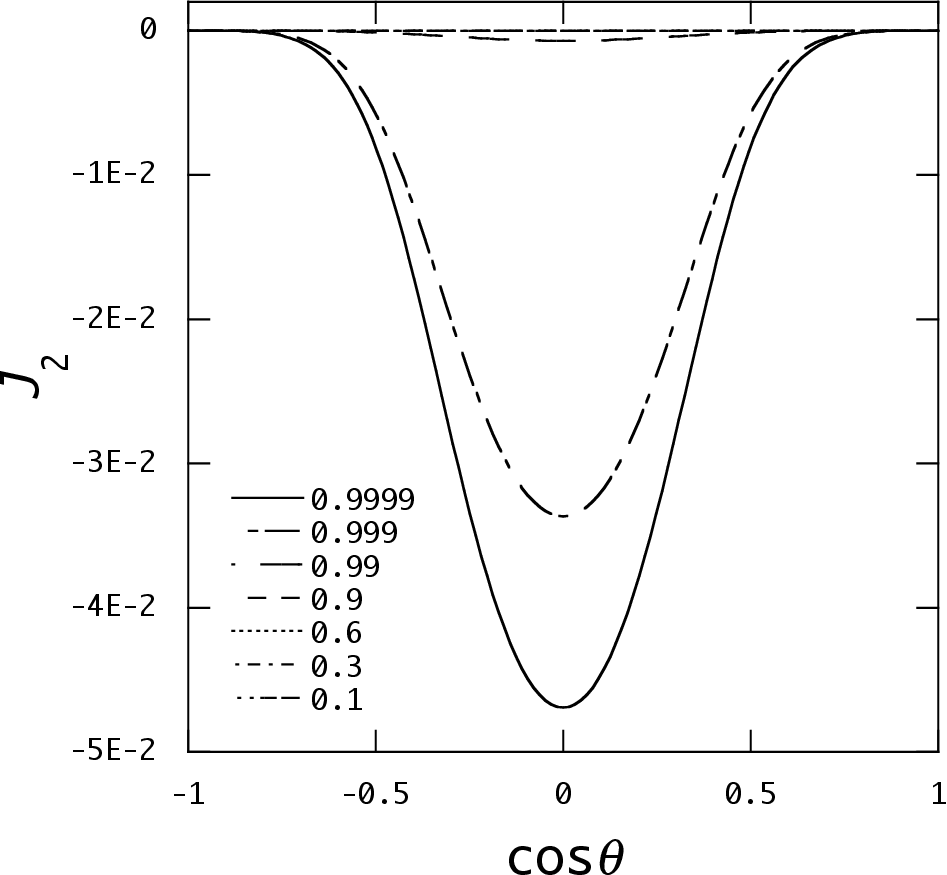}}
\hspace{0.3cm}
\resizebox{0.4\columnwidth}{!}{
\includegraphics{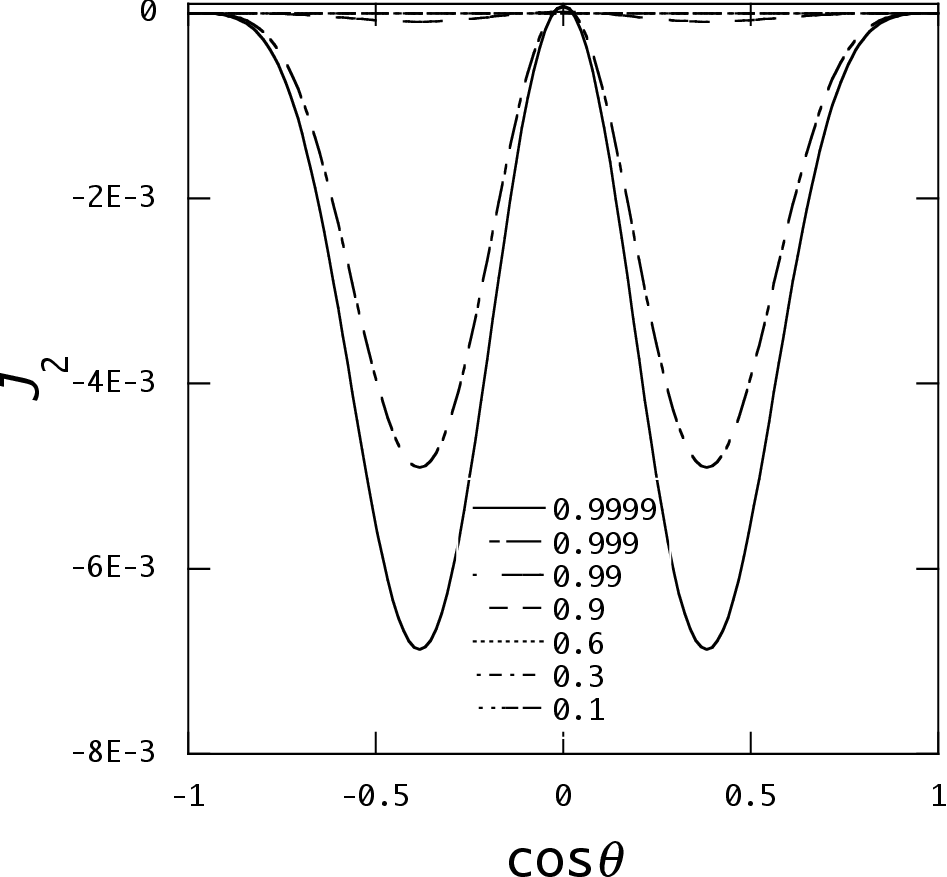}}
\caption{Inhomogeneous term $J_2$ at different radii is plotted versus $\cos\theta$ for the even prograde $g_{30}$-mode
(left panel) and for the odd $r_{30}$-mode (right panel) of $m=-2$ at $\bar\Omega=0.4$ where the linear modes are non-adiabatic and their amplitudes are normalized by $\epsilon_E=10^{-10}$.
The numbers in the legend indicate the fractional radii.
}
\label{fig:j2mm2gr30_nad}
\end{figure*}

Fig. \ref{fig:j2mm2gr30_nad} shows the inhomogeneous term $J_2=\sum_{j=1}^{k_{\rm max}}(\pmb{J}_2)_jY_{l_j}^0$ at different radii as a function of $\cos\theta$.
The $\cos\theta$ dependence of $J_2$ calculated for the non-adiabatic $g_{30}$- and $r_{30}$-modes is almost the same as that of $J_2-\alpha_TJ^s$ for the adiabatic modes although
the two peaks at mid latitudes for the non-adiabatic $r_{30}$-mode
are much sharper than for the adiabatic one.
We find that $J_2$ for the two non-adiabatic modes is strongly confined into the thin surface layer.

\begin{figure*}
\resizebox{0.4\columnwidth}{!}{
\includegraphics{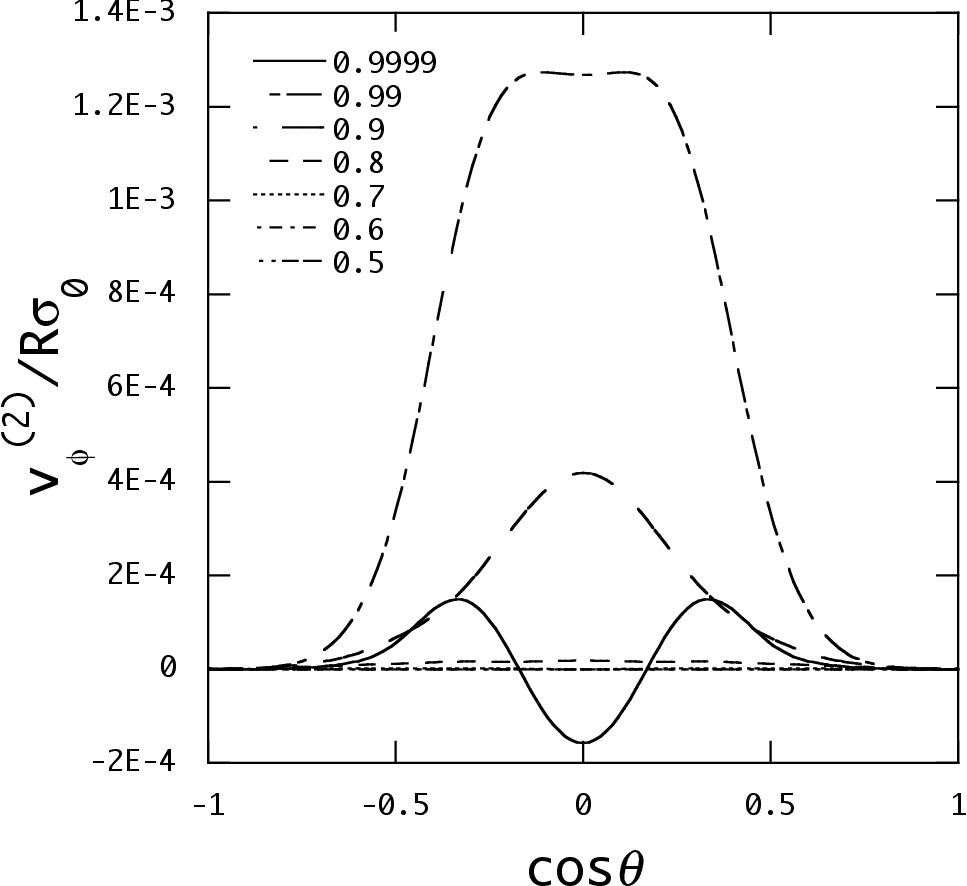}}
\hspace{0.3cm}
\resizebox{0.4\columnwidth}{!}{
\includegraphics{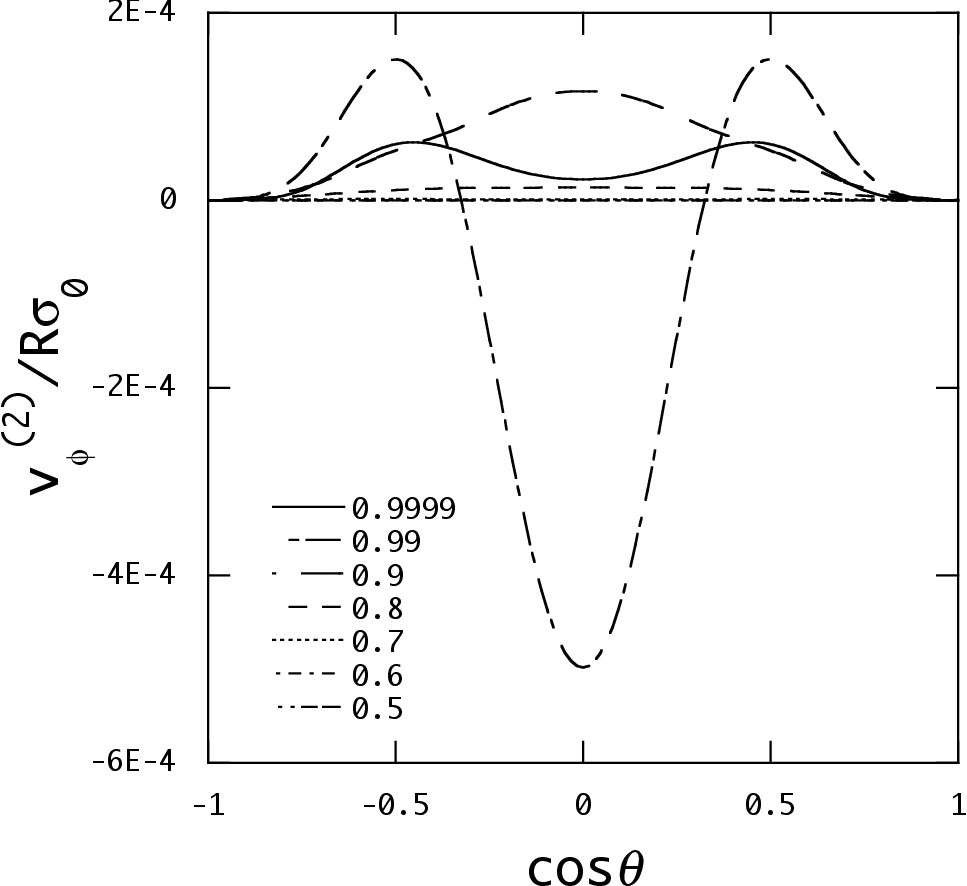}}
\caption{ Same as Fig. \ref{fig:j2mm2gr30_nad} but for $v_\phi^{(2)}/R\sigma_0$.
}
\label{fig:vphimm2gr30_nad_epsm6}
\end{figure*}

Although the behavior of $J_2$ as a function of $r$ and $\theta$ for the non-adiabatic modes
is quite similar to that for the adiabatic ones, non-adiabatic $v_\phi^{(2)}$ driven by non-adiabatic modes can be significantly different from adiabatic $v_\phi^{(2)}$.
In Fig. \ref{fig:vphimm2gr30_nad_epsm6}, non-adiabatic $v_\phi^{(2)}$ at different fractional radii for $\gamma=10^{-5}$ are plotted against $\cos\theta$ where
we have used the value $\gamma=10^{-5}$ so that $\gamma\sim-2\bar\sigma_{\rm I}$ for the non-adiabatic modes.
The non-adiabatic $v_\phi^{(2)}$ has appreciable amplitudes only in the outer envelope, and 
the amplitude confinement into the surface layers is much stronger than that for the adiabatic $v_\phi^{(2)}$.
The $\theta$ dependence of the non-adiabatic $v_\phi^{(2)}$ is similar to that for the adiabatic $v_\phi^{(2)}$ but the amplitudes of the non-adiabatic $v_\phi^{(2)}$ tend to be confined to the equatorial regions of the star.

\begin{figure*}
\resizebox{0.45\columnwidth}{!}{
\includegraphics{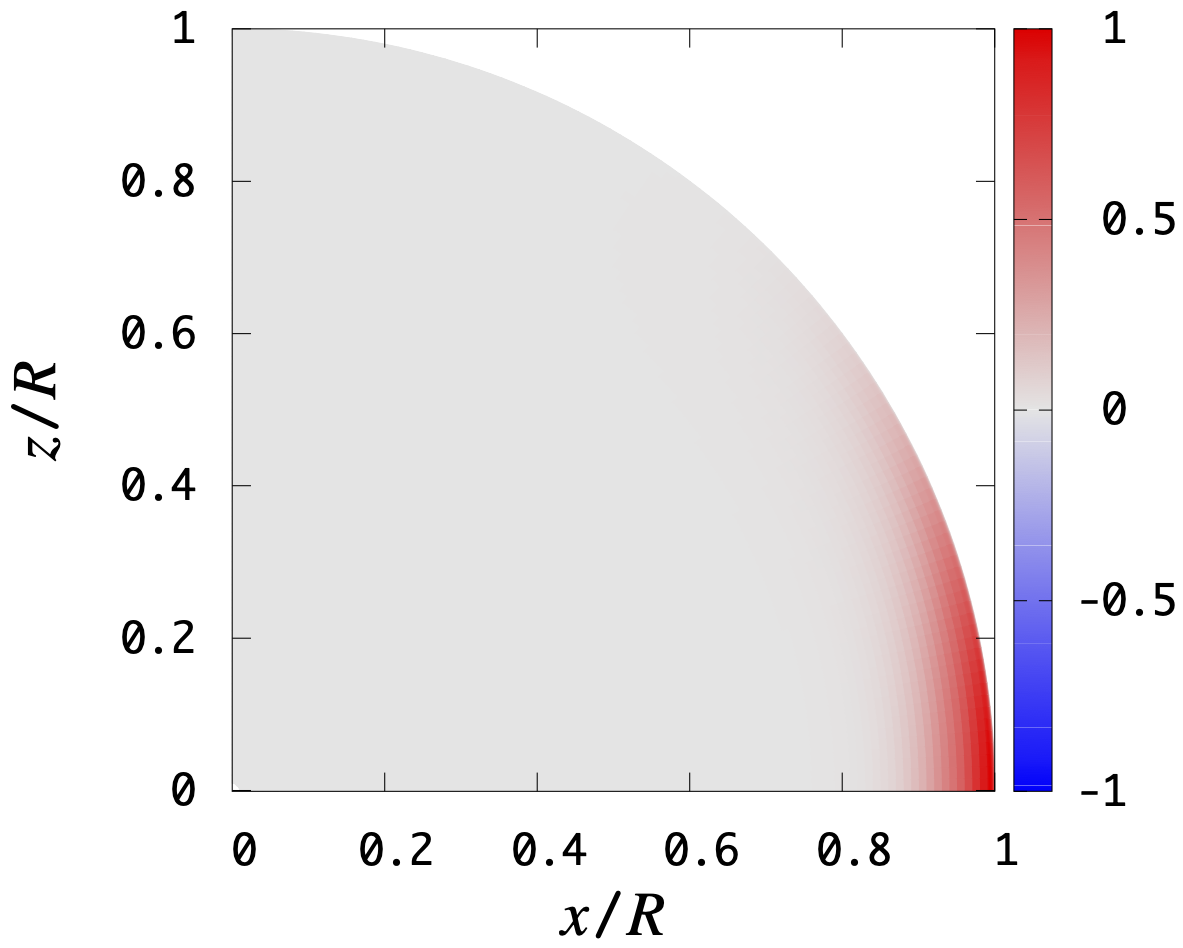}}
\hspace{0.3cm}
\resizebox{0.45\columnwidth}{!}{
\includegraphics{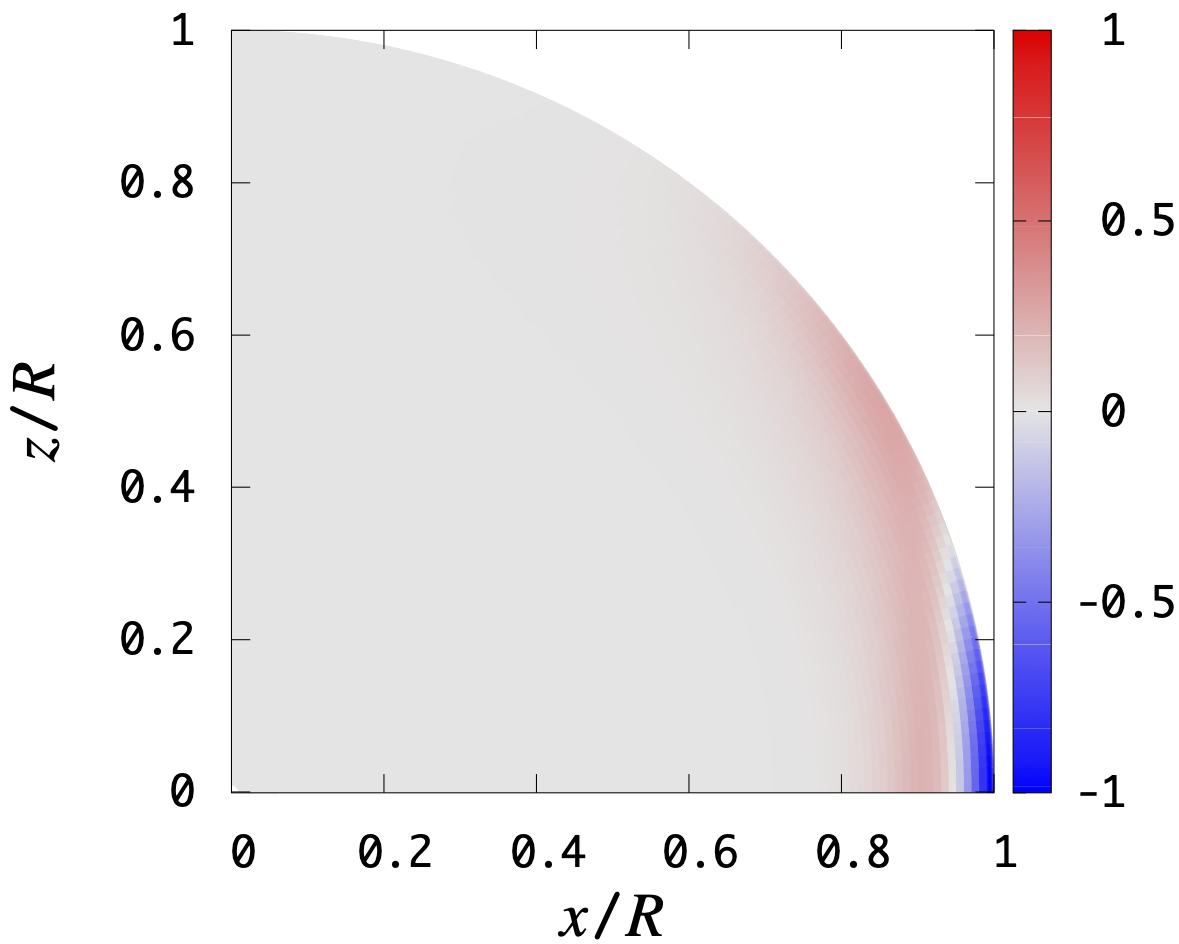}}
\caption{Color maps of $v_\phi^{(2)}$ driven by the even prograde $g_{30}$-mode
(left panel) and by the odd $r_{30}$-mode (right panel) of $m=-2$ at $\bar\Omega=0.4$ for $\gamma=10^{-5}$ where $v_\phi^{(2)}$ is normalized by its maximum value. Note that $\partial v_\phi^{(2)}/\partial \tau=\gamma v_\phi^{(2)}$.
}
\label{fig:vphimm2gr30_nadmap_epsm5}
\end{figure*}

Fig. \ref{fig:vphimm2gr30_nadmap_epsm5} shows the color maps of
non-adiabatic $v_\phi^{(2)}(r,\theta)$.
It is important to note that since $v_\phi^{(2)}\propto {\rm e}^{\gamma \tau}$, we have
$\partial v_\phi^{(2)}/\partial\tau=\gamma v_\phi^{(2)}$ and hence for $\gamma>0$, $v_\phi^{(2)}$ accelerates (decelerates) where $v_\phi^{(2)}$ is positive (negative).
We also note that
the acceleration and deceleration of $v_\phi^{(2)}$ take place because of the non-conservative (non-adiabatic) and non-steady effects of the oscillation modes.
Fig. \ref{fig:vphimm2gr30_nadmap_epsm5} suggests that acceleration of $v_\phi^{(2)}$ occurs in the surface equatorial regions for the $g_{30}$-mode,
and that the surface equatorial region is the place in which deceleration of $v_\phi^{(2)}$ takes place  for the $r_{30}$-mode although appreciable acceleration of $v_\phi^{(2)}$ also occurs inside the surface equatorial region.
When the amplitudes of non-adiabatic $v_\phi^{(2)}$ are strongly confined into the surface equatorial regions, $v_\phi^{(2)}$ shows the Taylor-Proudman (e.g., \citealt{Pedlosky1987}) like features.

\section{DISCUSSIONS}

\subsection{Angular momentum conservation equation of waves}

If we define the Lagrangian mean of the velocity field as $\overline{\pmb{v}}^L(\pmb{x},t)\equiv \overline{\pmb{v}(\pmb{x}+\pmb{\xi}(\pmb{x},t),t)}$ (\citealt{AndrewsMcintyre1978a})
, we obtain, assuming $\overline{\pmb{\xi}}=0$ and $a=O(|\pmb{\xi}|)\ll1$,
\begin{align}
\overline{\pmb{v}}^L(\pmb{x},t)=\pmb{v}^{(0)}+\overline{\delta\pmb{v}^{(2)}}+\cdots,
\end{align}
where $\overline{\delta\pmb{v}^{(2)}}$ is the zonally averaged Lagrangian velocity perturbation of second order, which is given as the sum of
the Stokes drift $\overline{\pmb{v}^{(2)}}^S$ (see the Appendix) and Eulerian mean velocity perturbation $\pmb{v}^{(2)}$ of second order, that is,
\begin{align}
\overline{\delta\pmb{v}^{(2)}}=\overline{\pmb{v}^{(2)}}^S+\pmb{v}^{(2)}.
\end{align}
We thus obtain for $\pmb{v}^{(0)}=v_\phi^{(0)}(\pmb{x})\pmb{e}_\phi$
\begin{align}
\overline{\pmb{v}}^L(\pmb{x},t)
=\overline{\delta v_r^{(2)}}\pmb{e}_r
+\overline{\delta v_\theta^{(2)}}\pmb{e}_\theta
+\left(v_\phi^{(0)}+\overline{\delta v_\phi^{(2)}}\right)\pmb{e}_\phi
=\left(\overline{v_r^{(2)}}^S+v_r^{(2)}\right)\pmb{e}_r+\left(\overline{v_\theta^{(2)}}^S+v_\theta^{(2)}\right)\pmb{e}_\theta+\left(v_\phi^{(0)}+\overline{v_\phi^{(2)}}^S+v_\phi^{(2)}\right)\pmb{e}_\phi.
\end{align}
Using $\overline{\pmb{v}}^L(\pmb{x},t)$ given above,
we define the mean material derivative $\overline{D}^L$ as (\citealt{AndrewsMcintyre1978a,Buhler2014})
\be
\overline{D}^L={\partial\over \partial t}+\overline{\pmb{v}}^L\cdot\nabla={\partial\over\partial t}+
\overline{\delta v_r^{(2)}}{\partial\over\partial r}+\overline{\delta v_\theta^{(2)}}{1\over r}{\partial\over\partial\theta}+\left(v_\phi^{(0)}+\overline{\delta v_\phi^{(2)}}\right){1\over r\sin\theta}{\partial\over\partial \phi}.
\ee

Since the partial time derivative of $\overline{v_\phi^{(2)}}^S$, the $\phi$-component of the Stokes drift (see equation (\ref{eq:StokesD00})), is given by
\begin{align}
{\partial \overline{v_\phi^{(2)}}^S\over\partial t}
&=-{1\over r\sin\theta\rho^{(0)}}\left(\overline{{\partial p^{(1)}\over\partial\phi}{\rho^{(1)}\over\rho^{(0)}}}-\overline{l^{(1)}{\cal D}\rho^{(1)}}\right)
-{\cal P}\overline{v_r^{(1)}{\rho^{(1)}\over\rho^{(0)}}}-{\cal Q}\overline{v_\theta^{(1)}{\rho^{(1)}\over\rho^{(0)}}}+{\cal D}H_\phi,
\end{align}
where $H_\phi$ is defined by equation (\ref{eq:hphi}), 
we can rewrite equation (\ref{eq:vphi_epf}) into
\begin{align}
{\partial\over\partial t}\left(v_\phi^{(2)}+\overline{v_\phi^{(2)}}^S\right)
&+{\cal P}\left(\hat v_r+\overline{v_r^{(1)}{\rho^{(1)}\over\rho^{(0)}}} \right)
+{\cal Q}\left(\hat v_\theta+\overline{v_\theta^{(1)}{\rho^{(1)}\over\rho^{(0)}}}\right)=-{1\over r\sin\theta\rho^{(0)}}\nabla\cdot\pmb{F}^{\rm EP}+{\cal D}H_\phi.
\label{eq:dvphidt_lag}
\end{align}
Since the $\phi$ component of $\overline{D}^L\overline{\pmb{v}}^L$, at the order of $a^2$, is given by
\be
\left(\overline{D}^L\overline{\pmb{v}}^L\right)_\phi={\partial\over\partial t}\overline{\delta v_\phi^{(2)}}
+\overline{\delta v_r^{(2)}}{\partial\over\partial r}v_\phi^{(0)}+\overline{\delta v_\theta^{(2)}}{1\over r}{\partial\over\partial\theta}v_\phi^{(0)}+{\overline{\delta v_r^{(2)}}v_\phi^{(0)}\over r}+{\overline{\delta v_\theta^{(2)}}v_\phi^{(0)}\over r}\cot\theta
={\partial\over\partial t}\overline{\delta v_\phi^{(2)}}
+{\cal P}\overline{\delta v_r^{(2)}}+{\cal Q}\overline{\delta v_\theta^{(2)}},
\ee
equation (\ref{eq:dvphidt_lag}) can be rewritten as
\begin{align}
\left(\overline{D}^L\overline{\pmb{v}}^L\right)_\phi
=-{1\over r\sin\theta\rho^{(0)}}\nabla\cdot\pmb{F}^{\rm EP}+{\cal D}\left(H_\phi+{\cal P}H_r+{\cal Q}H_\theta\right),
\label{eq:dvphidt_lag_b0}
\end{align}
where we have used equations (\ref{eq:vr_Stokes}) and (\ref{eq:vtheta_Stokes}) for $\overline{v_r^{(2)}}^S$ and $\overline{v_\theta^{(2)}}^S$, and equations (\ref{eq:hr}) and (\ref{eq:htheta}) for $H_r$ and $H_\theta$, respectively.
We then find that equation (\ref{eq:dvphidt_lag_b0}) reduces to
\be
\rho^{(0)}\overline{D}^L\bar l_\phi 
=-\nabla\cdot\pmb{F}^{\rm EP}+r\sin\theta\rho^{(0)}{\cal D}\left(H_\phi+{\cal P}H_r+{\cal Q}H_\theta\right),
\label{eq:lagmeanphi}
\ee
where $\bar l_\phi=l_\phi^{(0)}+l_\phi^{(2)}$ with $l_\phi^{(0)}=r\sin\theta v_\phi^{(0)}$ and
$l_\phi^{(2)}=r\sin\theta(v_\phi^{(2)}+\overline{v_\phi^{(2)}}^S)$ is the Eulerian mean specific angular momentum around the rotation axis possessed by the flow and waves.
Note that the right hand side of equation (\ref{eq:lagmeanphi}) is given solely by the sum of products of 
the eigenfunctions of oscillation modes.
Note also that for stationary waves the term ${\cal D}(\cdots)$ in equation (\ref{eq:lagmeanphi}) vanishes
and we simply obtain $\rho^{(0)}\overline{D}^L\bar l_\phi 
=-\nabla\cdot\pmb{F}^{\rm EP}$.

If we write $\pmb{l}(\pmb{x},t)=\pmb{x}\times\pmb{v}(\pmb{x},t)$, the Lagrangian mean of $\pmb{l}(\pmb{x},t)$ may be given by
\be
\overline{\pmb{l}}^L=\overline{\pmb{l}(\pmb{x}+\pmb{\xi}(\pmb{x},t),t)}=\pmb{x}\times\left(\pmb{v}^{(0)}+\overline{\delta \pmb{v}^{(2)}}\right)+\overline{\pmb{\xi}\times\delta\pmb{v}^{(1)}}
=\pmb{x}\times\left(\pmb{v}^{(0)}+\overline{\delta \pmb{v}^{(2)}}\right)+\overline{\pmb{\xi}\times\left(\pmb{v}^{(1)}+\pmb{\xi}\cdot\nabla\pmb{v}^{(0)}\right)}.
\ee
If we define $\overline{l}_\phi^L=\pmb{k}\cdot\overline{\pmb{l}}^L$ where $\pmb{k}=\cos\theta\pmb{e}_r-\sin\theta\pmb{e}_\theta$ is the unit vector along the rotation axis, we find that equation (\ref{eq:lagmeanphi}) can be rewritten into
\be
\rho^{(0)}\overline{D}^L\overline{l}_\phi^L=-\nabla\cdot\pmb{F}^W,
\label{eq:lagmeanphi_Lag}
\ee
where 
\be
\pmb{F}^W=\overline{\pmb{\xi}{\partial p^{(1)}\over\partial\phi}},
\ee
and we have used the relation given by
\begin{align}
\nabla\cdot\pmb{F}^{\rm EP}=
\nabla\cdot\pmb{F}^W+\nabla\cdot\pmb{F}^{\rm Z},
\label{eq:eptoz}
\end{align}
where
\begin{align}
\pmb{F}^{\rm Z}=r\sin\theta\rho^{(0)}\left({\cal D}\overline{\xi_r v_\phi^{(1)}}+{1\over 2}{\cal P}{\cal D}\overline{\xi_r\xi_r}\right)\pmb{e}_r
+r\sin\theta\rho^{(0)}\left({\cal D}\overline{\xi_\theta v_\phi^{(1)}}+{\cal P}{\cal D}\overline{\xi_\theta\xi_r}+{1\over 2}{\cal Q}{\cal D}\overline{\xi_\theta\xi_\theta}\right)\pmb{e}_\theta.
\end{align}
Note that $\pmb{k}\cdot[\pmb{x}\times(\pmb{v}^{(0)}+\overline{\delta\pmb{v}^{(2)}})]=l_\phi^{(0)}+l_\phi^{(2)}=\overline{l}_\phi$.
Equation (\ref{eq:lagmeanphi}) may be regarded as the Eulerian mean angular momentum conservation equation for non-axisymmetric waves, in which $\nabla\cdot\pmb{F}^{\rm EP}$ plays an essential role.
On the other hand, equation (\ref{eq:lagmeanphi_Lag}) is the Lagrangian mean angular momentum
conservation equation of non-axisymmetric waves and has the simpler form than the Eulerian mean equation
(\citealt{Lee13}).

\subsection{Eliassen-Palm flux $\pmb{F}^{\rm EP}$}

\citet{AndrewsMcintyre1976a} showed that the terms $|{\cal P}\hat v_r|$ and $|{\cal Q}\hat v_\theta|$ in equation (\ref{eq:vphi_epf}) is negligible compared to $|\nabla\cdot\pmb{F}^{\rm EP}/r\sin\theta\rho^{(0)}|$ for equatorial waves and the mid-latitude, quasi-geostrophic, vertically propagating Rossby waves in the atmosphere of the earth and that $\partial v_\phi^{(2)}/\partial t$ for the waves is in a good approximation determined by
\be
r\sin\theta\rho^{(0)}{\partial v_\phi^{(2)}\over\partial t}=-\nabla\cdot\pmb{F}^{\rm EP}.
\label{eq:andmci}
\ee
They also discussed that non-conservative and non-steady contributions of the waves to $\nabla\cdot\pmb{F}^{\rm EP}$ are essential to determine $\partial v_\phi^{(2)}/\partial t$.
For stellar $g$-modes and $r$-modes, however, the terms ${\cal P}\hat v_r$ and ${\cal Q}\hat v_\theta$
are not necessarily smaller than $\nabla\cdot\pmb{F}^{\rm EP}/r\sin\theta\rho^{(0)}$ and 
we cannot safely use equation (\ref{eq:andmci}) to discuss $\partial v_\phi^{(2)}/\partial t$.
But, it is interesting to note that $(\overline{D}^L\overline{\pmb{v}}^L)_\phi$, the $\phi$-component of the material derivative of the zonally averaged Lagrangian velocity perturbation of second order is in a good approximation given by $-\nabla\cdot\pmb{F}^{\rm EP}/\rho^{(0)}r\sin\theta$ as suggested by equation (\ref{eq:lagmeanphi}).

\begin{figure*}
\resizebox{0.45\columnwidth}{!}{
\includegraphics{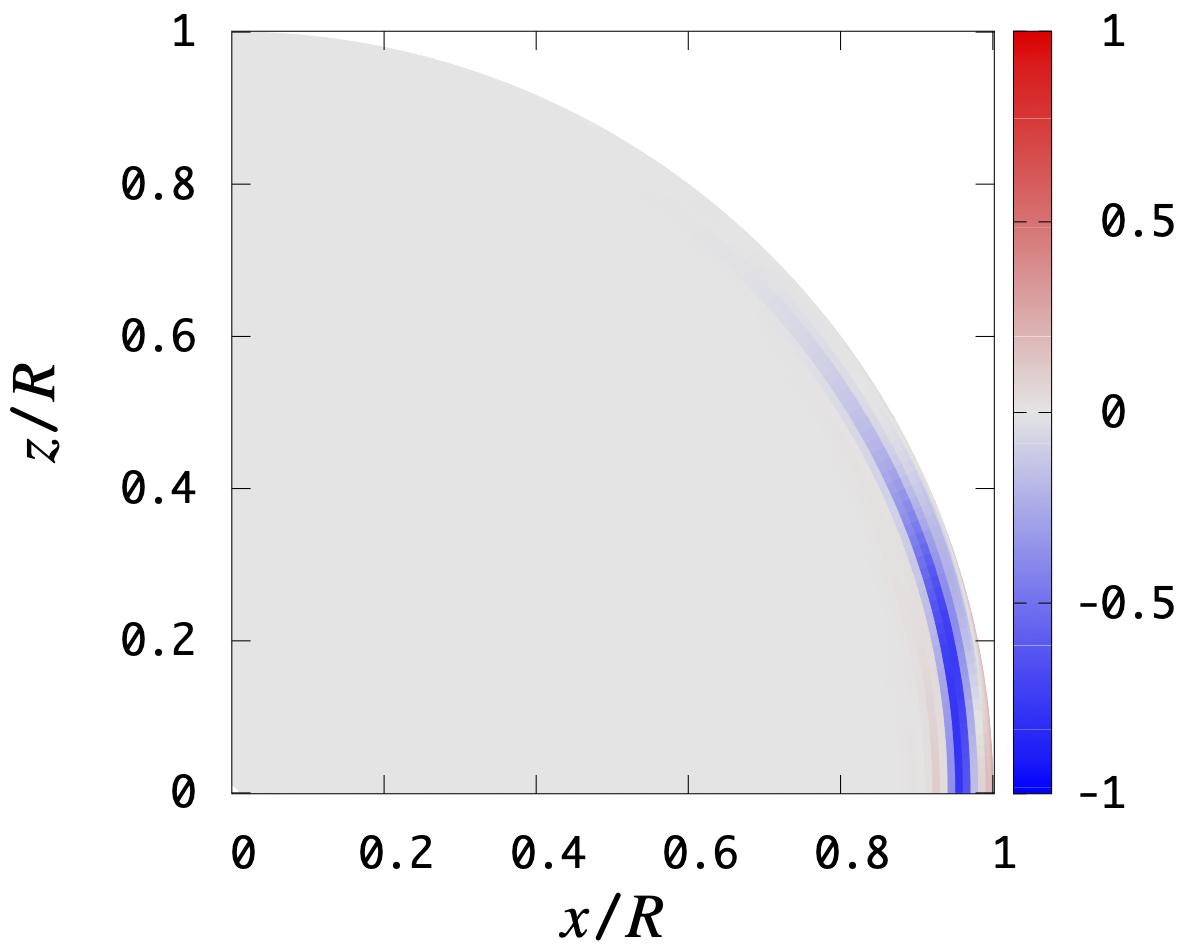}}
\hspace{0.3cm}
\resizebox{0.45\columnwidth}{!}{
\includegraphics{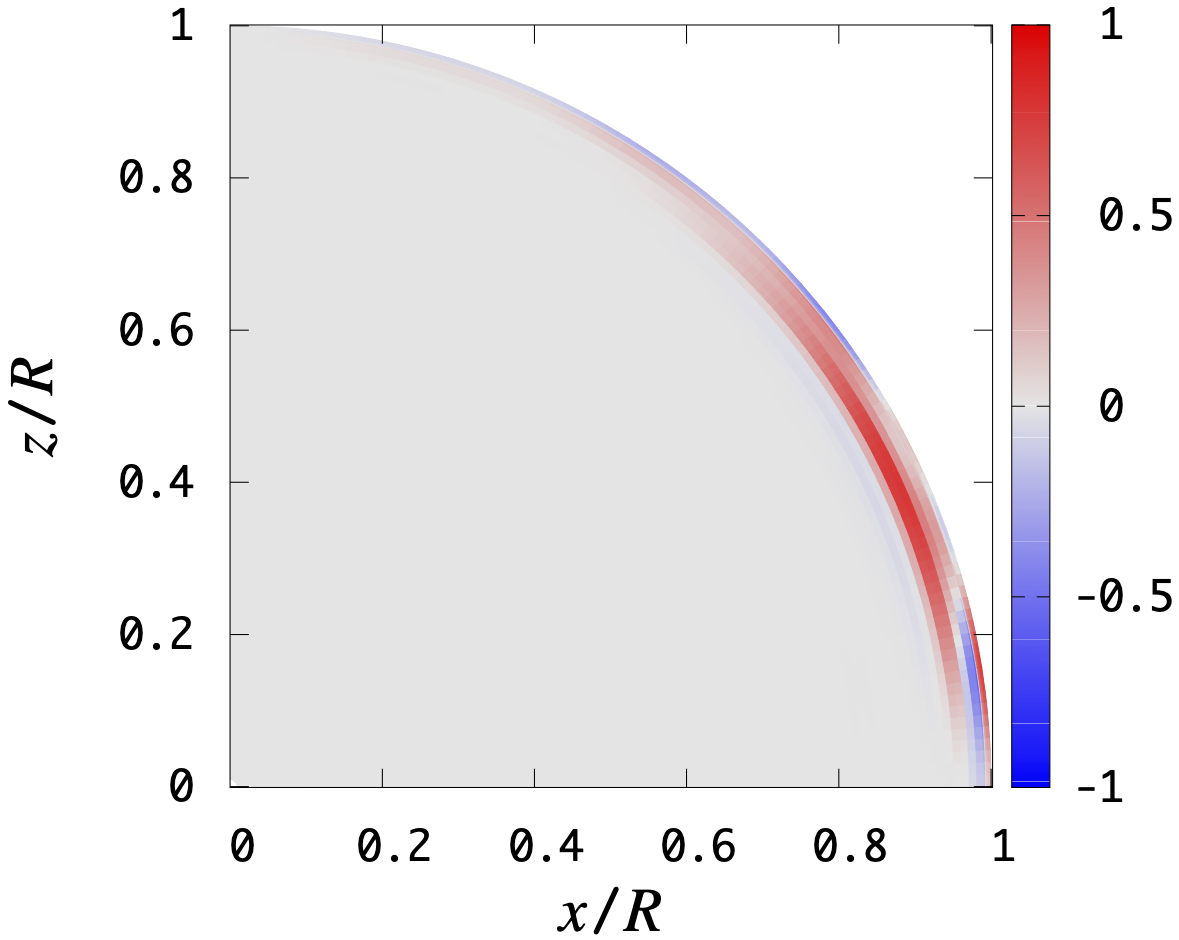}}
\caption{Color maps of $-\nabla\cdot\pmb{F}^{\rm EP}/\rho g r$ for the even prograde $g_{30}$-mode
(left panel) and for the odd $r_{30}$-mode (right panel) of $m=-2$ at $\bar\Omega=0.4$ for $\gamma=10^{-5}$ where $-\nabla\cdot\pmb{F}^{\rm EP}/\rho g r$ is normalized by its maximum value.
}
\label{fig:divfep}
\end{figure*}

In Fig. \ref{fig:divfep}, we show the color maps of $-\nabla\cdot\pmb{F}^{\rm EP}/\rho gr$ for
the prograde $g_{30}$-mode (left panel) and $r_{30}$-mode (right panel).
Since high radial order $g$- and $r$-modes suffer from strong cancelation between the terms $r^{-2}\partial (r\sin\theta\rho^{(0)}\overline{v_\phi^{(1)}v_r^{(1)}})/\partial r$
and $r^{-2}\partial(r\sin\theta\rho^{(0)}{\cal Q}\overline{\xi_rv_\theta^{(1)}})/\partial r$
when we numerically calculate $\nabla\cdot\pmb{F}^{\rm EP}$, 
we use the equation (\ref{eq:eptoz}).
Since 
${\cal D}\overline{a^{(1)} b^{(1)}}=-2\omega_{\rm I}\overline{a^{(1)} b^{(1)}}$,
we have $\nabla\cdot\pmb{F}^{\rm Z}\propto\omega_{\rm I}$, and for steady perturbations with $\omega_{\rm I}=0$ we may ignore $\nabla\cdot\pmb{F}^{\rm Z}$.
Note also that for the color maps in which $-\nabla\cdot\pmb{F}^{\rm EP}$ is plotted, $\nabla\cdot\pmb{F}^{\rm EP}$ is represented by
\be
\nabla\cdot\pmb{F}^{\rm EP}=\sum_l\left< Y_l^0(\theta){\nabla\cdot\pmb{F}^{\rm EP}}\right>Y_l^0(\theta),
\ee
where $\left<\cdots\right>=\int_0^\pi\int_0^{2\pi}\cdots \sin\theta d\theta d\phi$, and
$l=0,~2,~4,~\cdots$.
Since $\left<Y_0^0\right>=\sqrt{4\pi}$, and $\left<Y_l^0\right>=0$ for even values of $l$, we have 
\be
\left<{\nabla\cdot\pmb{F}^{\rm EP}}\right>={1\over r^2}{\partial \over\partial r}r^2\left<F_r^{\rm EP}\right>.
\label{eq:frtheta}
\ee
The arc like structures found in Fig. \ref{fig:divfep} are associated with the mode excitation zone due to the opacity bump mechanism operative at temperatures $T\sim 1.5\times 10^5{\rm K^\circ}$ and
the mode excitation zone in the outer envelope makes the term $-\nabla\cdot\pmb{F}^{\rm EP}/\rho gr$ negative for prograde modes and positive for retrograde modes (see also the next subsection).
Amplitudes of low frequency prograde $g$-modes tend to be confined in the equatorial regions and
those of $r$-modes which are retrograde modes tend to be large at middle latitudes.
Comparing Fig. \ref{fig:divfep} with Fig. \ref{fig:vphimm2gr30_nadmap_epsm5} in which the sign of $v_\phi^{(2)}$ is the same as that of $\partial v_\phi^{(2)}/\partial\tau=\gamma v_\phi^{(2)}$ for positive $\gamma$, we find that
equation (\ref{eq:andmci}) is not necessarily well satisfied, particularly for the $g$-mode, although
the sing of $\partial v_\phi^{(2)}/\partial\tau$ looks consistent with
that of $-\nabla\cdot\pmb{F}^{\rm EP}/\rho gr$ for the $r$-mode.
The comparison between the two figures may suggest that the terms ${\cal P}\hat v_r$ and ${\cal Q}\hat v_\theta$ are not necessarily negligible compared to $-\nabla\cdot\pmb{F}^{\rm EP}$ to determine
$\partial v_\phi^{(2)}/\partial\tau$.

\subsection{$\left<\nabla\cdot\pmb{F}^{\rm EP}\right>$ and $\left<\nabla\cdot\pmb{F}^W\right>$ as a function of $r$}

We integrate (\ref{eq:eptoz}) over the spherical surface of radius $r$ to obtain
\begin{align}
\epsilon^{\rm EPr}=\epsilon^W-\left<\nabla\cdot\pmb{F}^Z\right>,
\end{align}
where
\begin{align}
\epsilon^{\rm EPr}\equiv -\left<\nabla\cdot\pmb{F}^{\rm EP}\right>=-{1\over r^2}{\partial\over\partial r}r^2\left<F^{\rm EP}_r\right>,
\quad
\epsilon^W\equiv-\left<\nabla\cdot\pmb{F}^W\right>=-{1\over r^2}{\partial\over\partial r}r^2\left<\overline{\xi_r{\partial p^{(1)}\over\partial\phi}}\right>,
\label{eq:48b}
\end{align}
and
\be
\left<\nabla\cdot\pmb{F}^Z\right>={1\over r^2}{\partial\over\partial r}r^2\left<r\sin\theta\rho^{(0)}{\cal D}\left(\overline{\xi_rv_\phi^{(1)}}+{1\over 2}{\cal P}\overline{\xi_r\xi_r}\right)\right>.
\ee
Note that for stationary waves we have $\epsilon^{\rm EPr}=\epsilon^W$.
If we introduce the work function $W(r)$ defined by
\be
W(r)=-\pi r^2\Im\left(\left<p^{(1) *}\xi_r\right>\right)=\pi r^2\Im\left(\left<p^{(1) }\xi_r^*\right>\right),
\ee
we find
\be
\epsilon^W={m\over 2\pi r^2}{\partial W\over \partial r}.
\ee
We note that for $\omega_{\rm R}>0$ the regions in which $dW/dr>0$ ($dW/dr<0$) are excitation (damping) regions of
oscillation modes (e.g., \citealt{Unnoetal1989}).
If $\left<\rho^{(0)}\overline{D}^L\overline{l}_\phi\right>
\approx \left<r\sin\theta\rho^{(0)}{\partial v_\phi^{(2)}/\partial t}\right>$ and $\epsilon^W\approx\epsilon^{\rm EPr}$ hold in a good approximation, we find that
$\partial v_\phi^{(2)}/\partial t>0$ ($\partial v_\phi^{(2)}/\partial t<0$) in the damping (excitation) regions of prograde modes, while $\partial v_\phi^{(2)}/\partial t<0$ ($\partial v_\phi^{(2)}/\partial t>0$) occurs in the damping (excitation) regions of retrograde modes.

\begin{figure*}
\resizebox{0.45\columnwidth}{!}{
\includegraphics{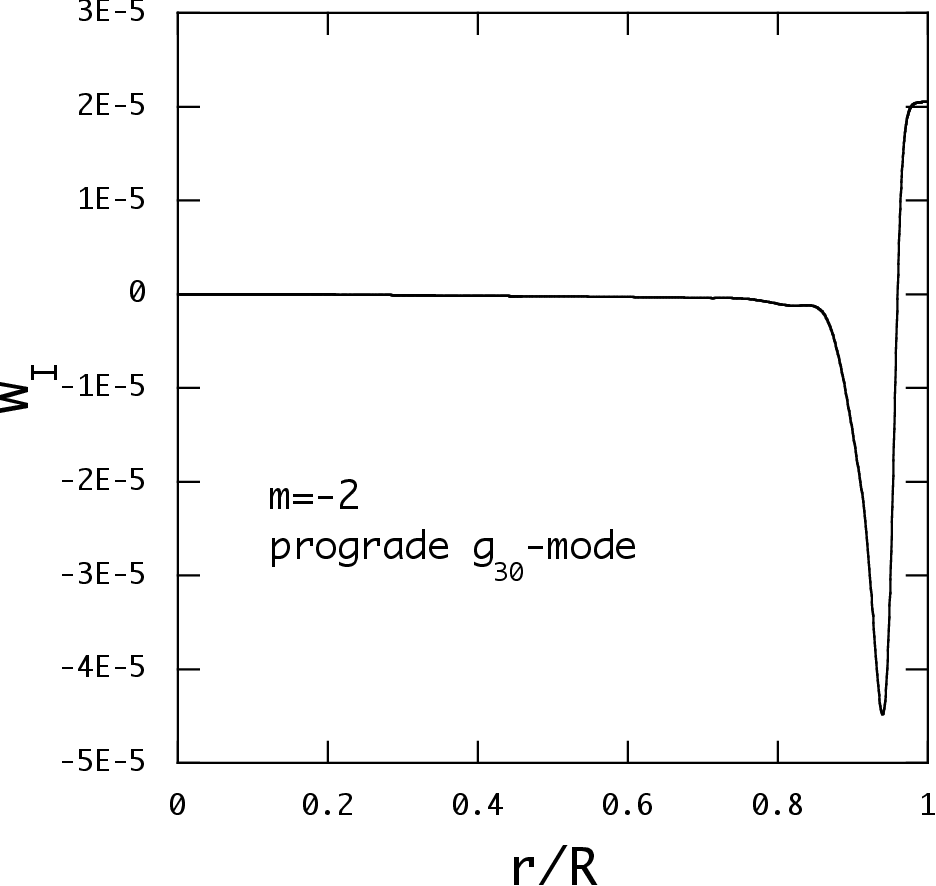}}
\hspace{0.3cm}
\resizebox{0.45\columnwidth}{!}{
\includegraphics{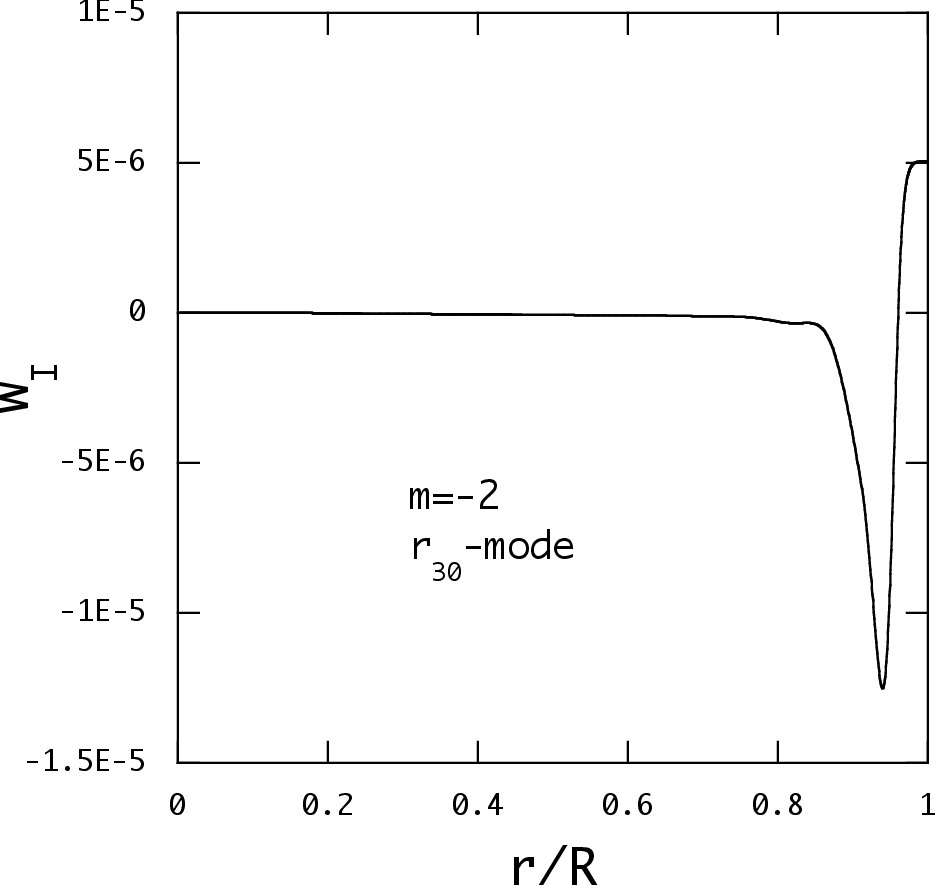}}
\caption{Work integral $w_{\rm I}$ as a function of $r/R$ for the $m=-2$ prograde $g_{30}$-mode
(left panel) and the $m=-2$ $r_{30}$-mode (right panel) at $\bar\Omega=0.4$ where the work integral
is normalized so that $w(R)\simeq-\omega_{\rm I}/\sigma_0$ at the stellar surface with $\omega_{\rm I}$
being the imaginary part of the eigenfrequency. Negative $\omega_{\rm I}$ indicates that the mode is unstable.
}
\label{fig:ws_mm2_g30r30}
\end{figure*}

To identify mode excitation and damping regions in a star, we may also use the work integral $w_{\rm I}$ defined by  (e.g., \citealt{Unnoetal1989})
\be
w_{\rm I}(r)={1\over 2\omega_{\rm R}E}\int_0^r\alpha_T\Im\left(\delta p{\delta s^*\over c_p}\right)dV,
\label{eq:workintegral}
\ee
where $E=\int\rho\pmb{\xi}^*\cdot\pmb{\xi}dV$, 
and the function $w_{\rm I}$ is normalized such that $w_{\rm I}(R)\simeq-\bar\omega_{\rm I}$ at the stellar surface.
A good property of the work integral (\ref{eq:workintegral}) is that we can use $w_{\rm I}$ 
irrespective of the sign of $\omega_{\rm R}$.
As in the case of the work function $W$, the regions of $dw_{\rm I}/dr>0$ ($dw_{\rm I}/dr<0$)
correspond to excitation (damping) regions of an oscillation mode.
Fig. \ref{fig:ws_mm2_g30r30} shows the work integral $w_{\rm I}(r)$ as a function of $r/R$
for the $m=-2$ prograde $g_{30}$-mode (left panel) and $r_{30}$-mode (right panel).
Both panels in the figure show that in the outer envelope there exists a strong excitation region where $dw_{\rm I}/dr>0$, below which there extends damping regions where $dw_{\rm I}/dr<0$, and that
the strong excitation of the modes
exceeds the total amount of damping below the excitation region,
which makes the modes pulsationally unstable as a whole.
The mode excitation is due to the opacity bump mechanism in which the positive opacity derivative $d\kappa_{ad}/dr>0$ at the opacity bump, produced by metal lines located at $T\sim 1.5\times 10^5{\rm K}^\circ$, plays an essential role.

\begin{figure*}
\resizebox{0.45\columnwidth}{!}{
\includegraphics{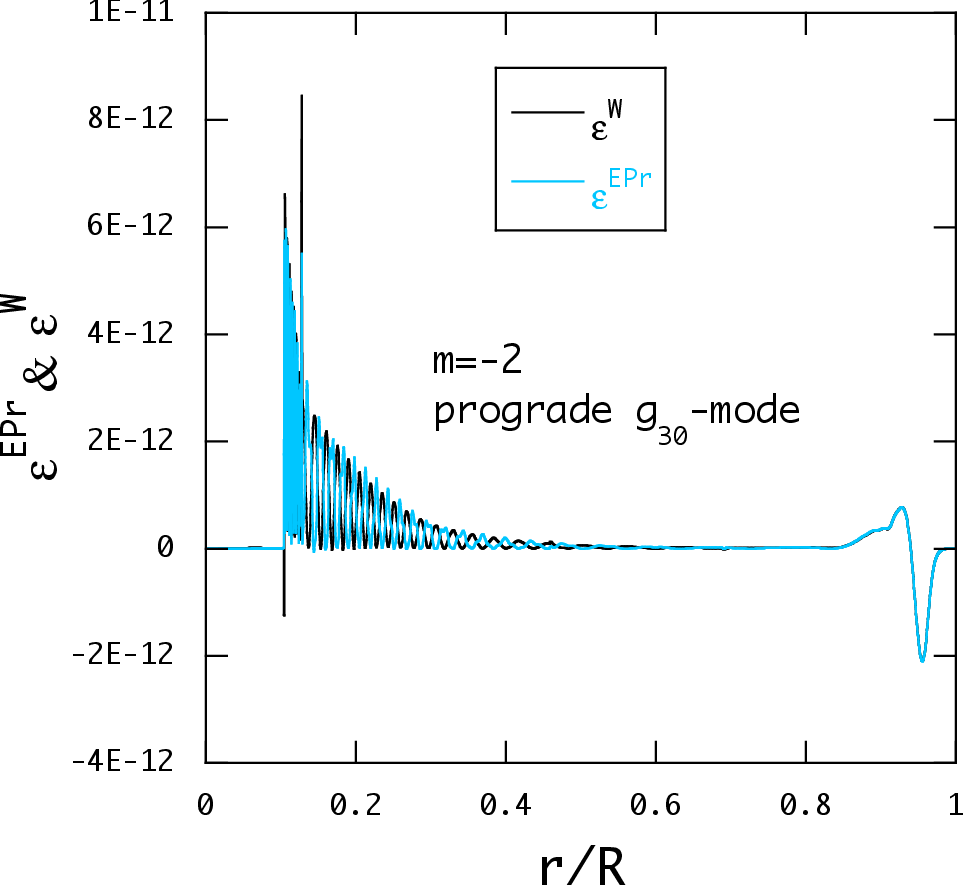}}
\hspace{0.3cm}
\resizebox{0.45\columnwidth}{!}{
\includegraphics{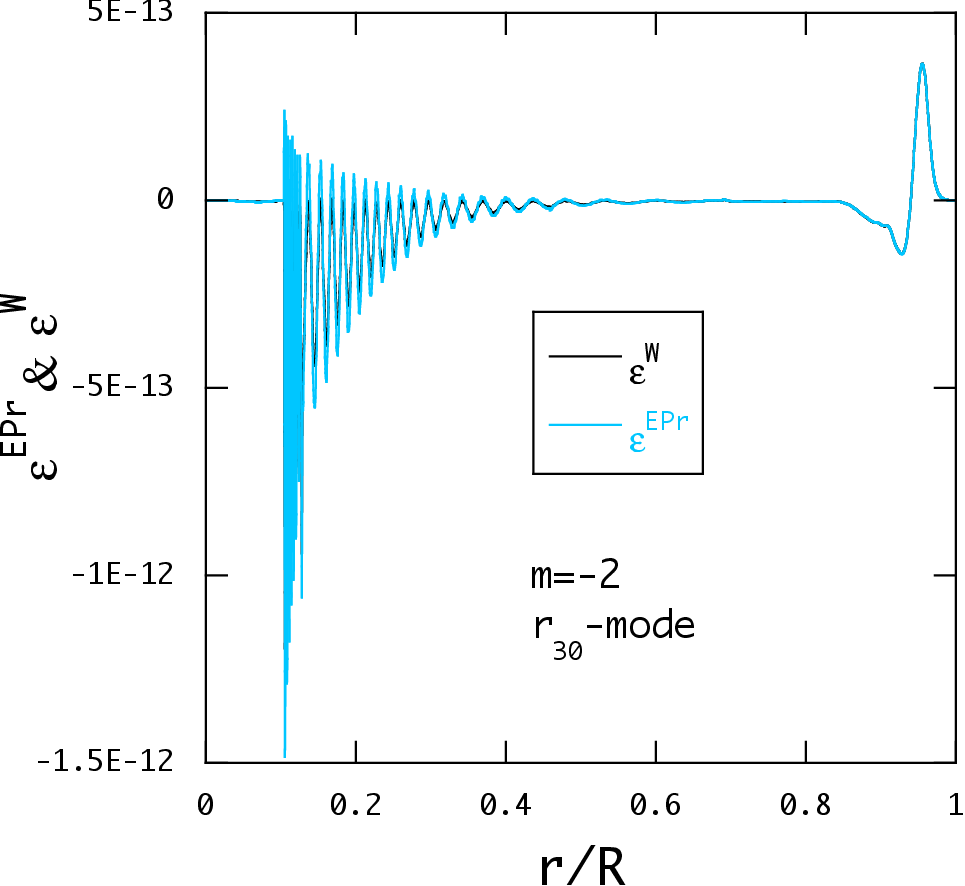}}
\caption{$\epsilon^{\rm EPr}$ and $\epsilon^W$ versus $r/R$ for the $m=-2$ prograde $g_{30}$-mode (left panel)
and the $m=-2$ $r_{30}$-mode (right panel) at $\bar\Omega=0.4$ for $\epsilon_{E}=10^{-10}$.
Both $\epsilon^{\rm EPr}$ and $\epsilon^W$ are normalized by $GM\bar\rho /R=3.7\times10^{14}{\rm erg/cm^3}$ with $\bar\rho=M/(4\pi R^3/3)=0.16{\rm g/cm^3}$.
}
\label{fig:epss_mm2_g30}
\end{figure*}

Fig. \ref{fig:epss_mm2_g30} shows $\epsilon^{\rm EPr}$ and $\epsilon^W$ versus $r/R$
for the $m=-2$ prograde $g_{30}$-mode (left panel) and $r_{30}$-mode (right panel).
We find that $\epsilon^{\rm EPr}$ and $\epsilon^W$ agree well in the outer envelope where non-adiabatic effects
are significant for both modes.
For the prograde $g$-mode, both $\epsilon^{\rm EPr}$ and $\epsilon^W$ are negative where
the mode excitation takes place due to the opacity bump and they are positive below the excitation region.
We also find that although both $\epsilon^{\rm EPr}$ and $\epsilon^W$ show spatial 
oscillations with short wavelengths and have almost the same amplitudes in the deep interior, the phases of the spatial oscillations differ by $\sim \pi$ between $\epsilon^{\rm EPr}$ and $\epsilon^W$.
Since the $r$-mode is a retrograde mode, the sign of the functions $\epsilon^{\rm EPr}$ and $\epsilon^W$
for the $r$-mode is in general opposite to that for the prograde $g_{30}$-mode.
In fact, both $\epsilon^{\rm EPr}$ and $\epsilon^W$ for the $r$-mode is positive at the excitation region of the mode and they are in general negative in the damping regions below although $\epsilon^{\rm EPr}$
can be slightly positive outside of the convective core.
As in the case of the $g$-mode, in the deep interior $\epsilon^{\rm EPr}$ and $\epsilon^W$ shows short spatial oscillations, which are in phase for the $r$-mode.

\begin{figure*}
\resizebox{0.45\columnwidth}{!}{
\includegraphics{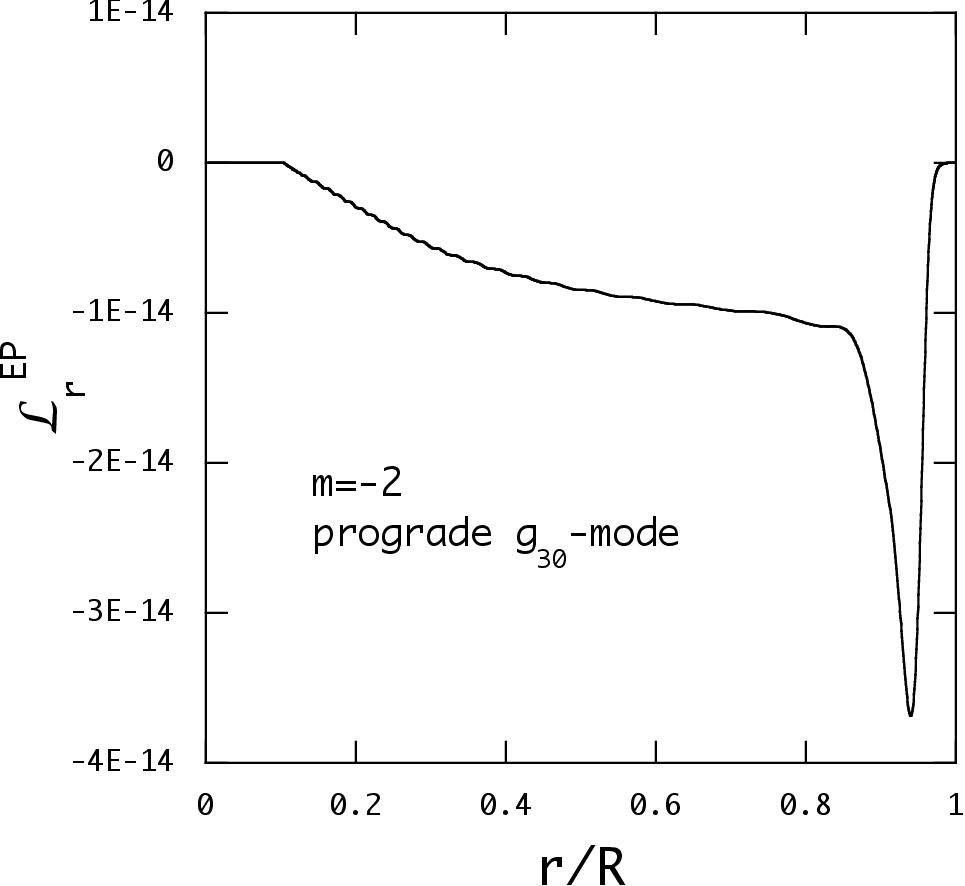}}
\hspace{0.3cm}
\resizebox{0.45\columnwidth}{!}{
\includegraphics{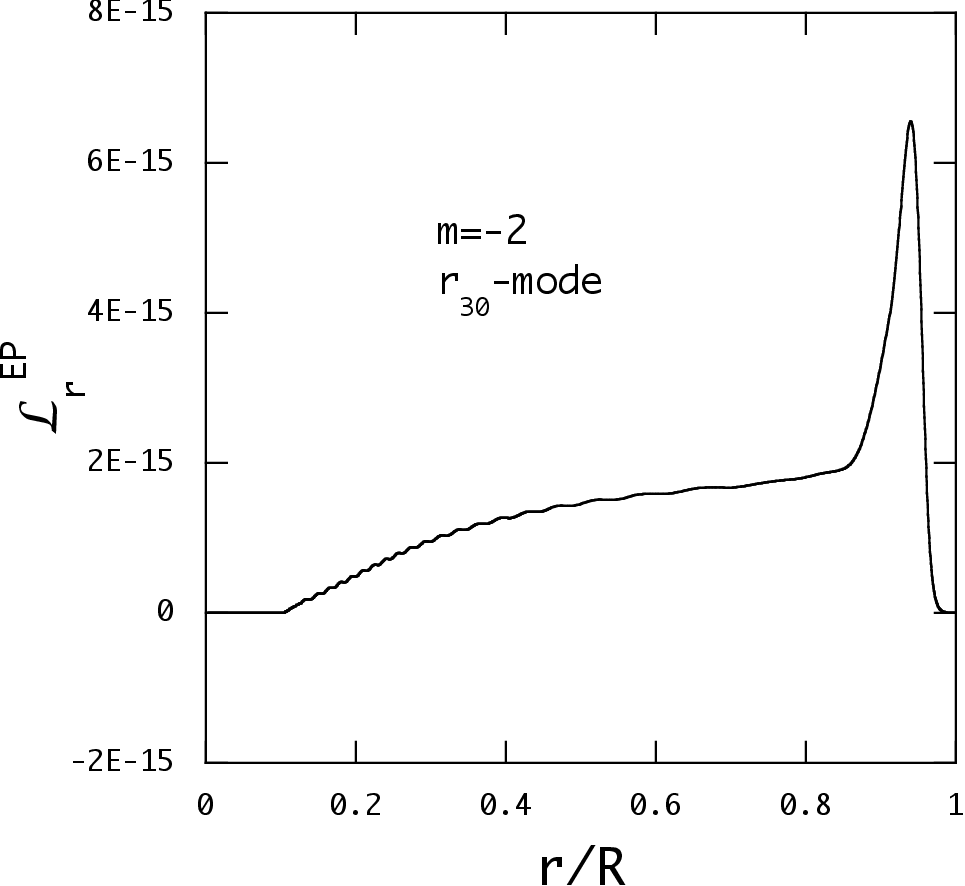}}
\caption{${\cal L}_r^{\rm EP}$ versus $r/R$ for the $m=-2$ prograde $g_{30}$-mode
(left panel) and for the $m=-2$ $r_{30}$-mode (right panel) at $\bar\Omega=0.4$ for $\epsilon_E=10^{-10}$.
${\cal L}_r^{\rm EP}$ is normalized by $GM^2/R=1.9\times10^{49}{\rm erg}$.
}
\label{fig:LAM_mm2_g30r30}
\end{figure*}

We define the angular momentum luminosity carried by waves as
\be
{\cal L}_r^{\rm EP}=4\pi r^2\left<F_r^{\rm EP}\right> \quad {\rm or} \quad {\cal L}_r^W= -2mW.
\ee
Fig. \ref{fig:LAM_mm2_g30r30} plots ${\cal L}_r^{\rm EP}$ versus $r/R$ for the $m=-2$ prograde $g_{30}$-mode (left panel) and $r_{30}$-mode (right panel).
Note that there appear no significant differences between ${\cal L}_r^{\rm EP}$ and ${\cal L}_r^{W}$ for the modes.
For high radial order low frequency modes, radiative dissipation associated with short wavelength waves is
the main damping mechanism in the deep interior where $|c_2\bar\omega_{\rm R}|\gg 1$ and non-adiabatic effects can be treated in the quasi-adiabatic approximation.
Since ${\cal L}_r^{\rm EP}$ is negative for the prograde $g$-mode, we consider that angular momentum is transported inwardly, that is, the mode extracts angular momentum from the outer envelope where the mode is excited  and deposits the inwardly transported angular momentum in the deep interior through dissipative processes.
On the other hand, since ${\cal L}_r^{\rm EP}$ is positive for the retrograde $r$-mode, we consider
outward transport of angular momentum in the star, that is, the angular momentum extracted from the inner regions through dissipative processes is transported outwardly and deposited in the surface regions where the mode is excited.

\subsection{Overstable convective modes}

When massive main sequence stars are rapidly rotating,
convective modes in the convective core become overstable and resonantly excite prograde $g$-modes in the envelope (e.g., \citealt{LeeSaio86,LeeSaio20,Lee2021}).
The overstable convective modes in the core resonantly coupled with envelope $g$-modes are simply called overstable convective (OsC) modes.
This coupling between the convective mode and envelope $g$-mode effectively takes place when the core rotates slightly faster than the envelope (\citealt{LeeSaio20}).
To see how OsC modes excite second order perturbations,
we use the $4M_\odot$ ZAMS model and assume that the core rotates faster by 20\% than the envelope, that is, $\Omega_c/\Omega_e=1.2$, where
$\Omega_e$ and $\Omega_c$ denote the rotation speed in the envelope and the core, respectively.
For this model, we calculate an $m=-2$ OsC mode at $\bar\Omega_e=0.318$ and the second order perturbations. 
No $g$-modes and $r$-modes are found unstable for the ZAMS model.
The radius of the convective core of the model is $R_c=0.17R$.
Note that the OsC mode is coupled with the $g_{32}$-mode in the envelope and has the complex eigenfrequency 
$\bar\sigma=\bar\sigma_{\rm R}+\rmi\bar\sigma_{\rm I}=(0.7648, -2.215\times10^{-5})$ at $\bar\Omega_e=0.318$.
Since $\bar\sigma_{\rm I}<0$, the mode is unstable.
For this OsC mode, we have $\delta L_r/L_r\approx 0.046$ at the surface for the amplitude given by
$\epsilon_E=10^{-10}$.

\begin{figure*}
\resizebox{0.45\columnwidth}{!}{
\includegraphics{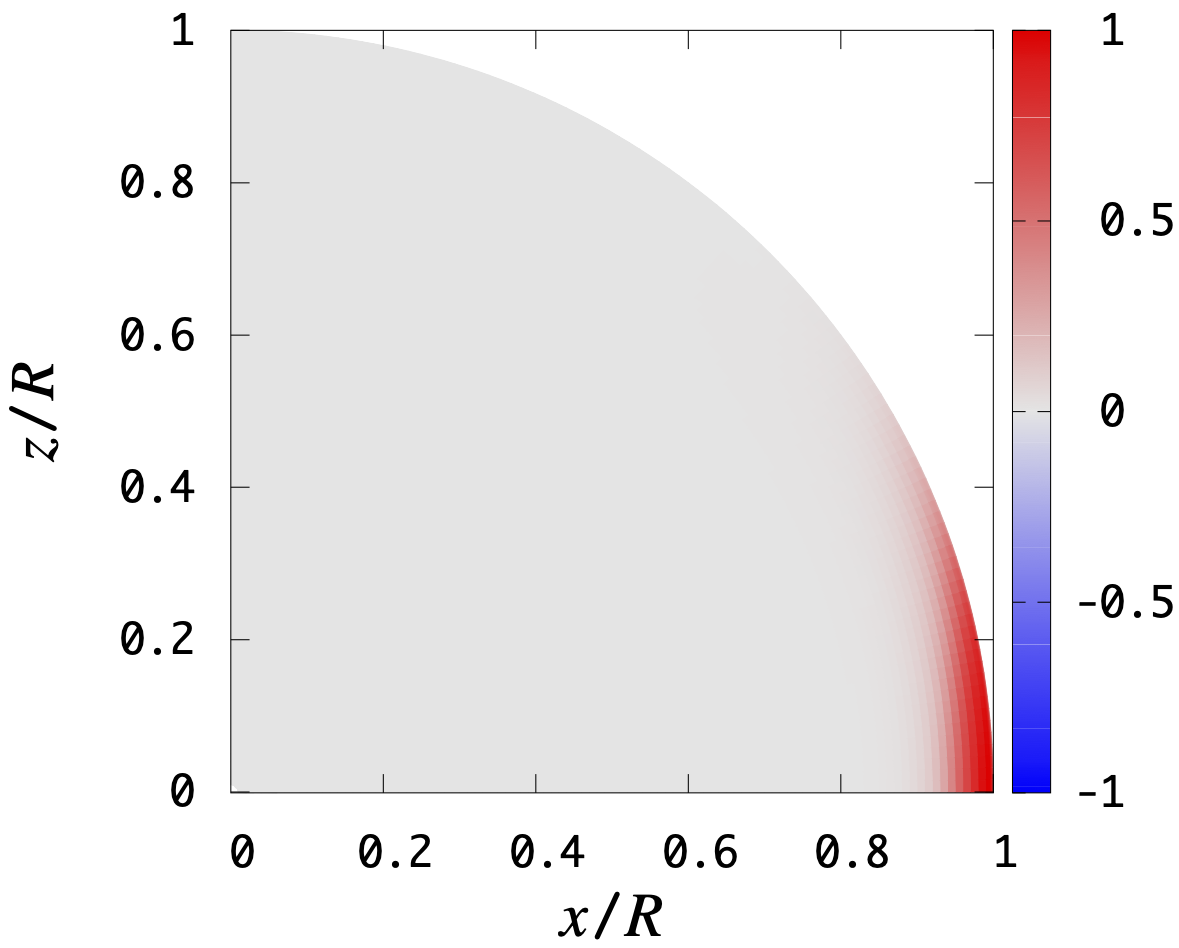}}
\hspace{0.3cm}
\resizebox{0.45\columnwidth}{!}{
\includegraphics{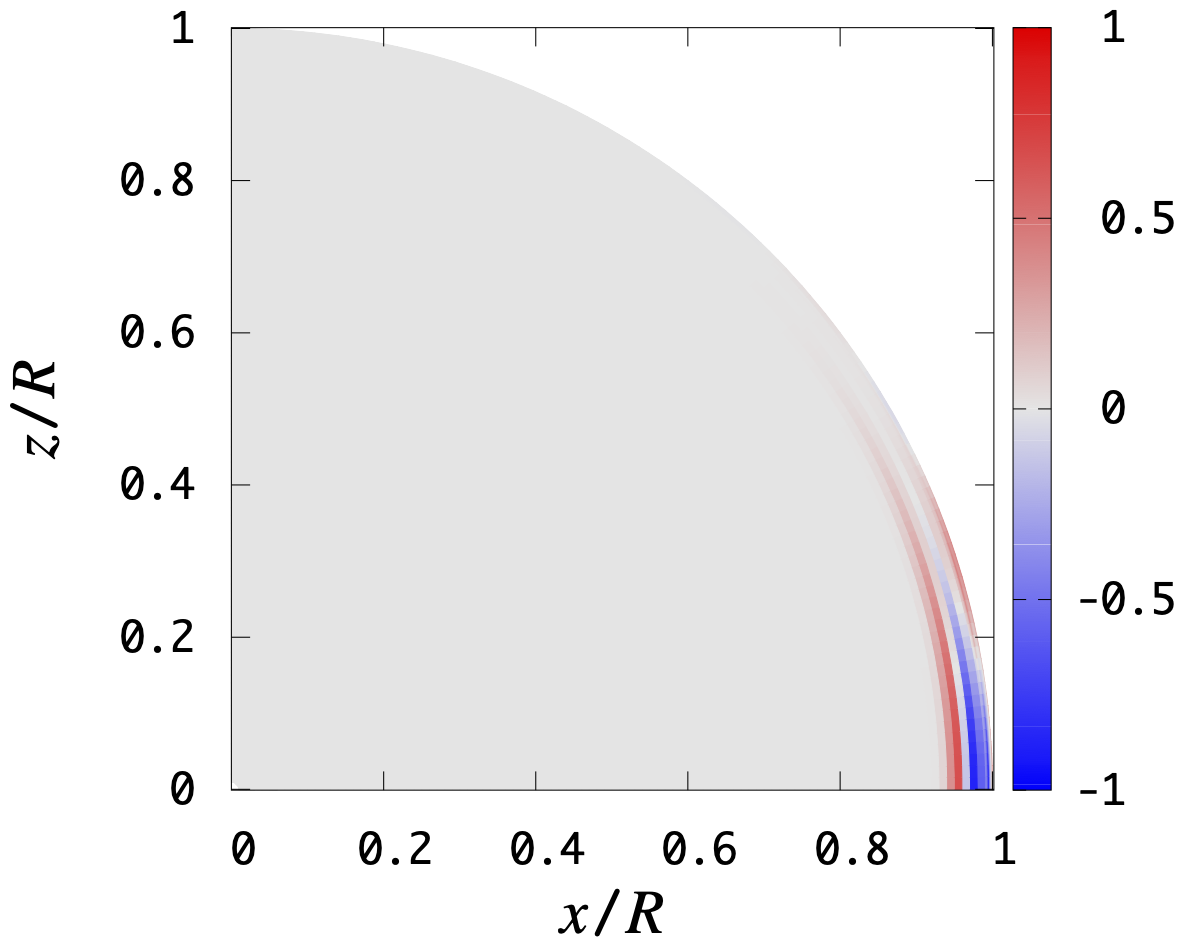}}
\caption{Color map of $v_\phi^{(2)}$ (left panel) and $-\nabla\cdot\pmb{F}^{\rm EP}/\rho g r$ (right panel) for the $m=-2$ OsC mode coupled with $g_{32}$-mode in the envelope at $\bar\Omega=0.318$
where $\gamma=10^{-5}$ is assumed. For the background model, we use the $4M_\odot$ ZAMS model, which has the convective core whose fractional radius is 0.17.
Each of $v_\phi^{(2)}$ and $-\nabla\cdot\pmb{F}^{\rm EP}/\rho g r$ is normalized by its maximum value.
}
\label{fig:vphidivfep_OsC}
\end{figure*}

Fig. \ref{fig:vphidivfep_OsC} shows the color maps of $v_\phi^{(2)}$ (left panel) and $-\nabla\cdot\pmb{F}^{\rm EP}/\rho g r$ (right panel) for the OsC mode where $\gamma=10^{-5}$ is assumed.
We find that for the OsC mode, $\partial v_\phi^{(2)}/\partial t>0$ in the surface equatorial regions as in the case of prograde $g$-modes excited by the opacity bump mechanism.
However, as the color map of $-\nabla\cdot\pmb{F}^{\rm EP}/\rho g r$ indicates, the relation (\ref{eq:andmci}) is not necessarily well satisfied, which is the same as found for the prograde $g_{30}$-mode.
We see that $-\nabla\cdot\pmb{F}^{\rm EP}<0$ in the equatorial surface layer and 
$-\nabla\cdot\pmb{F}^{\rm EP}>0$ in the thin layer immediately below the surface layer of $-\nabla\cdot\pmb{F}^{\rm EP}<0$.
Since the ZAMS model has a higher surface temperature than the slightly evolved main sequence model used in the previous sections, the mode excitation zone associated with the opacity bump for the ZAMS model is located in the layers shallower than that for the evolved model.

\begin{figure*}
\resizebox{0.45\columnwidth}{!}{
\includegraphics{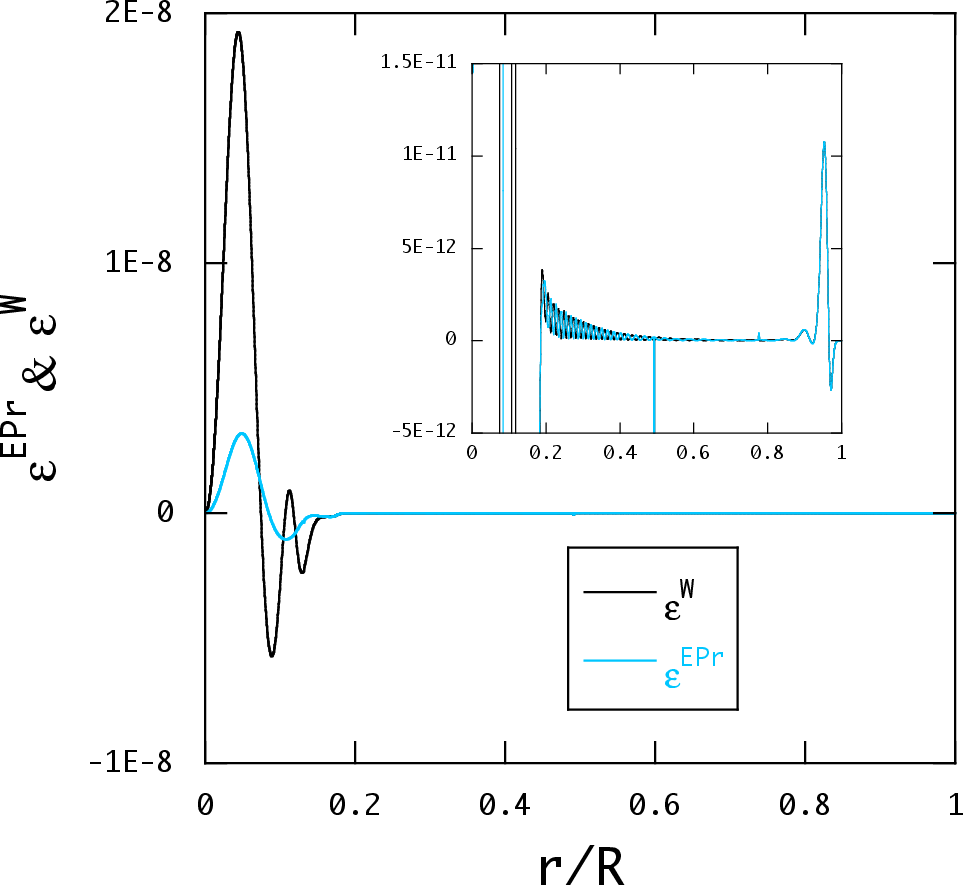}}
\hspace{0.3cm}
\resizebox{0.45\columnwidth}{!}{
\includegraphics{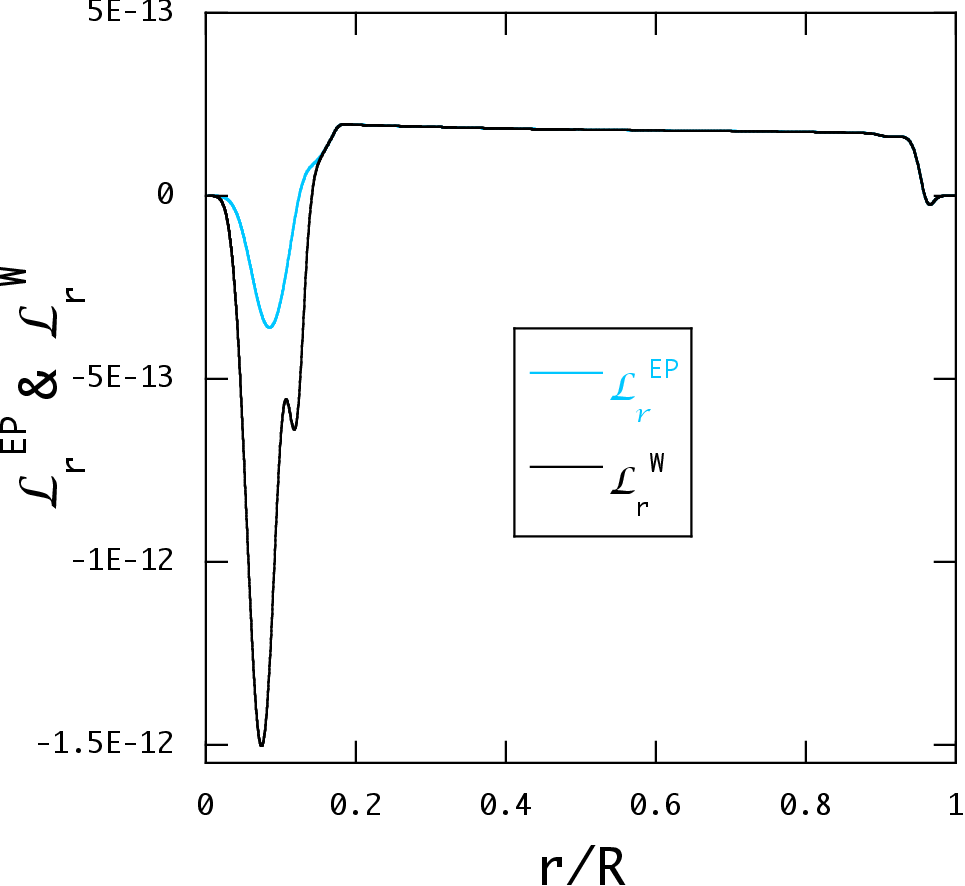}}
\caption{$\epsilon^{\rm EPr}$ and $\epsilon^W$ (left panel) and ${\cal L}_r^{\rm EP}$ and ${\cal L}^W_r$ (right panel) versus $r/R$ for
the $m=-2$ OsC mode at $\bar\Omega=0.318$ where $\epsilon_E=10^{-10}$ is assumed for the mode amplitudes. Both $\epsilon^{\rm EPr}$ and $\epsilon^W$ are normalized by $GM\bar\rho/R=1.3\times10^{15}{\rm erg/cm^3}$ with $\bar\rho=0.42{\rm g/cm^3}$, and ${\cal L}_r^{\rm EP}$ and ${\cal L}^W_r$ are normalized by $GM^2/R=2.5\times10^{49}{\rm erg}$.
}
\label{fig:epsEPrW+LAM_OsCmm2}
\end{figure*}

Fig. \ref{fig:epsEPrW+LAM_OsCmm2} shows $\epsilon^{\rm EPr}$ and $\epsilon^W$ (left panel) and ${\cal L}_r^{\rm EP}$ and ${\cal L}_r^W$ (right panel) versus $r/R$ for the OsC mode.
The magnitudes of $\epsilon^W$ and $\epsilon^{\rm EPr}$ in the core is much larger than those in the envelope. 
We find that $\epsilon^W$ behaves differently from $\epsilon^{\rm EPr}$ in the core although the behavior of $\epsilon^W$ and $\epsilon^{\rm EPr}$
in the radiative envelope is quite similar to that found for the prograde $g_{30}$-mode as shown by the inset of the left panel.
The right panel of the figure shows that ${\cal L}_r^{\rm EP}$ and ${\cal L}_r^W$ behave in a complex way in the core but they describe almost the same curves extending from the core-envelope boundary to the surface, that is,
they experiences a large increase near the outer boundary of the core and 
decrease gradually from the core-envelope boundary toward the outer envelope and suffer an abrupt and large drop just below the mode excitation zone.
This suggests that OsC modes can effectively transport angular momentum from the convective core to the outer envelope of rotating massive main sequence stars.
Note that since $g$-modes and $r$-modes have no appreciable amplitudes in the convective core,
they transport angular momentum between the inner part and outer parts of the radiative envelope.

\begin{figure*}
\resizebox{0.5\columnwidth}{!}{
\includegraphics{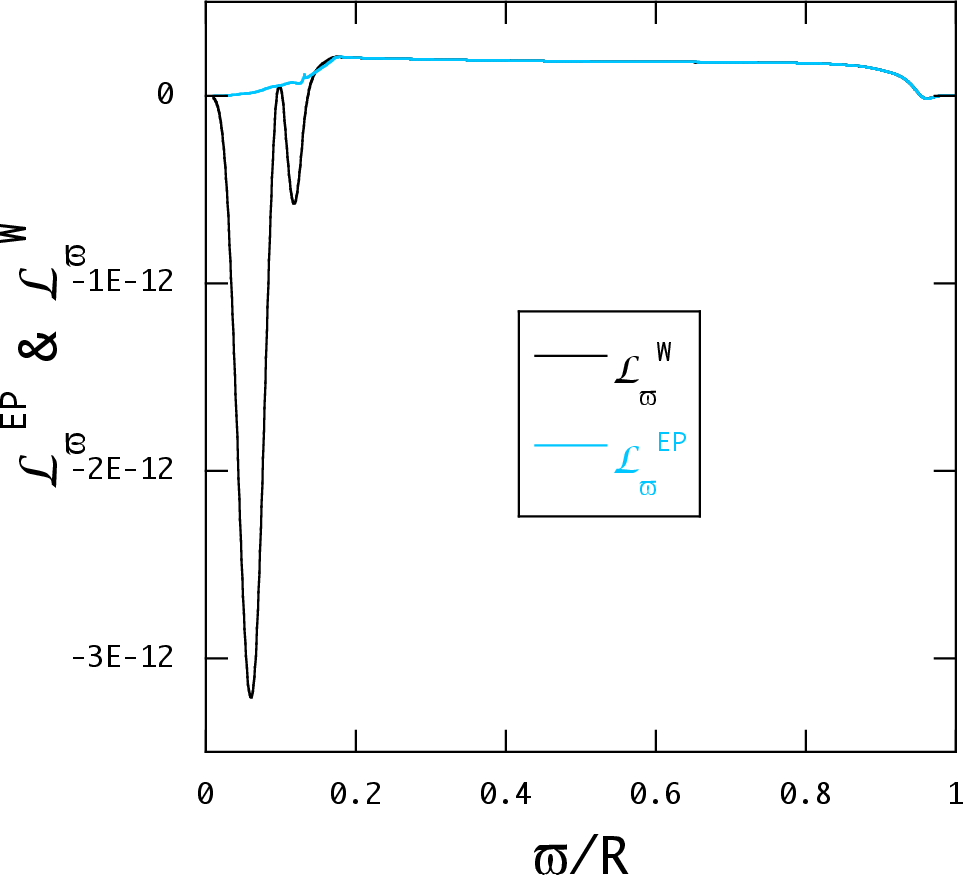}}
\caption{${\cal L}_\varpi^{\rm EP}$ and ${\cal L}^W_\varpi$ versus $\varpi/R$ for
the $m=-2$ OsC mode at $\bar\Omega=0.318$ for the $4M_\odot$ ZAMS model where $\epsilon_E=10^{-10}$ is assumed to determine the mode amplitudes.
Both ${\cal L}_\varpi^{\rm EP}$ and ${\cal L}^W_\varpi$ are normalized by $GM^2/R=2.5\times10^{49}{\rm erg}$.
}
\label{fig:LAM_OsCmm2_cyl}
\end{figure*}

It may be interesting to rewrite the Eliassen-Palm flux  
in cylindrical coordinates $(\varpi,\phi,z)$ where $\varpi=r\sin\theta$ and $z=r\cos\theta$.
If we let $F_\varpi^{\rm EP}$ and $F_z^{\rm EP}$ denote the $\varpi$- and $z$-components of the Eliassen-Palm flux, we have
\be
\nabla\cdot\pmb{F}^{\rm EP}={1\over \varpi}{\partial\over\partial \varpi}\varpi F_\varpi^{\rm EP}+{\partial F_z^{\rm EP}\over\partial z},
\ee
where
\be
F_\varpi^{\rm EP}=F_r^{\rm EP}\sin\theta+F_\theta^{\rm EP}\cos\theta=\rho^{(0)}\left(\varpi \overline{v_\phi^{(1)}v_\varpi^{(1)}}+\varpi^2{\partial\Omega\over\partial z}\overline{\xi_rv_\theta^{(1)}}\right), 
\ee
\be
F_z^{\rm EP}=F_r^{\rm EP}\cos\theta-F_\theta^{\rm EP}\sin\theta=\rho^{(0)}\left[\varpi\overline{v_\phi^{(1)}v_z^{(1)}}-2\varpi\Omega\left(1+{1\over 2}{\partial\ln\Omega\over\partial \ln \varpi}\right)\overline{\xi_rv_\theta^{(1)}}\right].
\ee
Integrating $\nabla\cdot\pmb{F}^{\rm EP}$ with respect to $z$ from $z=-z_\varpi$ to $z=+z_\varpi$ where $z_\varpi=R\sqrt{1-(\varpi/R)^2}$, we obtain
\be
\int_{-z_\varpi}^{z_\varpi}\nabla\cdot\pmb{F}^{\rm EP}dz
={1\over 2\pi \varpi}{\partial\over\partial \varpi}{\cal L}_\varpi^{\rm EP},
\ee
where 
\be
{\cal L}_\varpi^{\rm EP}=2\pi \varpi\int_{-z_\varpi}^{z_\varpi}F_\varpi^{\rm EP}dz.
\ee
Similarly, we have $\pmb{F}^W=F_r^W\pmb{e}_r+F_\theta^W\pmb{e}_\theta=F_\varpi^W\pmb{e}_\varpi+F_z^W\pmb{e}_z$ and ${\cal L}_\varpi^W=2\pi\varpi\int_{-z_\varpi}^{z_\varpi}F_\varpi^Wdz$.
Note that we have ignored rotational deformation of the star.
Fig. \ref{fig:LAM_OsCmm2_cyl} shows ${\cal L}_\varpi^{\rm EP}$ and ${\cal L}_r^W$ versus $\varpi/R$ for the $m=-2$ OsC mode at $\bar\Omega=0.318$. 
It is interesting to note that ${\cal L}_r^{\rm EP}\sim{\cal L}_\varpi^{\rm EP}$ in the radiative envelope, suggesting that $F_r^{\rm EP}$ dominates $F_\theta^{\rm EP}$ and
has large amplitudes only in the thin volume containing the equatorial plane in the envelope.
In the convective core, on the other hand, we find that ${\cal L}_\varpi^{\rm EP}\propto\varpi$ and hence
$\int_{-z_\varpi}^{z_\varpi}F_\varpi^{\rm EP}dz$ only weakly depends on $\varpi$.
If $|F_r^{\rm EP}|\sim|F_\theta^{\rm EP}|$ and $F_r^{\rm EP}$ has different sign from $F_\theta^{\rm EP}$, ${\cal L}_r^{\rm EP}$ and ${\cal L}_\varpi^{\rm EP}$ can have different signs in the core.
We also find that $|F_r^{W}|\sim|F_\theta^{W}|$ and $F_r^W$ has the same sign as $F_\theta^W$.

Since oscillations in the convective core are almost adiabatic in the sense $|c_2\bar\omega_{\rm R}|\gg1$,
it is unlikely that the large values of ${\cal L}_r$ and ${\cal L}_\varpi$ within the core are caused by non-adiabatic effects, which are usually significant in the outer envelope where $|c_2\bar\omega_{\rm R}|\sim 1$ and the opacity bump mechanism can be
effective to excite oscillation modes.
For ${\cal L}_\varpi^{\rm EP}$ to have large values, for example, we need an appropriate phase difference $\delta_p$ between $v_\phi^{(1)}$ and $v_\varpi^{(1)}$ such that $\delta_p-\pi/2\not=0$. 
This difference $\delta_p-\pi/2$ in the core may be due to the overstability of the convective modes 
caused by the core rotation and also by the resonant coupling with envelope $g$-modes.
At rapid rotation speeds, OsC modes in the core are excited as prograde modes and are resonantly coupled with prograde $g$-modes, which suffer damping in the envelope. 
Excitation of a prograde mode in the core leads to deceleration of the core rotation and its decay in the envelope results in acceleration of the envelope rotation.
In the outer envelope, on the other hand, the phase difference between $v_\phi^{(1)}$ and $v_\varpi^{(1)}$
is mainly due to non-adiabatic effects and is closely related to the excitation mechanism of $g$-modes.

\subsubsection{Formation mechanism of decretion discs around Be stars}

Be stars are rapidly rotating B type stars showing Balmer emission lines, which originate from a circumferential gaseous decretion disc around the stars (e.g., \citealt{PorterRivinius2003,RiviniusCarciofiMartayan13}).
Although decretion discs around Be stars are now believed to be viscous discs (\citealt{LeeSaioOsaki91}), 
their formation mechanisms are not necessarily well understood.
\citet{Lee13} calculated steady decretion discs around rotating stars assuming that there occurs 
angular momentum deposition in the subsurface layers of the stars (see also \citealt{Martinetal2025}).
For decretion discs, we may measure the angular momentum deposition rates, i.e., the torques applied at the inner boundary of the disc, in terms of $\dot M\sqrt{GMR}$ where $\dot M$ is the mass decretion rate in the discs.
\citet{Lee13} obtained steady decretion disc solutions having the outer radius $\sim 10 R$ for deposition rates of order of $\sim10 \dot M\sqrt{GMR}$.

Although steady decretion discs are not necessarily a realistic model for Be star discs, 
we may use ${\cal N}_{\rm disc}=10 \dot M\sqrt{GMR}$ to represent the lower limit of the torque necessary to support decretion discs of the radius $\sim 10R$.
Typical mass decretion rates $\dot M$ through decretion discs around Be stars is between $ 10^{-12}$ and $10^{-9}{\rm M_\odot yr^{-1}}$ (e.g., \citealt{Vieiraetal2017}).
When we assume $\dot M=10^{-10}{\rm M_\odot yr^{-1}}$ as the typical rate, we have ${\cal N}_{\rm disc}\sim 6\times 10^{35}{\rm erg}$ for the mass $M$ and the radius $R$ of the $4M_\odot$ ZAMS model.
If we assume that the angular momentum transport by OsC modes from the core to the surface is
responsible for decretion disc formation around Be stars,
the magnitude of torque $\Delta {\cal L}_\varpi$ acting on the subsurface layers between $0.9\lesssim \varpi/R\lesssim 1$ is of order of $10^{-13}GM^2/R=3\times10^{36}{\rm erg}$, as suggested by Fig. \ref{fig:LAM_OsCmm2_cyl}.
Note that the mode amplitude is determined by assuming $\epsilon_E=10^{-10}$, which leads to a rather large amplitude $\delta L_r/L_r\approx 0.046$ at the surface.
This rough estimation suggests that if OsC modes are able to provide the subsurface layer of a Be star with the necessary torque $\Delta{\cal L}_\varpi\gtrsim {\cal N}_{\rm disc}$ to support a decretion disc of radius $\sim 10R$,
OsC modes can be a possible candidate for the disc formation mechanism.

\subsubsection{Braking of the core rotation}

We may obtain a rough estimate of the change rate of the rotation speed of the convective core by estimating
\be
\int_0^{R_c}4\pi r^2\left<\rho^{(0)}\overline{D}^L\overline{l}_\phi^L\right>dr=-{\cal L}_r^W(R_c).
\ee
Since ${\cal L}_r^W(R_c)>0$ for the OsC mode as shown by Fig. \ref{fig:epsEPrW+LAM_OsCmm2} and Fig. \ref{fig:LAM_OsCmm2_cyl}, the excitation of the OsC mode contributes to
braking the core rotation.
When we define the core angular momentum as $J_c\sim M_cR_c^2\Omega_c$, the time scale $\tau_c$ for the change of the core angular rotation velocity is given by $\tau_c\sim J_c/{\cal L}_r^W(R_c)$.
For the $4M_\odot$ ZAMS model, we have $M_c/M=0.22$, $R_c/R=0.17$, $\overline\Omega_c=\Omega_c/\sigma_0=0.38$, and
$\sigma_0=3.4\times10^{-4}$ where $M_c$ is the core mass.
For the OsC mode, we read ${\cal L}_r^W(R_c)\sim 10^{-13}GM^2/R$ from Fig. \ref{fig:epsEPrW+LAM_OsCmm2} and then we obtain $\tau_c\sim 2.3\times10^6{\rm yr}$.
The lifetime $\tau_{\rm MS}$ of the star on the main sequence may be estimated as $\tau_{\rm MS}\approx1.2\times10^{10}(M/M_\odot)^{-3}{\rm yr}$ (e.g., \citealt{Clayton1983}), which gives $\tau_{\rm MS}\sim2\times10^8{\rm yr}$ for the $4M_\odot$ star. 
The amplitudes of the OsC mode we assume here to give $\delta L_r/L_r\approx 0.046$ and $\tau_c\ll\tau_{\rm MS}$ may be too high. 
If $\tau_c\lesssim\tau_{\rm MS}$
for observationally reasonable mode amplitudes, we expect that OsC modes work for braking the core rotation of massive main sequence stars.
Note that if the ratio $\Omega_c/\Omega_e$ decreases as a result of braking the core rotation, OsC modes will stop exciting envelope $g$-mode and hence braking the core rotation.
However, since the convective core continues to contract and hence 
the ratio $\Omega_c/\Omega_e$ tends to increase during the main sequence evolution, we expect that OsC modes will start again exciting envelope $g$-modes to brake the core.

The results discussed here for OsC modes are consistent with those obtained by \citet{Lee2021}, who used equation
(\ref{eq:lagmeanphi}) with ${\cal D}(\cdots)=0$.
Since ${\cal L}_r^W\approx{\cal L}_r^{\rm EP}$ around $r= R_c$ and also in the envelope, we think that the results obtained for OsC modes based on equation (\ref{eq:lagmeanphi}) are almost the same as those based on equation (\ref{eq:lagmeanphi_Lag}).

\section{conclusions}

Expanding the physical quantities of a rotating star to the
second order of the amplitude parameter $a$ and applying the azimuthal averaging to the quantities,
we have derived the set of linear partial differential equations governing 
the second order perturbations
where the parameter $a$ represents the typical amplitude of non-axisymmetric oscillation modes of the star and the zeroth- and second-order quantities are assumed axisymmetric.
The set of differential equations contain inhomogeneous terms which
are given by azimuthally averaged products of the eigenfunctions of the oscillation modes.
The solutions to the inhomogeneous set of linear partial differential equations may be given by the sum of 
complementary solutions and particular solutions.
Since complementary solutions are found decaying solutions with time,
we have calculated only the particular
solutions to the set of differential equations for $g$- and $r$-modes and for
OsC modes.
We have confirmed that adiabatic and stationary perturbations do not produce non-zero meridional
velocity fields of second order.
Carrying out non-adiabatic calculations for the first order and second order perturbations for a $4M_\odot$ main sequence model, we have found that $\partial v_\phi^{(2)}/\partial t>0$ for unstable prograde $g$-modes and $\partial v_\phi^{(2)}/\partial t<0$ for retrograde $r$-modes
in the surface equatorial regions of the stars.

Employing the transformed Eulerian mean theory for non-axisymmetric waves, 
we have derived the angular momentum conservation equation for waves, which may be represented by equation
 (\ref{eq:lagmeanphi}) or by equation (\ref{eq:lagmeanphi_Lag}).
Equation (\ref{eq:lagmeanphi}) suggests that the divergence of the Eliassen-Palm flux $\nabla\cdot\pmb{F}^{\rm EP}$ plays an important role in angular momentum transport by waves.
However, we have also found that
$\partial v_\phi^{(2)}/\partial t$ is not necessarily solely determined by $-\nabla\cdot\pmb{F}^{\rm EP}$.
Equation (\ref{eq:lagmeanphi_Lag}) gives another expression of angular momentum conservation by waves
and has a more compact form than equation (\ref{eq:lagmeanphi}).
For $g$-modes and $r$-modes excited by the opacity bump mechanism in rotating stars, we have argued that $r$-modes, which are
retrograde modes, transfer angular momentum from the deep interior to the surface while prograde $g$-modes transfer angular momentum from the surface layers
to the deep interior (e.g., \citealt{TownsendGoldsteinZweibel18}).

We have suggested that OsC modes can be an efficient mechanism for transporting angular momentum from the core to the surface layers and that the magnitudes of the torque supplied by the angular momentum transport can be large enough to support a decretion disc
around a rapidly rotating Be star.
This angular momentum extraction from the core to the envelope by OsC modes
can also be a mechanism to brake the core rotation of massive main sequence stars.

In this paper, regarding the steady state $Q^{(0)}$ of a rotating star as the zeroth order state,
we have derived the inhomogeneous set of linear partial differential equations for axisymmetric second order perturbations $Q^{(2)}$ and we have looked for the particular solutions driven by unstable linear modes where we have used zonal averaging to pick up the second order perturbations.
Applying the zonal averaging, however, we may also derive non-linear partial differential equations for mean flows as defined by $Q^{(0)}+Q^{(2)}$, but
the non-linear differential equations for the mean flows are definitely much more difficult to solve.
Probably, a quasi-linear approach in which $v_\phi^{(0)}+v_\phi^{(2)}$, instead of $v_\phi^{(2)}$, is treated as a dependent variable
can be a next step we may take (e.g., \citealt{Matsuno1971}).


\begin{appendix}

\section{Stokes drifts}

We may define the Eulerian perturbation $Q'$ and Lagrangian perturbation $\delta Q$
of a quantity $Q$
as
\be
Q'(\pmb{x},t)=Q(\pmb{x},t)-Q^{(0)}(\pmb{x}),
\ee
and 
\be
\delta Q(\pmb{x},t)=Q(\pmb{x}+\pmb{\xi}(\pmb{x},t),t)-Q^{(0)}(\pmb{x}),
\ee
where $Q(\pmb{x},t)$ and $Q^{(0)}(\pmb{x})$ represent the quantity in a perturbed state and in the equilibrium state, respectively,
and $\pmb{\xi}$ is the displacement vector connecting fluid elements in the equilibrium state
with those in the perturbed state.
Assuming the amplitudes of $\pmb{\xi}$ is small, we may expand the perturbed quantity $Q(\pmb{x},t)$ in terms of the small parameter $a\sim O(|\pmb{\xi}|)$, such that
\be
Q(\pmb{x},t)=Q^{(0)}(\pmb{x})+Q^{(1)}(\pmb{x},t)+Q^{(2)}(\pmb{x},t)+\cdots,
\ee
where the amplitudes of $Q^{(1)}$ and $Q^{(2)}$ are of order of $a$ and $a^2$, respectively.
The first order and second order Eulerian perturbations of $Q$ are simply given by $Q^{(1)}$ and $Q^{(2)}$, respectively.
The first order Lagrangian perturbation $\delta Q^{(1)}$ is
\begin{align}
\delta Q^{(1)}(\pmb{x},t)=Q^{(1)}(\pmb{x},t)+\pmb{\xi}\cdot\nabla Q^{(0)}(\pmb{x}),
\end{align}
and the difference between the first order Lagrangian and Eulerian perturbations is given by
\be
\delta Q^{(1)}-Q^{(1)}= \pmb{\xi}\cdot\nabla Q^{(0)}.
\ee
In this paper, the first order perturbations $Q^{(1)}$ and $\delta Q^{(1)}$ as well as the displacement $\pmb{\xi}$ are given by linear oscillation modes that satisfy the set of linearized equations.

The second order Lagrangian perturbation may be given by
\begin{align}
\delta Q^{(2)}(\pmb{x},t)=Q^{(2)}+ \pmb{\xi}\cdot\nabla Q^{(1)}+{1\over 2}\xi_i\xi_j{\nabla_j\nabla_iQ^{(0)}},
\end{align}
where $\nabla_j$ denotes the covariant derivative with respect to the coordinate $x^j$.
The difference between the Lagrangian and the Eulerian perturbation of second order
is given by
\be
\delta Q^{(2)}-Q^{(2)}=\pmb{\xi}\cdot\nabla Q^{(1)}+{1\over 2}\xi_i\xi_j{\nabla_j\nabla_iQ^{(0)}},
\ee
where the repeated indices imply summation over the indices.

Applying the zonal averaging to the second order Lagrangian perturbations, we obtain
\be
\overline{\delta Q^{(2)}}=\overline{Q^{(2)}}+\overline{\pmb{\xi}\cdot\nabla Q^{(1)}}+{1\over 2}\overline{\xi_i\xi_j}{\nabla_j\nabla_iQ^{(0)}},
\ee
and the difference $\overline{Q^{(2)}}^S\equiv\overline{\delta Q^{(2)}}-\overline{Q^{(2)}}$ is called the Stokes correction, that is,
\be
\overline{Q^{(2)}}^S
=\overline{\pmb{\xi}\cdot\nabla Q^{(1)}}+{1\over 2}\overline{\xi_i\xi_j}\nabla_j\nabla_iQ^{(0)}.
\ee
For spherical polar coordinates $(r,\theta,\phi)$, for example, we have
\be
\overline{\xi_i\xi_j}\nabla_j\nabla_iQ^{(0)}=\overline{\xi_r\xi_r}{\partial^2Q^{(0)}\over\partial r^2}
+\left(\overline{\xi_\theta\xi_\theta}+\overline{\xi_\phi\xi_\phi}\right){1\over r}{\partial Q^{(0)}\over\partial r},
\ee
where we have assumed $Q^{(0)}$ depends only on $r$.
The Stokes corrections can be calculated by using linear solutions of the perturbations.
To obtain $\overline{\delta Q^{(2)}}$, however, the quantity $\overline{Q^{(2)}}$ have to be computed by solving the inhomogeneous set of 
differential equations for the second order perturbations.

The Stokes correction to the velocity field $\pmb{v}(\pmb{x},t)$
is called the Stokes drift, which is given by
\be
\overline{\pmb{v}^{(2)}}^S=\overline{\pmb{\xi}\cdot\nabla\pmb{v}^{(1)}}+{1\over 2}\overline{\xi_i\xi_j}\nabla_i\nabla_j\pmb{v}^{(0)}.
\label{eq:StokesD00}
\ee
In spherical polar coordinates $(r,\theta,\phi)$, we have
\begin{align}
\overline{\pmb{\xi}\cdot\nabla\pmb{v}^{(1)}}
=\left(\overline{\pmb{\xi}\cdot\nabla v^{(1)}_r}-\overline{{v^{(1)}_\theta\xi_\theta\over r}}
-\overline{{v^{(1)}_\phi \xi_\phi\over r}}\right)\pmb{e}_r
+\left(\overline{\pmb{\xi}\cdot\nabla v^{(1)}_\theta}+\overline{{v^{(1)}_r\xi_\theta\over r}}-\overline{{v^{(1)}_\phi\xi_\phi\cot\theta\over r}}\right)\pmb {e}_\theta
+\left(\overline{\pmb{\xi}\cdot\nabla v^{(1)}_\phi}+\overline{{v^{(1)}_r\xi_\phi\over r}}+\overline{{v^{(1)}_\theta\xi_\phi\cot\theta\over r}}\right)\pmb{ e}_\phi,
\label{eq:StokesD}
\end{align}
and if we write ${1\over 2}\sum_{i,j}\overline{\xi_i\xi_j}\nabla_i\nabla_j\pmb{v}^{(0)}={1\over 2}\sum_{i,j,k}\overline{\xi_i\xi_j}v^k_{ij}\pmb{e}_k$, the non-zero elements of the tensor $v^k_{ij}$ for $\pmb{v}^{(0)}=v_\phi^{(0)}\pmb{e}_\phi$ with $v_\phi^{(0)}=r\sin\theta\Omega$ are
\be
v^r_{r\phi}=-{1\over r}\left({\cal P}-2{v_\phi^{(0)}\over r}\right)=-\sin\theta{\partial\Omega\over\partial r}, \quad v^r_{\theta\phi}=-{1\over r}\left({\cal Q}-2\cot\theta{v_\phi^{(0)}\over r}\right)=-\sin\theta{1\over r}{\partial\Omega\over\partial\theta},
\ee
\be
v^\theta_{r\phi}=-{\cot\theta\over r}\left({\cal P}-2{v_\phi^{(0)}\over r}\right)=-\cos\theta{\partial\Omega\over\partial r}, \quad
v^\theta_{\theta\phi}=-{\cot\theta\over r}\left({\cal Q}-2\cot\theta{v_\phi^{(0)}\over r}\right)
=-\cos\theta{1\over r}{\partial\Omega\over\partial\theta},
\ee
\be
v^\phi_{rr}={\partial{\cal P}\over\partial r}-{{\cal P}\over r}+2{v_\phi^{(0)}\over r^2}, \quad
v^\phi_{r\theta}={1\over r}{\partial{\cal P}\over\partial\theta}-2{{\cal Q}\over r}+2\cot\theta{v_\phi^{(0)}\over r^2}, 
\ee
\be
v^\phi_{\theta\theta}={1\over r}{\partial{\cal Q}\over\partial \theta}+{{\cal P}\over r}-\cot\theta{{\cal Q}\over r}+2\cot^2\theta{v_\phi^{(0)}\over r^2}, \quad
v^\phi_{\phi\phi}={{\cal P}\over r}+\cot\theta{{\cal Q}\over r}-2{v_\phi^{(0)}\over r^2\sin^2\theta}.
\ee
Note that $v^k_{ij}=v^k_{ji}$ and that $v^k_{ij}$ vanishes for uniform rotation, for which ${\cal P}=2\sin\theta\Omega$ and ${\cal Q}=2\cos\theta\Omega$ with constant $\Omega$.
We also note that
\be
{1\over 2}\overline{\xi_i\xi_j}v^r_{ij}=-\sin\theta\overline{\xi_\phi(\pmb{\xi}\cdot\nabla\Omega)}, \quad
{1\over 2}\overline{\xi_i\xi_j}v^\theta_{ij}=-\cos\theta\overline{\xi_\phi(\pmb{\xi}\cdot\nabla\Omega)}.
\ee
Using equation (\ref{eq:vdxi}), we rewrite the three components of $\overline{\pmb{ v}^{(2)}}^S$ as
\begin{align}
\overline{v_r^{(2)}}^S
={1\over\rho^{(0)}}\nabla\cdot(\rho^{(0)}\overline{v_r^{(1)}\pmb{\xi})}
+{1\over\rho^{(0)}}\overline{\rho^{(1)}v_r^{(1)}}
-\overline{v_\theta^{(1)}\xi_\theta\over r}-\overline{v_\phi^{(1)}\xi_\phi\over r}
-\sin\theta\overline{\xi_\phi(\pmb{\xi}\cdot\nabla\Omega)}
=\widetilde v_r+{1\over\rho^{(0)}}\overline{\rho^{(1)}v_r^{(1)}}+{\cal D}H_r,
\label{eq:vr_Stokes}
\end{align}
\begin{align}
\overline{v_\theta^{(2)}}^S
={1\over\rho^{(0)}}\nabla\cdot(\rho^{(0)}\overline{v_\theta^{(1)}\pmb{\xi})}
+{1\over\rho^{(0)}}\overline{\rho^{(1)}v_\theta^{(1)}}
+\overline{v_r^{(1)}\xi_\theta\over r}-\overline{v_\phi^{(1)}\xi_\phi\over r}\cot\theta
-\cos\theta\overline{\xi_\phi(\pmb{\xi}\cdot\nabla\Omega)}
=\widetilde v_\theta+{1\over\rho^{(0)}}\overline{\rho^{(1)}v_\theta^{(1)}}+{\cal D}H_\theta,
\label{eq:vtheta_Stokes}
\end{align}
\begin{align}
\overline{v_\phi^{(2)}}^S={1\over\rho^{(0)}}\nabla\cdot(\rho^{(0)}\overline{v_\phi^{(1)}\pmb{\xi})}
+{1\over\rho^{(0)}}\overline{\rho^{(1)}v_\phi^{(1)}}
+{\overline{v_r^{(1)}\xi_\phi}\over r}+{\overline{v_\theta^{(1)}\xi_\phi}\over r}\cot\theta+{1\over 2}\overline{\xi_i\xi_j}v_{ij}^\phi={1\over\rho^{(0)}}\overline{\rho^{(1)}v_\phi^{(1)}}+H_\phi,
\label{eq:vphi_Stokes}
\end{align}
where $\widetilde v_r$ and $\widetilde v_\theta$ are given by equation (\ref{eq:vtilde}) for the stream function (\ref{eq:aphi_svtheta}), and
\be
H_r={1\over\rho^{(0)}}\left({1\over r^2}{\partial\over\partial r}r^2\rho^{(0)}{\overline{\xi_r\xi_r}\over 2}
+{1\over r\sin\theta}{\partial\over\partial\theta}\sin\theta\rho^{(0)}\overline{\xi_r\xi_\theta}\right)
-{\overline{\xi_\theta\xi_\theta}\over 2r}-{\overline{\xi_\phi\xi_\phi}\over 2r},
\label{eq:hr}
\ee
\be
H_\theta={1\over\rho^{(0)}}{1\over r\sin\theta}{\partial\over\partial\theta}\sin\theta\rho^{(0)}{\overline{\xi_\theta\xi_\theta} \over 2}
+{\overline{\xi_r\xi_\theta}\over r}-{\overline{\xi_\phi\xi_\phi}\over 2r}\cot\theta,
\label{eq:htheta}
\ee
\begin{align}
H_\phi={1\over\rho^{(0)}}\left({1\over r^2}{\partial\over\partial r}r^2\rho^{(0)}\overline{\xi_r v_\phi^{(1)}}+{1\over r\sin\theta}{\partial\over\partial\theta}\sin\theta\rho^{(0)}\overline{\xi_\theta v_\phi^{(1)}}\right)+{\overline{v_r^{(1)}\xi_\phi}\over r}+
{\overline{v_\theta^{(1)}\xi_\phi}\over r}\cot\theta+{1\over 2}\overline{\xi_i\xi_j}v_{ij}^\phi.
\label{eq:hphi}
\end{align}
Note that
\be
v_r^{(2)}+\overline{v_r^{(2)}}^S=\hat v_r+{1\over\rho^{(0)}}\overline{\rho^{(1)}v_r^{(1)}}+{\cal D}H_r,\quad
v_\theta^{(2)}+\overline{v_\theta^{(2)}}^S=\hat v_\theta+{1\over\rho^{(0)}}\overline{\rho^{(1)}v_\theta^{(1)}}+{\cal D}H_\theta,
\ee
where $\hat v_r$ and $\hat v_\theta$ are defined by equation (\ref{eq:vhat}).
For steady perturbations for which $\omega_{\rm I}=0$ and ${\cal D}\overline{a'b'}=0$, we obtain
\begin{align}
\overline{v_r^{(2)}}^S
=-{1\over r\sin\theta}{\partial\over\partial\theta}\sin\theta 
\overline{{\xi_r}v_\theta^{(1)}}+{1\over\rho^{(0)}}\overline{\rho^{(1)}v_r^{(1)}}, 
\end{align}
\begin{align}
\overline{v_\theta^{(2)}}^S={1\over\rho^{(0)}}{1\over r}{\partial\over\partial r}r\rho^{(0)}\overline{v_\theta^{(1)}\xi_r}+{1\over\rho^{(0)}}\overline{\rho^{(1)}v_\theta^{(1)}}.
\end{align}
If we further assume the anelastic approximation $\rho^{(1)}=-\nabla\cdot(\rho^{(0)}\pmb{\xi})=0$, the meridional velocity field defined by
\be
\overline{\pmb{v}^{(2)}_M}^S=\overline{v_r^{(2)}}^S\pmb{e}_r+\overline{v_\theta^{(2)}}^S\pmb{e}_\theta,
\ee
reduces to $\widetilde{\pmb{v}}=\widetilde v_r\pmb{e}_r+\widetilde v_\theta\pmb{e}_\theta$ and satisfies
$
\nabla\cdot\left(\rho^{(0)}\overline{\pmb{v}^{(2)}_M}^S\right)=0.
$

\end{appendix}


\bibliographystyle{mnras}
\bibliography{myref}

\end{document}